\documentclass[reprint,amsmath,amssymb,aip]{revtex4-2}
\usepackage{graphicx}
\usepackage{dcolumn}
\usepackage{bm}
\usepackage[table]{xcolor} 
\usepackage{braket}
\usepackage[normalem]{ulem}
\renewcommand\bra[1]{{\langle{#1}|}}
\renewcommand\ket[1]{{|{#1}\rangle}}

\def \be {\begin{equation}}
	\def \ee {\end{equation}}
\def \ben {\begin{eqnarray}}
	\def \een {\end{eqnarray}}
\def \bi {\begin{itemize}}
	\def \ei {\end{itemize}}
\def\re#1{{\color{blue} {#1}}}
\def\rc#1{{\color{black} {#1}}}
\usepackage{multirow}
\usepackage{ctable}

\definecolor{dkgreen}{rgb}{0,0.6,0}
\definecolor{gray}{rgb}{0.5,0.5,0.5}
\definecolor{mauve}{rgb}{0.58,0,0.82}
\usepackage{listings}

\graphicspath{./Figures}
\begin{document}
	\bibliographystyle{prsty}
	
	\title{A simple fourth order propagator based on the Magnus expansion in the Liouville space: Application to a $\Lambda$-system and assessment of the rotating wave approximation}
	\author{Taner M. Ture}
	\affiliation{Department of Chemistry and Biochemistry, Queens College, City University of New York, 65-30 Kissena Boulevard, Queens, New York 11367, USA \& PhD Programs in Chemistry and Physics, Graduate Center of the City University of New York, New York, New York 10016, USA}
	\author{Changbong Hyeon} 
	\email{hyeoncb@kias.re.kr}
	\affiliation{Korea Institute for Advanced Study, Hoegiro 85, Dongdaemun-gu, Seoul 02455, South Korea}
	\author{Seogjoo J. Jang}
	\email[ ]{seogjoo.jang@qc.cuny.edu}
	\affiliation{Department of Chemistry and Biochemistry, Queens College, City University of New York, 65-30 Kissena Boulevard, Queens, New York 11367, USA \& PhD Programs in Chemistry and Physics, Graduate Center of the City University of New York, New York 10016, USA}
	\affiliation{Korea Institute for Advanced Study, Hoegiro 85, Dongdaemun-gu, Seoul 02455, Korea}	
	
	\date{Accepted for publication in {\it the Journal of Chemical Physics} on January 19, 2026}
	
	\begin{abstract}
		A simple 4th order propagator [Ture and Jang, {\it J. Phys. Chem. A.} {\bf 128}, 2871 (2024)] based on the Magnus expansion (ME)  is extended to the Liouville space for both closed-system and Lindbladian open-system  quantum dynamics.  For both dynamics, commutator free versions of 4th order propagators are provided as well. These propagators are then applied to the dynamics of a driven $\Lambda$-system, where Lindblad terms represent the effect of a photonic bath.  For both dynamics, the accuracy of the rotating wave approximation (RWA) for the matter-radiation interaction is assessed.  We confirmed reasonable performance of RWA for weak and resonant fields. However, small errors appear for moderate fields and substantial errors can be found for strong fields where coherent population trapping can still be expected.  We also found that the presence of bath for open system quantum dynamics consistently reduces the errors of the RWA.  These results provide a quantitative information on how the RWA breaks down beyond weak field or for non-resonant cases.  Major results are benchmarked against results of our 6th order ME-based propagator.  We also provide numerical comparison of our algorithms with other 4th order algorithms for the $\Lambda$-system. These confirm reasonable performance of our simple propagators and the improvement gained through commutator-free expressions. 	
	\end{abstract}
	
	\maketitle
	
	\section{Introduction}
Many quantum calculations involve time evolution of states or observables under time dependent Hamiltonian by nature or design. Well-known examples include nuclear magnetic resonance,\cite{slichter,mananga-pr609,ganguly-jcp159,giscard-prr2} quantum control (QC),\cite{shapiro-qc,bergmann-rmp70,singh-jcp150,hu-jcp152,cao-prl80,jang_con}  \rc{quantum thermodynamics,\cite{gemmer,hasegawa-prl126,hasegawa-ncomms14,xu-njp18}} the time-dependent Kohn-Sham equation,\cite{castro-jcp121,goings-wcms2017} and more recently quantum computation\cite{watkins-prx5,cao-prr7,berry-quantum4,nghiem-pra112} and quantum sensing (QS).\cite{degen-rmp89-2017,pang-nc8,herb-prl133,zhou-prl131}   Recent advances in the formulation of driven open system quantum dynamics,\cite{grifoni-pr304,xu-njp18,jang-jcp157,jang-exciton,montoya-castillo-jcp143,mulvihill-jcp156,devega-rmp89} also brings the possibility that the time dependences are not only limited to Hamiltonians but can also be incorporated into the terms involving relaxation and decoherence. 
However, except for very few model systems and those with periodic time dependence, analytical solutions are not available when the system has general time dependence.  For these cases, numerical solutions are inevitable. 

Although there are many well established methods that can be used for the numerical integration of quantum  evolution, most of them have been originally developed for time independent cases, making them ill-suited or untested for  cases with general time dependence.
Propagators based on the Magnus expansion (ME)\cite{magnus-cpam7,pechukas-jcp44,burum-prb24,prato-jcp106,blanes-pr470,ture-jang-jpca124} are expected to be advantageous in this respect since ME by construction preserves important formal properties of quantum dynamics for each term.  By assuming an exponential form for the propagator and rewriting the differential equation in terms of the exponent, the derived series is guaranteed to remain within the Lie algebra at any order of truncation.\cite{magnus-cpam7,Iserles-acn9} This property guarantees that any finite truncation of ME preserves the unitarity of the original dynamics.   Indeed, the merit of ME has enabled successful applications\cite{ture-jang-jpca124,mananga-pr609,singh-jcp150,burum-prb24,prato-jcp106,blanes-pr470} including the cases with stochastic fluctuations.\cite{blanes-pr470,burrage-physica-d133,lord-sjna46,sun-cpl735}  
	
Our major focus here is application of ME propagators for driven open-system quantum dynamics using the Lindblad equation,\cite{lindblad-cmp48} for which the dynamics is no longer unitary and there have been relatively few studies.\cite{begzjav-rp17,alvermann-njp14} The Lindblad equation satisfies the semi-group property, guaranteeing  a completely positive trace-preserving (CPTP) map. On the other hand, it is known that only the leading term of the ME defines a CPTP map, while subsequent higher order terms containing commutators break the CPTP map.\cite{mizuta-pra103,Blanes-imajna41}    In this situation, the commutator free Magnus Expansion (CFME)\cite{blanes-anm56,thalhammer-SIAMjna44,alvermann-jcp230,blanes-cpc220} is an interesting alternative.  Thus, at the cost of increasing the number of exponentials per time step, the CFME is able to produce more satisfactory commutator-free higher order propagators.  Indeed, applications\cite{alvermann-njp14} of an optimized fourth order CFME to driven Lindblad equations have shown promising outcomes.   We here provide similar extensions based on our recently developed fourth order ME propagator that utilizes minimal discrete points.\cite{ture-jang-jpca124}
	
As the model system, we here consider a well-known $\Lambda$-system, which  represents a simple three state system consisting of an excited state radiatively coupled to two low lying states that are not directly coupled  with each other. 
The $\Lambda$-system has long served as a prototypical model for the description of  laser driven quantum dynamical processes.  Well known examples include QC,\cite{shapiro-qc,bergmann-rmp70,vasilev-pra80} coherent population trapping (CPT),\cite{bergmann-rmp70,gray-ol3,fu-prl95} and electromagnetically induced transparency (EIT).\cite{bolier-prl66,fleischhauer-pra65,ma-jo19,carreo-pra71,abbumalikov-prl104,singh-jpa56}  More recently, $\Lambda$-system or extended versions have also been used as important model systems for QS,\cite{degen-rmp89-2017,zhang-oe29,rembold-avs-qs2} for which more complicated pulses or time dependent fields are needed in general.  \rc{Many of these studies have relied on the rotating wave approximation (RWA), which has been well tested for weak matter-radiation interaction and resonant conditions.  However, beyond these well-established regimes, the errors of RWA are poorly understood even for the $\Lambda$-system.  Thus, we provide an assessment of RWA in this work.}

The sections are organized as follows.  Section II describes details of our main numerical methods.  Section III presents the model and summarizes the RWA.    Section IV provides results of calculation and analyses. 	The paper concludes with discussion in Sec. V.

\section{Numerical propagation based on Magnus expansion method \label{method}}
Considering the importance of open system dynamics under time-dependent Hamiltonian, either arising from an external driving or from noisy environments, extension of ME-based methods for open system dynamics have significant implications.  To this end, we first review a simple fourth order method based on ME, which was derived recently,\cite{ture-jang-jpca124} and extend it to the application for an open system dynamics described by the Lindblad equation.\cite{lindblad-cmp48} 

\subsection{Unitary dynamics in the Hilbert space}
In a recent work,\cite{ture-jang-jpca124} we have developed new ME-based fourth and sixth order expressions for the real time propagator of  general time dependent Hamiltonian.  In particular, we identified a simple fourth order algorithm that involves evaluation of the Hamiltonian only at one mid-point in addition to two end points.  

For a closed system governed by a time dependent Hamiltonian operator $\hat H(t)$, the corresponding time evolution operator $\hat U(t,0)$  is obtained by solving the following time dependent  Schr\"{o}dinger equation: 
	\be
	\frac{\partial}{\partial t}\hat U(t,0)=-\frac{i}{\hbar}\hat H (t) \hat U(t,0) ,
	\label{eq:schrodinger}
	\ee
where $\hat U(0,0)=\hat I$, the identity operator of the system Hilbert space.  
Introducing $\hat H_k=\hat H(t_k)$, our simplest fourth order approximation\cite{ture-jang-jpca124} for $\hat U(t)$ is expressed as
	\ben
	\hat U^{(4)}(t_{k+1},t_k) &=& \exp \left \{ -\frac{i}{\hbar} \frac{\delta t}{6} (\hat H_k+4\hat H_{k+1/2}+ \hat H_{k+1}) \right .  \nonumber \\
	&&\hspace{.3in}\left . -\frac{\delta t^2}{12\hbar^2} [\hat H_{k+1},\hat H_k] \right\} , \label{eq:u-fourth}
	\een
where $\delta t=t_{k+1}-t_k$.  

In Eq. (\ref{eq:u-fourth}), the last term in the exponent involves a commutator, which introduces additional numerical errors. It is possible to remove this at the expense of increasing the number of exponentials, employing the Baker-Campbell-Hausdorff formula,\cite{alvermann-jcp230,blanes-cpc220} \rc{as described  in \rc{App. }\ref{sec:cfme-der}.}
The resulting CFME version of Eq. (\ref{eq:u-fourth}) is 
\rc{
\ben
&&\hat U^{(4)}_{_{\rm CF}}(t_{k+1},t_k) = \exp\Bigl \{-\frac{i \delta t}{12\hbar}\left(3
\hat H_{k+1}+4\hat H_{k+1/2}-\hat H_k\right ) \Bigl\} \nonumber \\
&&\hspace{.6in}\times \exp\Bigl \{\frac{i \delta t}{12 \hbar}\left (\hat  H_{k+1} -4\hat H_{k+1/2}-3\hat H_{k}\right ) \Bigl \} . \label{eq:cfme-h}
\een}

Note that our expressions, Eqs. (\ref{eq:u-fourth}) and (\ref{eq:cfme-h}), satisfy both time-reversal symmetry and the positivity condition, which produces a stable scheme.\cite{blanes-imajna38} They also preserve the CPTP property of the original unitary dynamics.  In addition, note that our expressions are for equally spaced points including the boundary terms, which are more convenient  to use widely than those using Gauss-Legendre points inside the time interval,\cite{blanes-pr470,alvermann-jcp230} in that our method does not require extra calculation for the evaluation of physical quantities at each time step.  

\subsection{Dynamics in the Liouville space}
We here extend the  ME-based propagators to the Liouville space, for the following  open system dynamics of Lindblad form:\cite{lindblad-cmp48}
\be
\frac{d}{dt}\hat{\rho}(t) = -i{\mathcal L} (t)\hat{\rho} (t) +{\mathcal D}_L \hat \rho (t) ,  \label{eq:evol-rhos}
\ee
where $\hat \rho(t)$ is the density operator of the system, ${\mathcal L}(t)$ is the Liouville super-operator for the unitary dynamics, ${\mathcal L}(t)(\cdot) = [\hat{H}(t), (\cdot)]$,  and ${\mathcal D}_L$ is a Lindbladian that accounts for the interaction with the environment.   Thus, we assume the following generic form:
\be
{\mathcal D}_L \hat \rho(t)  =\sum_j c_j \left (\hat L_j  \hat{\rho}(t) \hat L_j^\dagger -\frac{1}{2} \left \{\hat L_j^\dagger\hat L_j,\hat \rho(t) \right\}_+ \right ) , \label{eq:lindblad-1}
\ee
where $c_j$ is a real positive number, $\hat L_j$ is an operator representing a certain transition, and the subscript $+$ represents the anticommutator.   

For numerical propagation, it is convenient to propagate the density operator  in the Liouville space, where it is viewed as a vector.  
The explicit form of ${\mathcal L}(t)$ in the Liouville space can be found from the corresponding definition in the Hilbert space  by using the superket triple product identity,\cite{gyamfi-ejp41} as defined in App. \ref{sec:diff_propagators}, and can be expressed as
\be
\mathcal{L}(t) = \frac{1}{\hbar}\left(\hat{H}(t)\otimes \hat I-\hat I\otimes \hat{H}(t)^T\right), 
\label{eq:closed_Liouville}
\ee
where $\hat I$ is the identity operator in the Hilbert space of the system.  Thus, in the absence of the Lindbladian, the  density operator in the Liouville space at time $t$ is given by 
\be
|\rho(t) \rangle\rangle= {\mathcal U} (t,0)|\rho(0)\rangle\rangle=e_{(+)}^{-i\int_0^{t}\mathcal{L}(\tau)d\tau}|\rho(0)\rangle\rangle, 
\ee
where the second equality defines the time-evolution superoperator $\mathcal{U}(t,0)$ in terms of $\mathcal{L}(\tau)$ given by Eq. (\ref{eq:closed_Liouville}).
Appendix \ref{sec:positive_definite} also provides a proof that any finite truncation of ME for the Liouvillian in the Liouville space corresponds to a CPTP map. 

For ${\mathcal U}(t,0)$, the extension of Eq. (\ref{eq:u-fourth}), namely our simplest 4th order propagator\cite{ture-jang-jpca124} based on ME, is 
	\ben
	{\mathcal U}^{(4)}(t_{k+1},t_k) &=& \exp\left\{-i\frac{\delta t}{6}\left (\mathcal{L}_{k}+4\mathcal{L}_{k+1/2} + \mathcal{L}_{k+1} \right) \right . \nonumber \\ 
	&&\hspace{.2in}\left . - \frac{\delta t^2}{12}\left[\mathcal{L}_{k+1},\mathcal{L}_{k}\right]\right\} , \label{eq:us-4th}
	\een
where ${\mathcal L}_{k}={\mathcal L}(t_k)$.  The above expression is equivalent to using Eq. (\ref{eq:u-fourth}) in the Hilbert space.   The commutator free version of the above propagator has the same expression as Eq. (\ref{eq:cfme-h}) except that \rc{$\hat H_{k}$} is replaced with ${\mathcal L}_{k}$. 

For the Lindbladian given by \rc{Eq. (\ref{eq:lindblad-1})}, the expression in the Liouville space is 
\be
{\mathcal D}_L = \sum_j c_j\left ( \hat L_j\otimes \hat L^*_j-\frac{1}{2}\left(\hat L^\dagger_j \hat L_j\otimes \hat I + \hat I \otimes \hat L^T_j\hat L^*_j \right)\right) .
\label{eq:open_Liouville}
\ee
With this expression, the time-evolution of the system density operator vector in the Liouville space  is expressed as 
\be
|\rho(t)\rangle\rangle = {\mathcal U}_{L} (t,0)|\rho(0)\rangle\rangle=e_{(+)}^{\int_0^{t}(-i\mathcal{L}(\tau) + {\mathcal D}_L)d\tau}|\rho(0)\rangle\rangle , \label{eq:superunitary}
\ee
where the second equality serves as the definition of ${\mathcal U}_{L} (t,0)$.  Although $\mathcal{U}_{L}$ plays the role of the propagator in Eq. (\ref{eq:superunitary}) and is therefore analogous to the unitary operators $\hat{U}$ and $\mathcal{U}$, it is not unitary.  As yet, the 4th order propagator given by  Eq. (\ref{eq:us-4th}) can still be generalized to this case without affecting conditions validating the accuracy up to the 4th order.  Thus, we obtain 
	\ben
	&&{\mathcal U}_{L}^{(4)}(t_{k+1},t_k) = \exp\biggl\{-i\frac{\delta t}{6}\left (\mathcal{L}_{k}+4\mathcal{L}_{k+1/2} +
	\mathcal{L}_{k+1}\right) \nonumber \\ 
	&&\hspace{.4in} +\delta t \mathcal{D}_L-\frac{\delta t^2}{12}\left[\mathcal{L}_{k+1}-i\mathcal{D}_L,\mathcal{L}_{k}-i\mathcal{D}_L \right]\biggr\} .  \label{eq:usl-4th}
	\een
By going through a procedure similar to that for obtaining Eq. (\ref{eq:cfme-h}), it is straightforward to obtain a commutator free-version of the above expression as follows:
\ben
&&{\mathcal U}_{L}^{\rm CF (4)}(t_{k+1},t_k) \nonumber \\
&&= \exp\Bigl \{-\frac{i \delta t}{12} \bigl(3\mathcal{L}_{k+1}+4\mathcal{L}_{k+1/2} -\mathcal{L}_{k} + 6i\mathcal{D}_L\bigr) \Bigl\} \nonumber \\ 
&&\times \exp\Bigl \{\frac{i \delta t}{12}\left ( \mathcal{L}_{k+1} -4\mathcal{L}_{k+1/2}-3\mathcal{L}_{k} -6i\mathcal{D}_L\right )\Bigl\}. 
	\label{eq:usl-4th-cf}
\een

Equations (\ref{eq:usl-4th}) and (\ref{eq:usl-4th-cf}) represent main results of this work.  Although these are for non-unitary Lindblad dynamics, their apparent simplicity and symmetry make these attractive numerical methods.  In the next section, we provide applications and tests of these to a driven $\Lambda$-system that has served as a prototypical model for EIT.   
	
\section{Applications to $\Lambda$-system}
\subsection{Model Hamiltonian and Lindbladian} 	
\begin{figure}
	
\includegraphics[width=0.95\columnwidth]{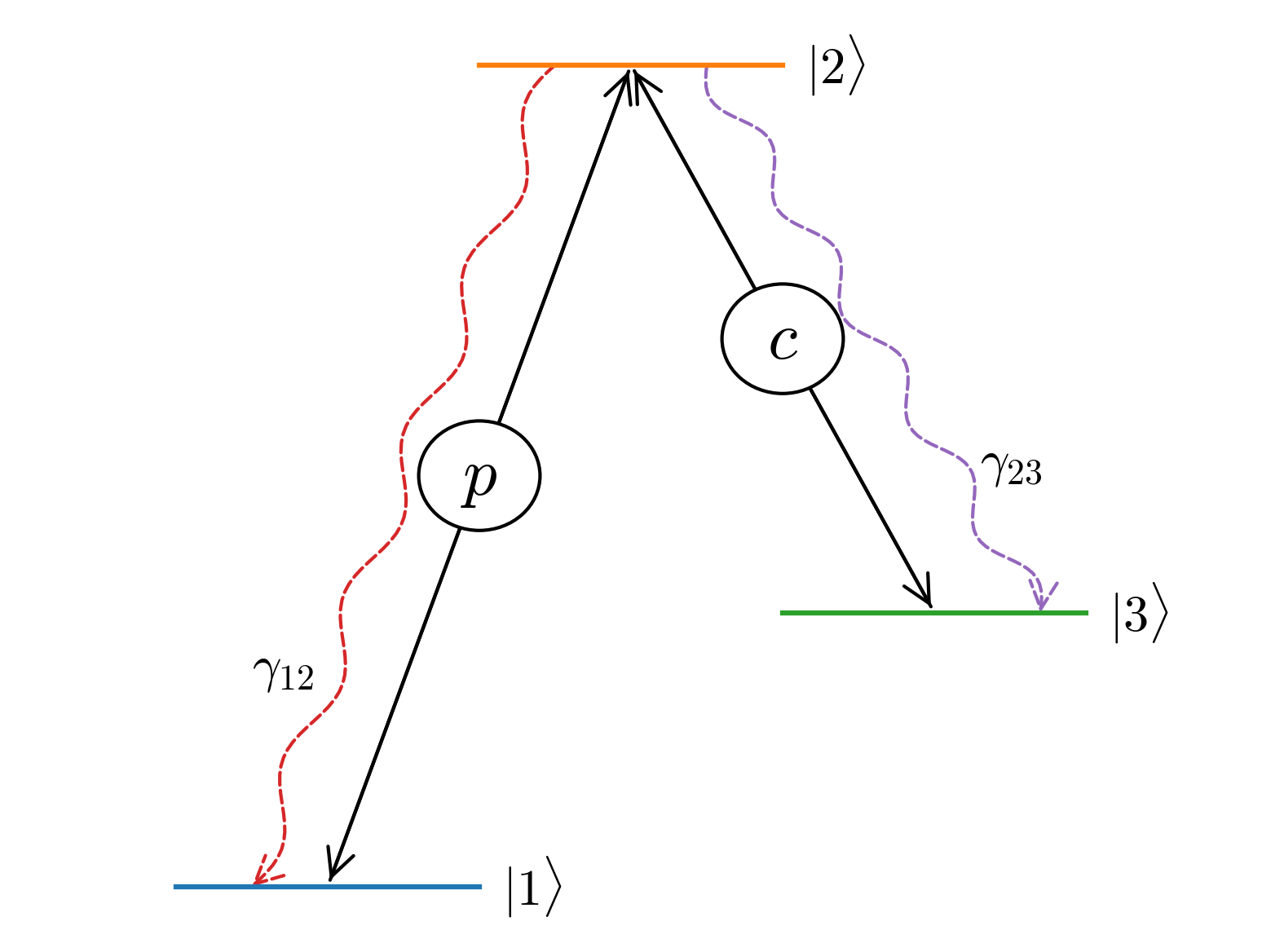} 
\caption{Depiction of a three-level $\Lambda$-system. Solid lines represent coupling between states due to radiation fields while dashed lines represent spontaneous emission from the excited state. The probe (control) pulse coupling states $|1\rangle$ ($|3\rangle$) and $|2\rangle$ is parameterized by $\omega_p,\Omega_p$ ($\omega_c,\Omega_c$). }
	\label{fig:lambda}
\end{figure}

The $\Lambda$-system consists of three orthogonal states $|1\rangle$, $|2\rangle$, and $|3\rangle$ interacting through only radiation. The total system Hamiltonian is $\hat H(t)=\hat H_0+\hat H_{mr}(t)$.  The zeroth order term $\hat H_0$ is defined as\footnote{We use a labeling convention of the $\Lambda$-system states that are more commonly used for quantum control and quantum sensing.} 
	\be
	\hat H_0= E_1|1\rangle\langle 1|+E_2|2\rangle\langle 2 |+E_3|3\rangle\langle 3| , \label{eq:hs0}
	\ee
where $|2\rangle$ has the highest energy. Thus, the assumption of a $\Lambda$-system implies that $|E_2-E_1|, |E_2-E_3| > |E_1-E_3|$ (see Figure \ref{fig:lambda}).    
The matter-radiation term $\hat H_{mr}(t)$ represents the coupling between the state $|2\rangle$ with states $|1\rangle$  and $|3\rangle$, and has the following expression: 
		\ben
		\hat H_{mr}(t)&=&-\hbar\Omega_pf_p(t)\cos(w_pt)\left (|2\rangle \langle 1|+|1\rangle\langle 2|\right ) \nonumber \\
		&& -\hbar\Omega_cf_c(t)\cos(w_ct) \left (|2\rangle \langle 3|+|3\rangle\langle 2|\right ) , 
		\een
where  $\omega_p$ and $\omega_c$ are frequencies of probe and control fields, $\Omega_p$ and $\Omega_c$ are their respective Rabi frequencies, and  $f_p(t)$ and $f_c(t)$ are envelope functions. We here assume the simplest case of unit envelope functions, namely,  $f_p(t)=f_c(t)=1$.  It is also assumed here that matter-radiation terms not shown in the above Hamiltonian, {\it i.e.}, the coupling of the probe field to the transition between $|2\rangle$ and $|3\rangle$ and that of the control field between $|1\rangle$ and $|2\rangle$, are precluded through either spin selection rules and/or experimental design of polarization directions.  

In the presence of dissipation, the dynamics is no longer unitary and it is in general necessary to consider the time evolution at the level of a density operator.   Under the assumption that all the dissipations are caused by weak and Markovian baths, the effects can be accounted for by a Lindblad equation.\cite{lindblad-cmp48}    In a recent work,\cite{singh-jpa56} such a Lindblad equation was used for the study of EIT.  The corresponding time evolution equation for the system density operator $\hat \rho(t)$ is given by Eq. (\ref{eq:evol-rhos}), with the following definition for the Lindbladian: 
\ben
&&{\mathcal D}_L \hat \rho(t)\nonumber \\
&&\hspace{.1in}=\gamma_{12}(\tilde{n}_{12}+1)\left(\ket{1}\bra{2}\hat{\rho}(t)\ket{2}\bra{1}-\frac{1}{2}\{\ket{2}\bra{2},\hat{\rho}(t)\}_{+}\right) \nonumber \\ 
&&\hspace{.2in}+\gamma_{12}\tilde{n}_{12}\left(\ket{2}\bra{1}\hat{\rho}(t)\ket{1}\bra{2}-\frac{1}{2}\{\ket{1}\bra{1},\hat{\rho}(t)\}_{+}\right) \nonumber \\ 
&&\hspace{.2in}+\gamma_{23}(\tilde{n}_{23}+1)\Bigl( \ket{3}\bra{2}\hat{\rho}(t)\ket{2}\bra{3}-\frac{1}{2}\{\ket{2}\bra{2},\hat{\rho}(t)\}_{+}\Bigr) \nonumber \\ 
&&\hspace{.2in}+\gamma_{23}\tilde{n}_{23}\left(\ket{2}\bra{3}\hat{\rho}(t)\ket{3}\bra{2}-\frac{1}{2}\{\ket{3}\bra{3},\hat{\rho}(t)\}_{+}\right) . \label{eq:lindblad} 
\een

\subsection{Rotating wave approximation }		
Most of calculations and experimental interpretations involving the $\Lambda$-system were made \rc{under the RWA}, within which $\hat H_{mr}(t)$ is approximated as
\ben
&&\hat H_{mr}^{{\small \rm RWA}}(t)=-\frac{\hbar\Omega_p}{2}f_p(t)\left (e^{-i\omega_p t }|2\rangle \langle 1|+e^{i\omega_p t} |1\rangle\langle 2|\right ) \nonumber \\
&&\hspace{.5in} -\frac{\hbar\Omega_c}{2}  f_c(t)\left ( e^{-i\omega_c t} |2 \rangle \langle 3|+e^{i\omega_c t} |3\rangle\langle 2|\right ) . \label{eq:hsr-rwa}
	\een
We denote the time evolution operator corresponding to $\hat H^{\rm \small RWA}(t)=\hat H_0+\hat H_{mr}^{\rm \small RWA}(t)$ as $\hat U^{\rm \small RWA}(t,0)$.  

The RWA allows obtaining closed form expressions for eigenvalues of the Hamiltonian in the rotating wave frame (RWF), as detailed in Appendixes \ref{sec:rotating_frame} and \ref{RWA-solution}, and has been used widely. 
It is widely accepted that the RWA is reasonable for weak and resonant condition. For the steady state case of $f_p(t)=f_c(t)=1$ considered here, this is easy to understand since the terms omitted in the above expression, namely, the counter-rotating terms, can be ignored as long as $\Omega_{\gamma} \ll \omega_{\gamma}$, where $\gamma=p$ or $c$, as follows:
		\be
		\left |\Omega_{\gamma} \int^{1/2\Omega_{\gamma}}_{-1/2\Omega_{\gamma}} dt e^{\pm 2i\omega_{\gamma} t} \right | =\left |\frac{\sin (\omega_{\gamma}/\Omega_{\gamma})}{\omega_{\gamma}/\Omega_{\gamma}} \right |  \leq \frac{\Omega_{\gamma}}{\omega_{\gamma}} \ll 1 . \label{eq:rwa-jst}
		\ee
However, for intermediate and strong coupling strengths, the above condition breaks down and  the RWA is expected to become inaccurate.  
\rc{While this is expected, quantitative details of how such breakdown occurs are not clearly known even for the simple $\Lambda$-system considered here. Thus, we here provide such an assessment employing our ME-based propagators. }
 
For  \rc{a quantitative assessment of the accuracy of the RWA}, we employ the following measure of error:
\be
\mbox{error}=\frac{|{\hat \rho }^{\rm RWA}(t)-{\hat \rho}(t)|_F}{|{\hat \rho}(t)|_F} , \label{eq:error_frob}
\ee
where ${\hat \rho }^{\rm RWA}(t)$  is the system density operator \rc{that evolves according to} the RWA Hamiltonian, Eq. (\ref{eq:hsr-rwa}).  \rc{The subscript $F$ in Eq. (\ref{eq:error_frob})} denotes the Frobenius measure \rc{, defined as $|\hat A|_F=\sqrt{\sum_{i,j} |A_{ij}|^2}$ for an operator $\hat A$}, with $i$ and $j$ \rc{indicating the matrix element} in the basis of $|1\rangle$, $|2\rangle$, and $|3\rangle$.

\subsection{Numerical results for EIT and TPR regimes} 
\begin{table}
	\caption{List of key parameters of $\Lambda$-system and its interactions with radiation and bath.} 
	\begin{tabular}{ccccccc}
		\hline
		\hline
		\makebox[.5in]{Case} &\makebox[.5in]{ $E_2-E_1$}&\makebox[.5in]{$E_3-E_1$}&\makebox[.3in]{$\Omega_p$}&\makebox[.3in]{$\Omega_c$}&\makebox[.3in]{$\gamma_{12}$}&\makebox[.3in]{$\gamma_{13}$} \\
		\hline
		A-I & $6$ &$2$&$0.20$&$0.20$&  $0$& $0$\\
		A-II & $6$ &$2$&$0.20$&$0.20$&  $0.9$& $1.0$\\
		\hline
		B-I & $6$ &$2$&$1.00$&$1.12$&  $0$& $0$\\
		B-II & $6$ &$2$&$1.00$&$1.12$&  $0.9$& $1.0$\\
		\hline
		C-I & $6$ &$2$&$1.00$&$10$&  $0$& $0$\\
		C-II & $6$ &$2$&$1.00$&$10$&  $0.9$& $1.0$\\
		\hline
		\hline
		\label{table-1}
	\end{tabular}
\end{table}

\begin{figure}
	
	\includegraphics[width=0.4\columnwidth]{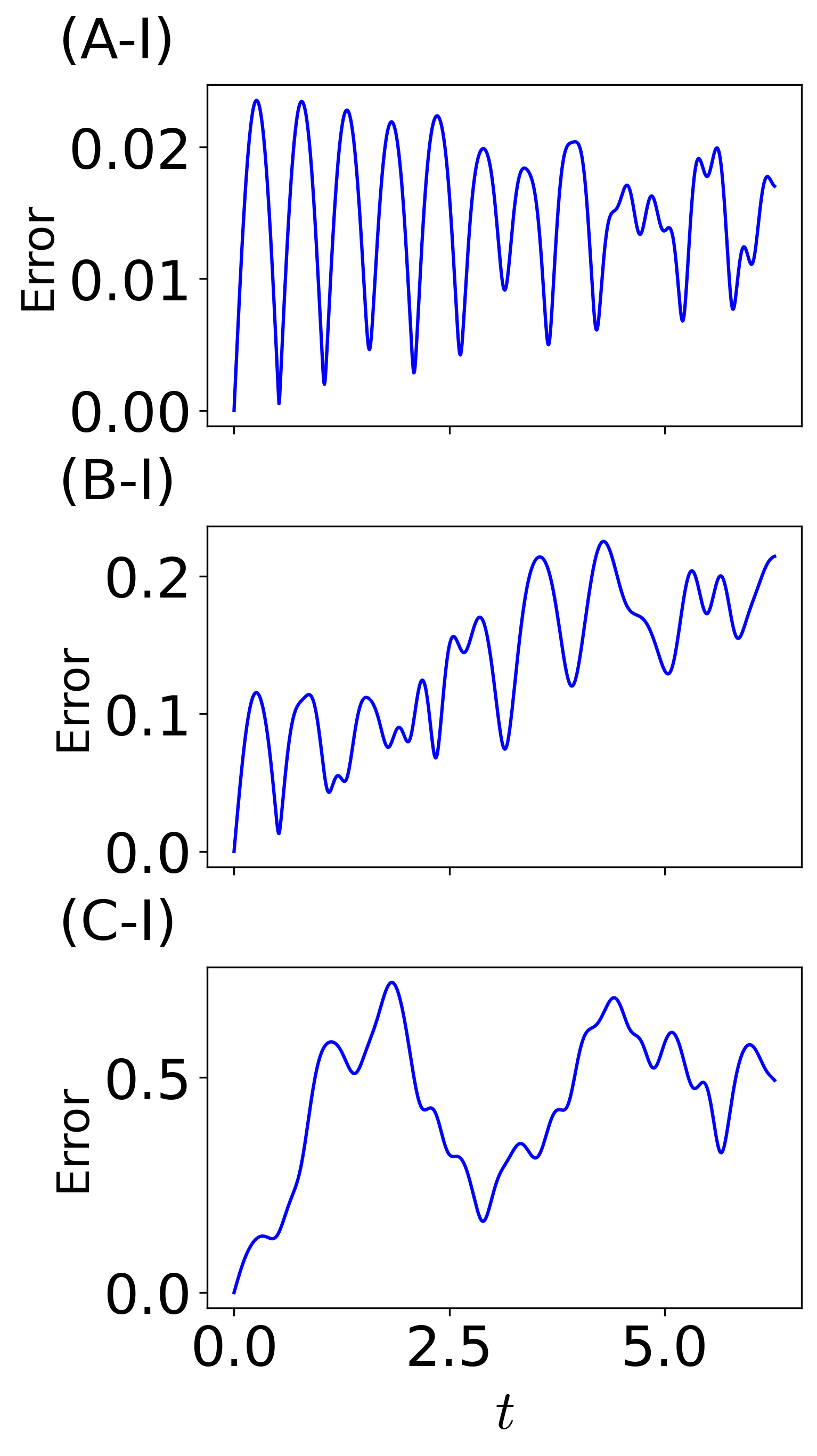}\makebox[.05in]{ }  \includegraphics[width=0.4\columnwidth]{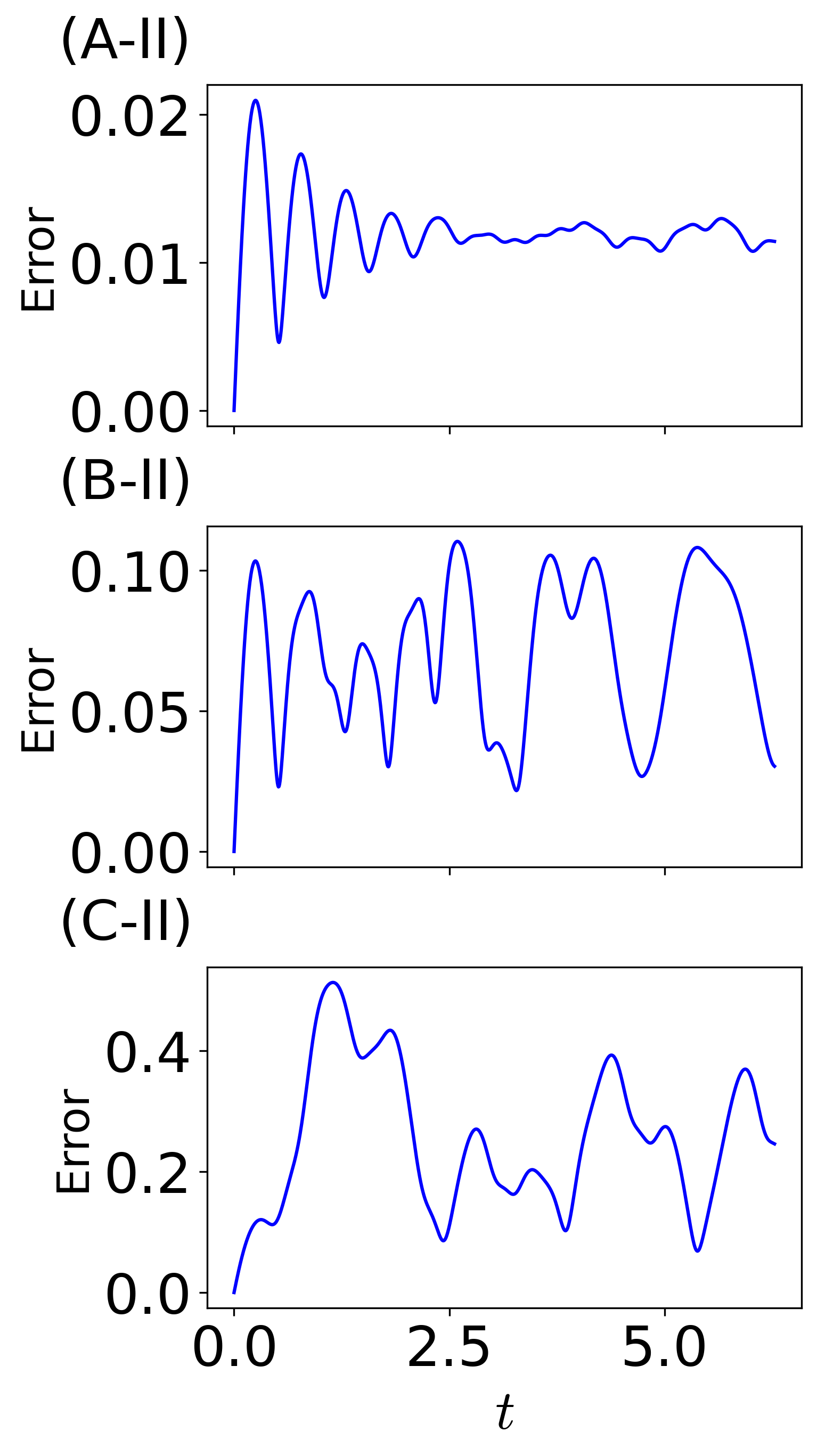} \\

	\caption{Errors of RWA calculated according to Eq. (\ref{eq:error_frob}) for the six different cases listed in Table \ref{table-1}.}
	\label{fig:error}
\end{figure}

We conducted numerical calculations for three cases of the parameters of the Hamiltonian.  Case A corresponds to weak control and probe fields, whereas case B represents moderate control and probe fields.  For case C, the control field is strong, breaking the condition for the RWA, whereas the probe field is moderate.  For each case, we considered both closed unitary system dynamics without bath (I) and open system non-unitary dynamics (II) with couplings to photonic bath,  for which the density operator evolves according to Eq. (\ref{eq:evol-rhos}).  The complete set of parameters are listed in Table \ref{table-1}.  The units were chosen such that $\hbar=1$.   
Unless stated otherwise, we assume the two-photon resonance (TPR) condition with zero detuning\cite{boradjiev-pra81} where $\omega_p=E_2-E_1$ and $\omega_c=E_2-E_3$. \rc{For the initial condition, we chose} $ \hat\rho(0)=|1\rangle\langle1|$.

\rc{If the Hamiltonian is periodic in time with period $T$, explicit calculation of the propagator up to $t=T$ is sufficient to generate the dynamics for arbitrarily long time, as  described in App. \ref{periodic}.   Indeed, for all the parameters of Table \ref{table-1} and under the zero-detuning TPR condition, the Hamiltonian is periodic in time, with $T=2\pi$.   Thus, we use the approach in App. \ref{periodic} to generate the dynamics for $t>T$ including the steady state limits.  This way, we can calculate accurate enough steady state limit results without resorting to Floquet approaches.\cite{casas-jpa34,grifoni-pr304,mori-arcmp14,eckardt-njp17,eckardt-rmp89,kuwahara-ap367,engelhardt-prl123} } 

For \rc{all }the calculations in this subsection, we used our simplest 4th order propagators, Eqs. (\ref{eq:us-4th}) and (\ref{eq:usl-4th}) in Sec.  \ref{method}, with a time step, $\delta t=T/(2^{10}-1)$.  This choice was confirmed to give sufficiently accurate results when benchmarked against the results of our 6th order ME propagator\cite{ture-jang-jpca124} extended to the Liouville space.  Thus, results presented in this subsection can be viewed virtually exact.  Regarding our 6th order propagator, we would like to note that it is the original expression with commutators.\cite{ture-jang-jpca124} This is because a stable CFME for open system dynamics beyond 4th order with \rc{real coefficients} was shown to be impossible.\cite{hofstatter-nm141}  Some of benchmark calculations are provided in the Supplementary Materials (SM). 

Figure \ref{fig:error} shows errors of the RWA calculated according to Eq. (\ref{eq:error_frob}) for all six cases during the initial dynamics.   As expected, the errors for cases A-I and A-II, which correspond to weak control and probe fields,  are smallest of the three cases and are negligible. \rc{Cases B-I and B-II exhibit about an order of magnitude larger errors although small}.  It is interesting to note that errors for the open system dynamics (B-II) are about half of those for closed system dynamics (B-I). On the other hand, the errors for the cases C-I and C-II grow rapidly during the initial stage and remain substantial. The presence of the bath for this case (C-II) is also shown to reduce the overall error, \rc{although}  less effective than the case B-II.   These are expected results since the strong control field renders the RWA to be inaccurate even though the probe field is moderate.  Nonetheless, quantitative details provided here \rc{offer a} meaningful reference for assessing the applicability of RWA.  The supplementary material shows errors in the long time limits  for the three open system quantum dynamics, which confirm these assessments in the presence of bath.

\begin{figure}
	(a)\makebox[0.8\columnwidth]{ }\\
	\includegraphics[width=0.65\columnwidth]{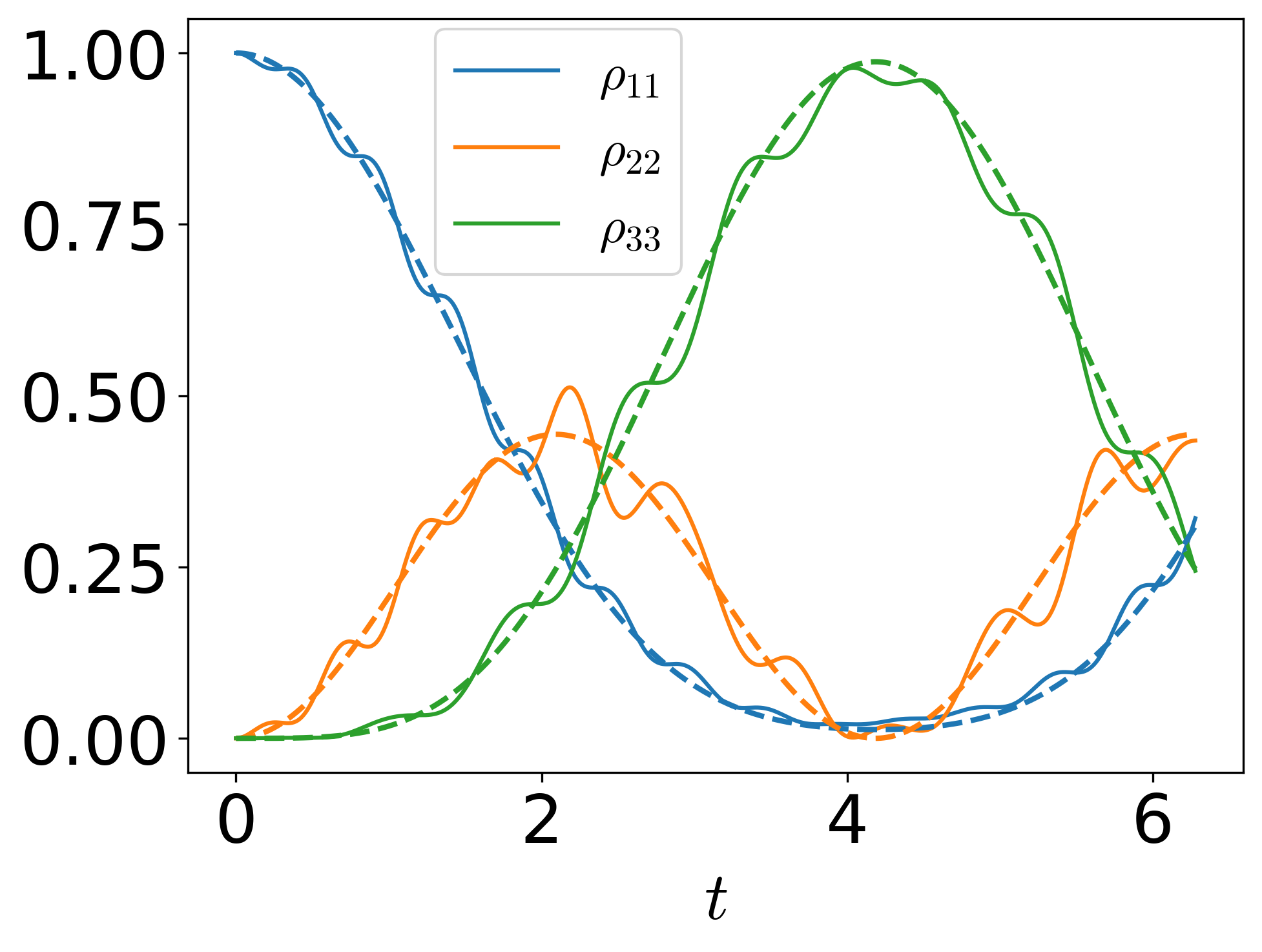}\\
	(b)\makebox[0.8\columnwidth]{ }\\
	\includegraphics[width=0.65\columnwidth]{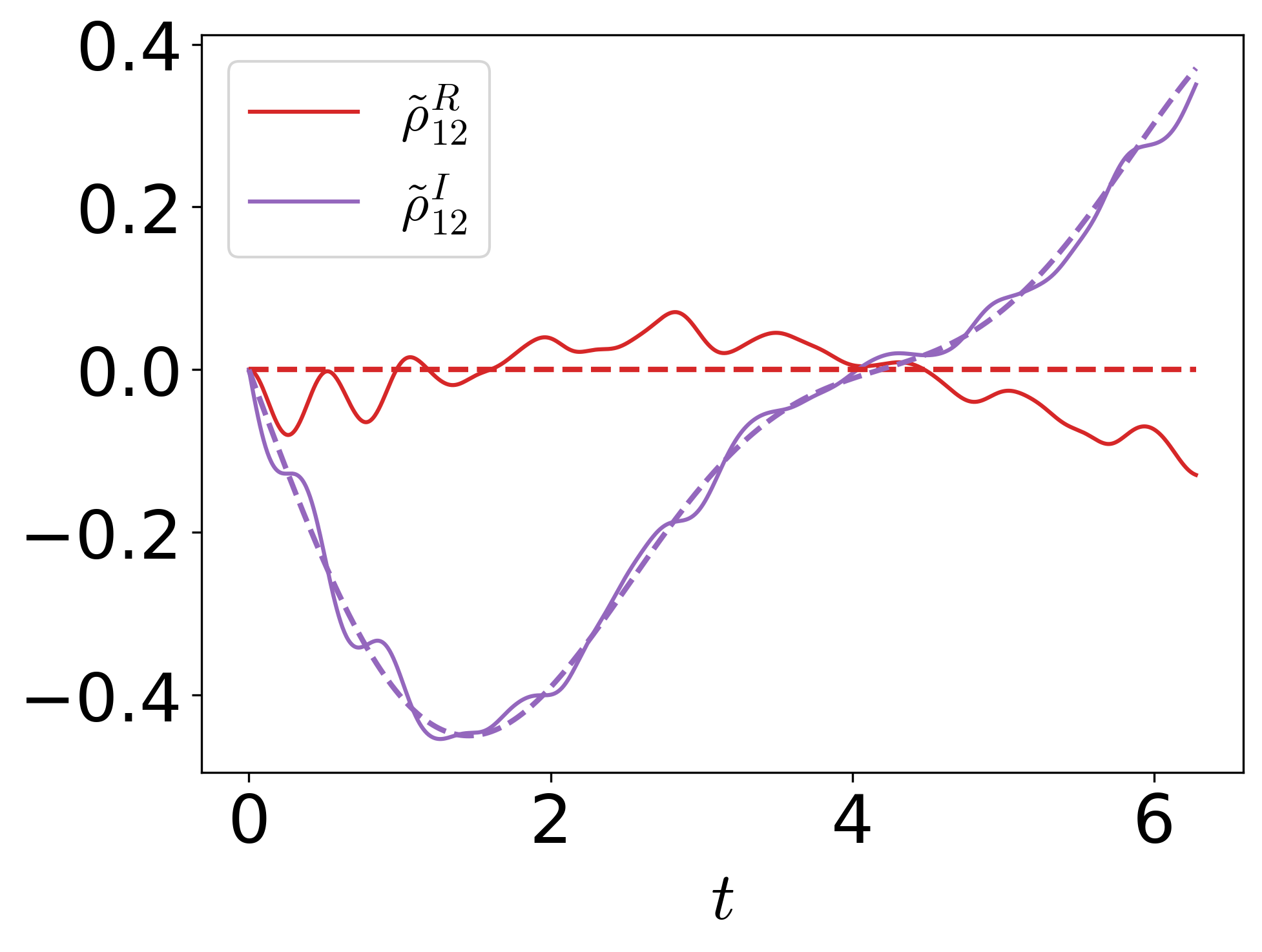} \\
	(c)\makebox[0.8\columnwidth]{ }\\
	\includegraphics[width=0.65\columnwidth]{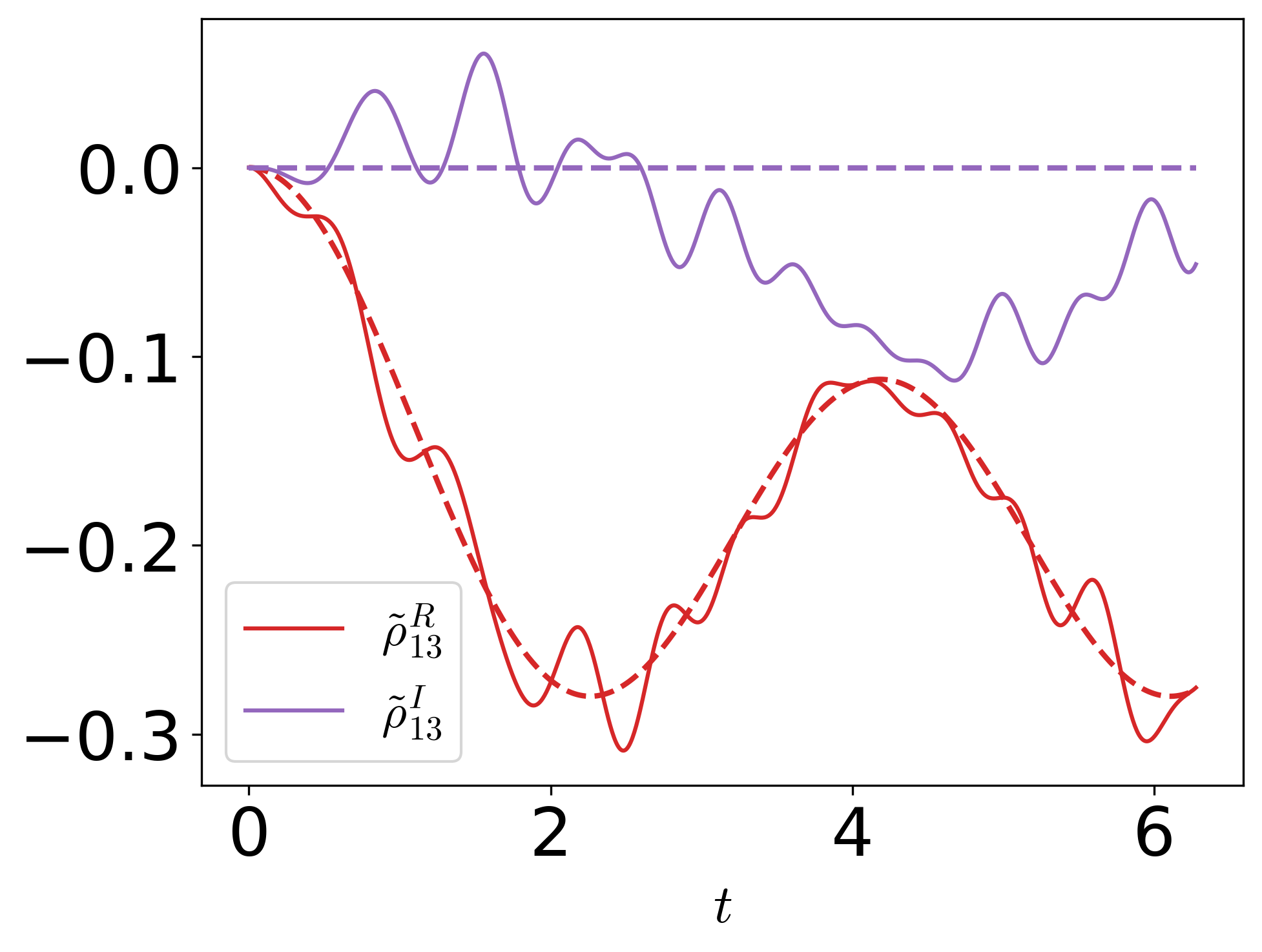}\\
	(d)\makebox[0.8\columnwidth]{ }\\
	\includegraphics[width=0.65\columnwidth]{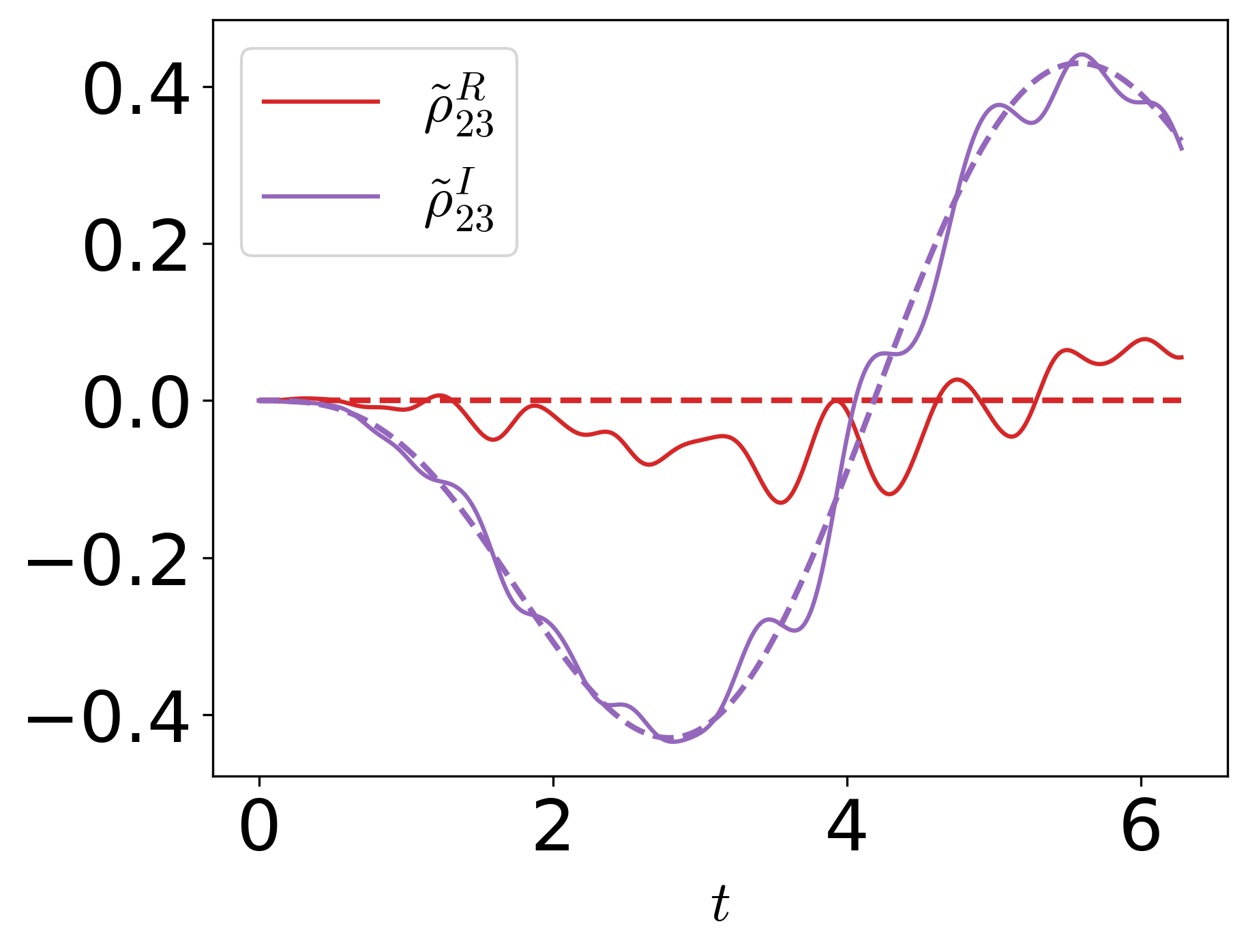}
	\caption{Elements of the time dependent system density operators for the case B-I,  closed system unitary dynamics, with full Hamiltonian (solid) and the RWA(dashed), for which the Hamiltonians are respectively $\hat H(t)$ and $\hat H^{\rm \small RWA}(t)$.  Both were calculated using the fourth-order ME-propagator with commutator, Eq. (\ref{eq:us-4th}).}
	\label{fig:closed_both}
\end{figure}

\begin{figure}
	(a)\makebox[0.8\columnwidth]{ }\\
	\includegraphics[width=0.65\columnwidth]{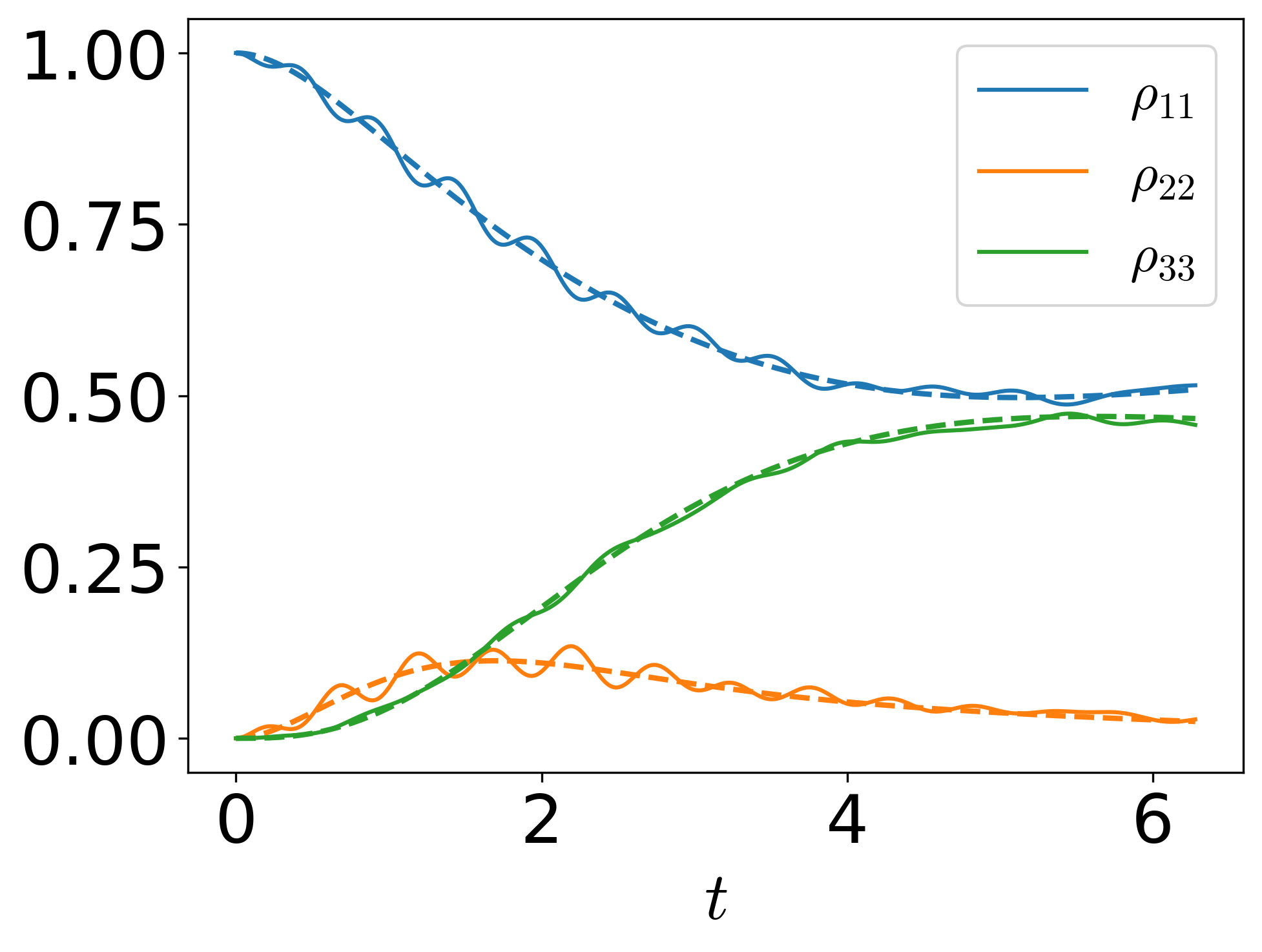} \\
	(b)\makebox[0.8\columnwidth]{ }\\
	\includegraphics[width=0.65\columnwidth]{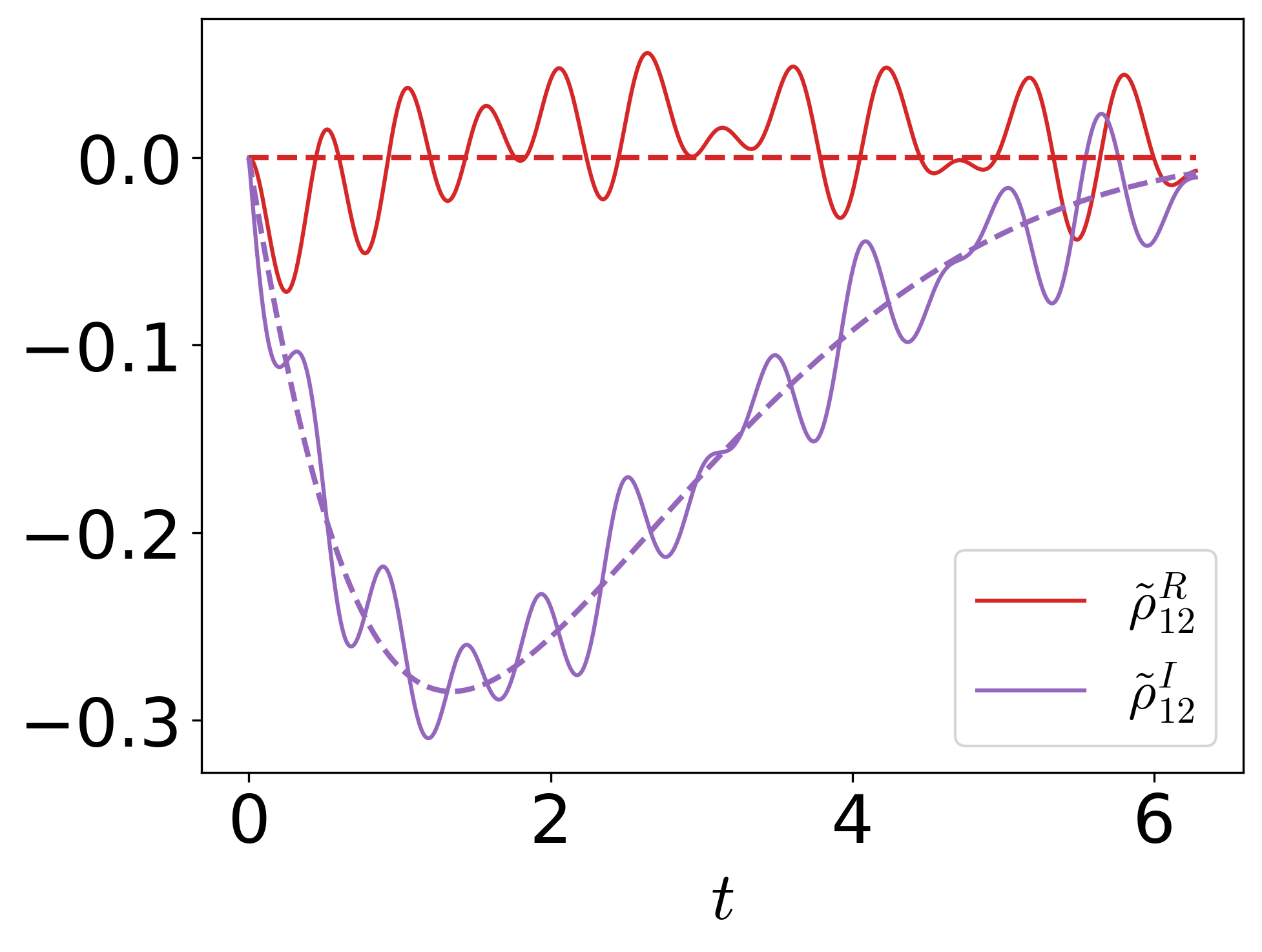} \\
	(c)\makebox[0.8\columnwidth]{ }\\
	\includegraphics[width=0.65\columnwidth]{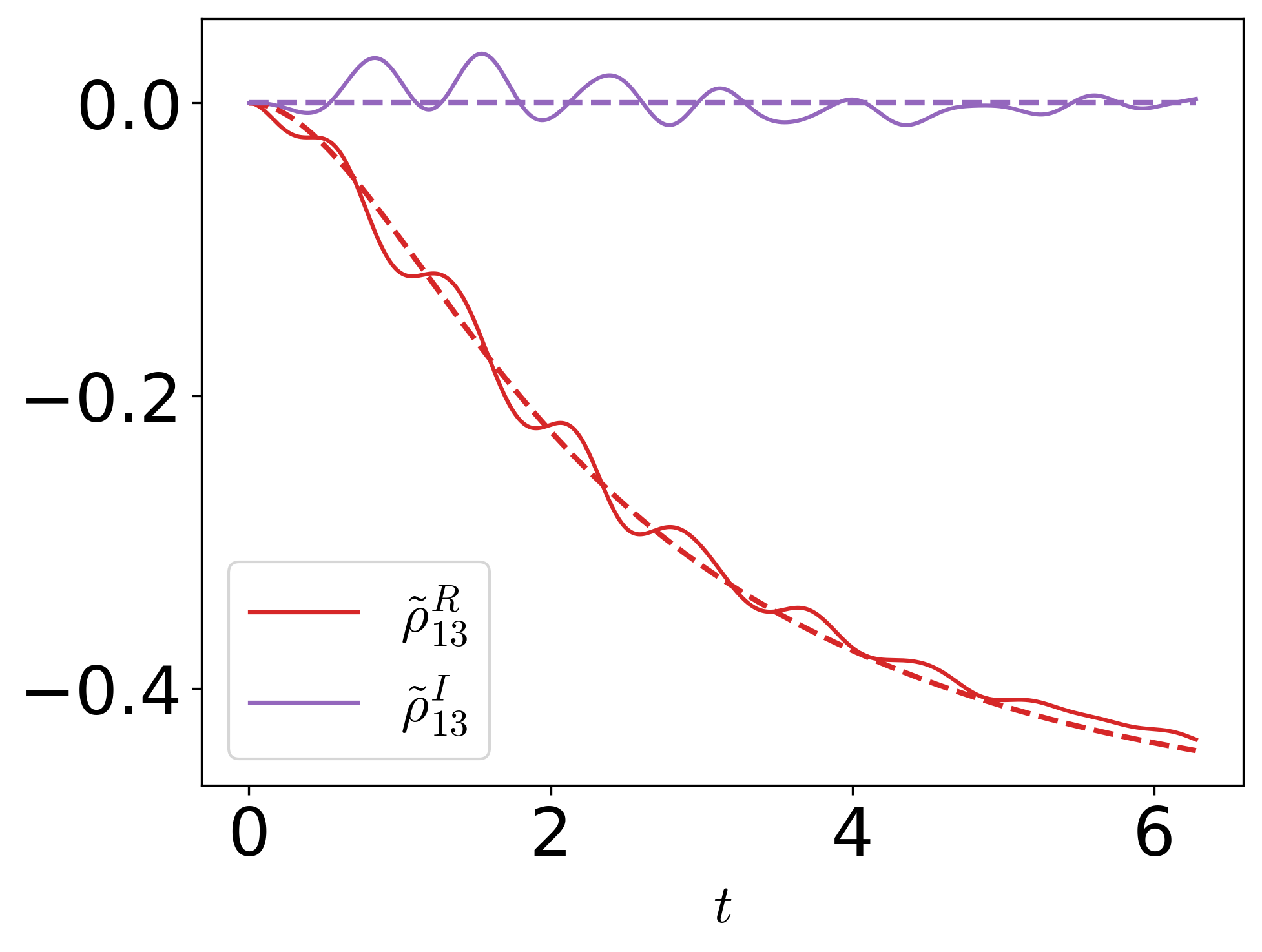}\\
	(d)\makebox[0.8\columnwidth]{ }\\
	\includegraphics[width=0.65\columnwidth]{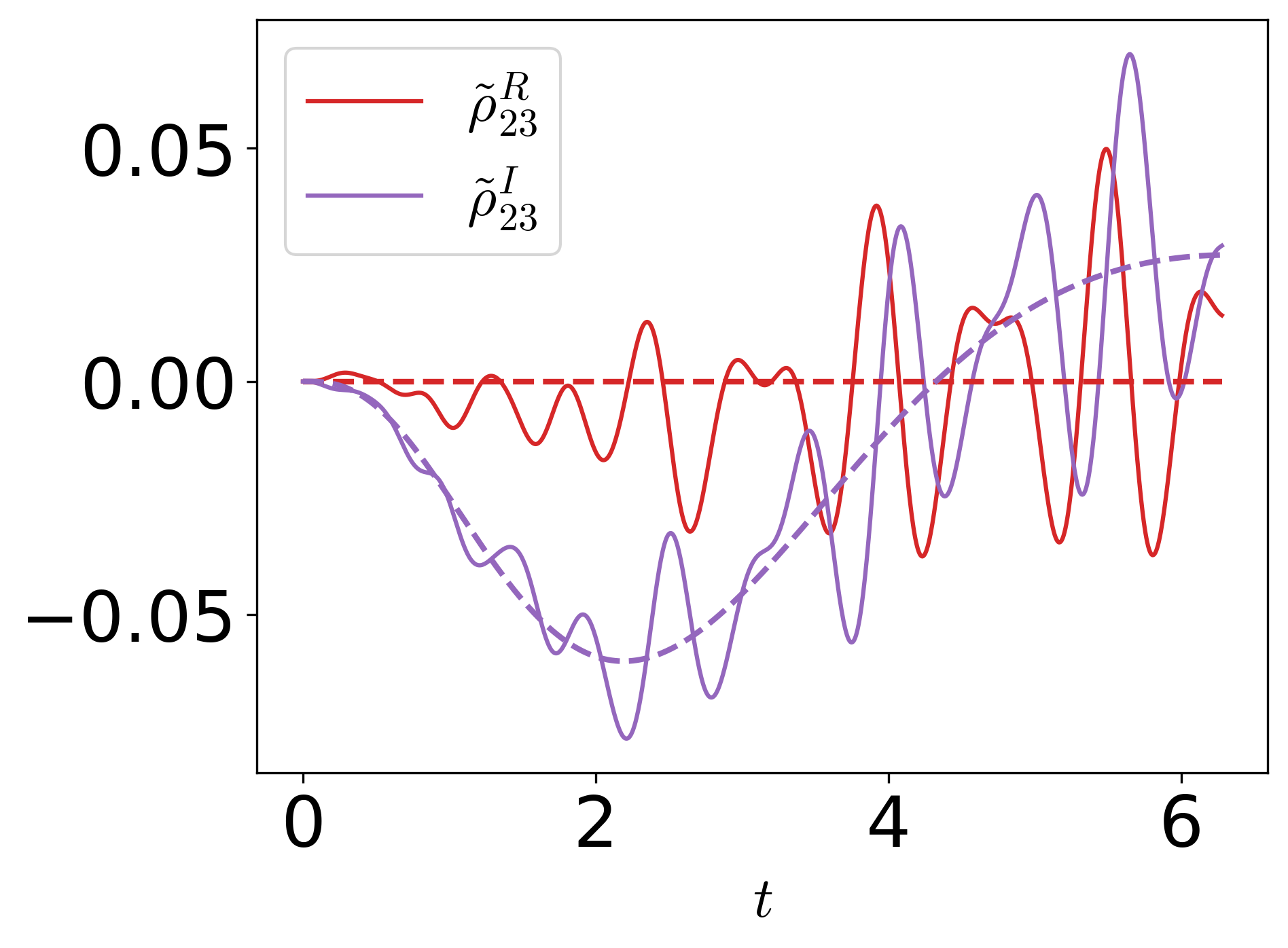}
	\caption{Elements of the time dependent system density operator for the case B-II, the open system non-unitary dynamics, with full Hamiltonian (solid) and the RWA (dashed), for which the Hamiltonians are respectively $\hat H (t)$ and $\hat H^{\rm \small RWA}(t)$.  Both were calculated using the fourth-order ME-propagator with commutator, Eq. (\ref{eq:usl-4th}).}
	\label{fig:open_both}
\end{figure}

As a representative example, \rc{we next provide further details of the difference in density operators for the full and RWA Hamiltonians for  cases B-I and B-II}. 
\rc{The results for cases A-I and A-II, where the RWA works well, and cases C-I and C-II, where the RWA clearly breaks down, are provided} in Figs. S1-S4 in the SM. 

Figure \ref{fig:closed_both} shows elements of time evolving density operator for the case B-I in Table \ref{table-1}, which corresponds to a representative example of EIT with perfect resonance but without the effects of bath. 
We use $\tilde\rho_{ij}$ to represent the off-diagonal elements in the RWF (see App. \ref{sec:rotating_frame}). 
\rc{Overall, the results for the RWA Hamiltonian are in reasonable agreement with those for the full Hamiltonian.}
In particular, all the diagonal elements of the density operator based on the RWA reproduce those of 
\rc{of the full Hamiltonian}, except for small wiggles that originate from non-resonant terms of the Hamiltonian. The errors in off-diagonal elements of the density operator are 
small but slightly worse than \rc{those for} diagonal elements.  In particular, the real parts of the off-diagonal elements of the density operator are shown to exhibit larger deviation.  

Figure \ref{fig:open_both} shows results for the case B-II in Table \ref{table-1}, for which EIT is observed with full resonance condition in the presence of photonic bath.   The \rc{time evolution of the }density operator in this case is governed by Eq. (\ref{eq:evol-rhos}). Overall, the presence of bath significantly dampens coherent behavior and reduces the discrepancies between the results for the full and RWA Hamiltonians.  It is interesting to note that both real and imaginary parts of  $\tilde \rho_{23}$ for this open system dynamics are smaller than those for the closed system dynamics by an order of magnitude, whereas the errors of the RWA from the exact dynamics for these off-diagonal terms remain of similar order.  As a result, the relative error of RWA for these off-diagonal elements of the density operator turn out to be larger than other elements.  

\begin{figure}
(a)\makebox[0.4\columnwidth]{ }(b)\makebox[0.4\columnwidth]{ }\\
\includegraphics[width=0.4\columnwidth]{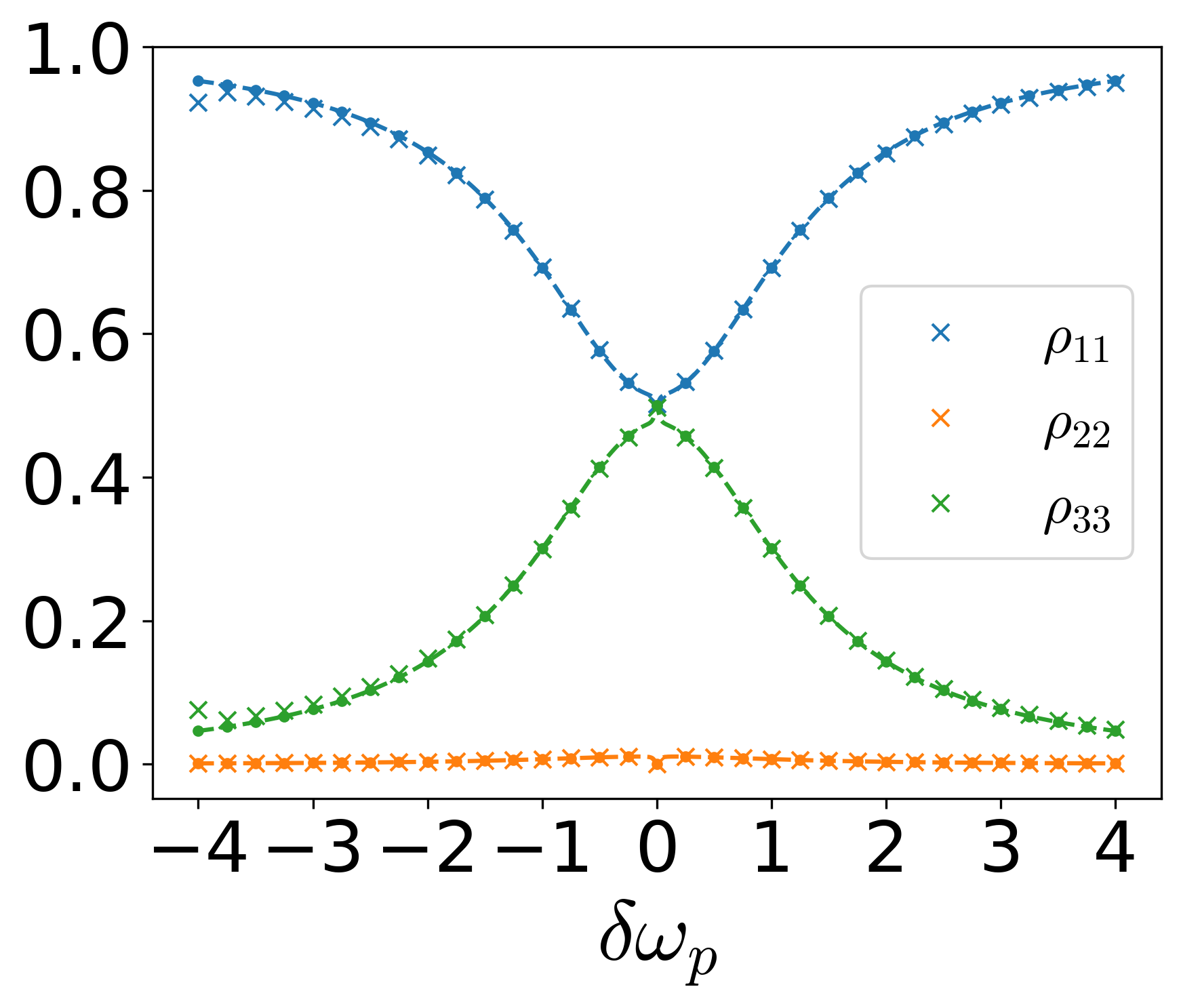} \hspace{.1in}\includegraphics[width=0.43\columnwidth]{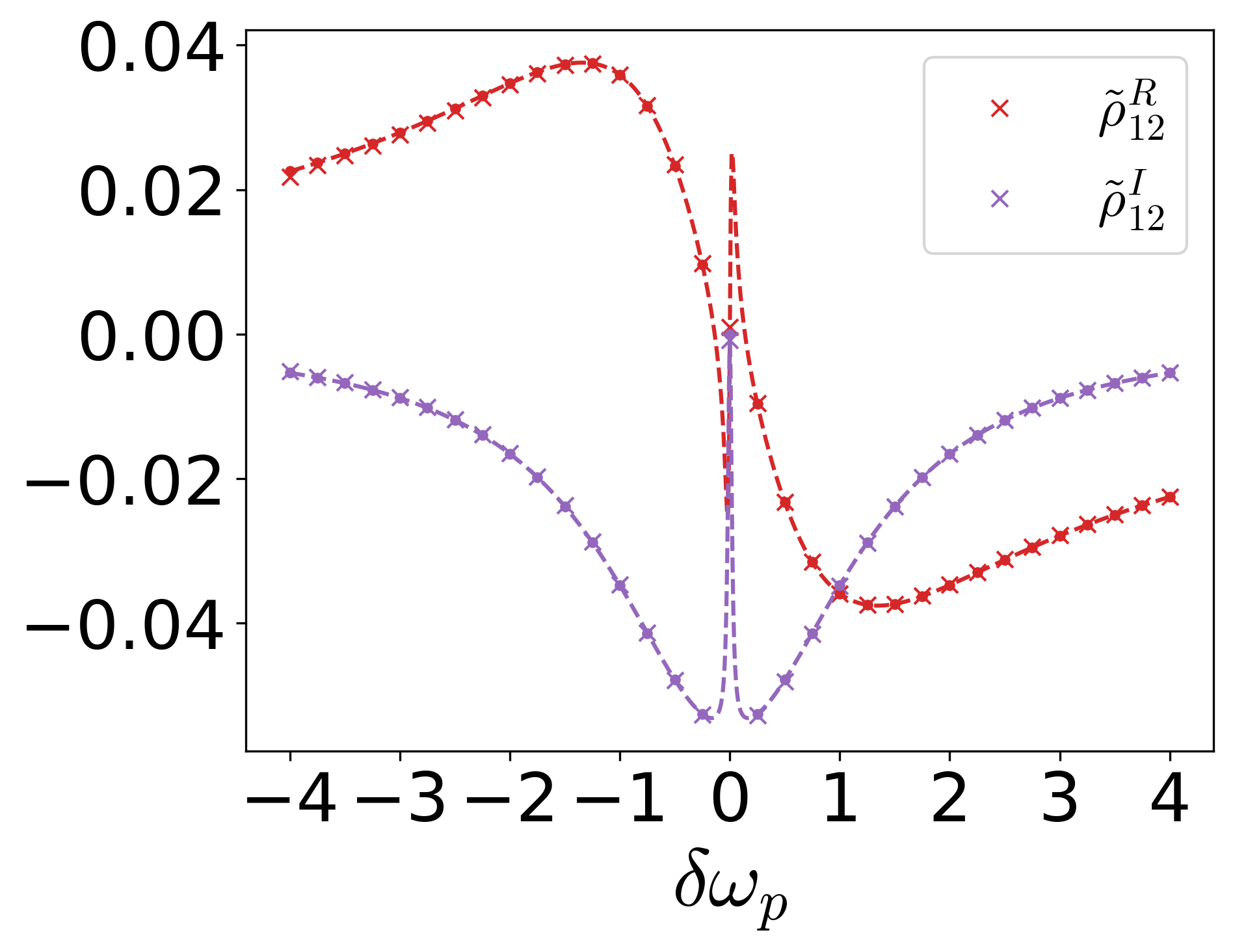} \\
(c)\makebox[0.4\columnwidth]{ }(d)\makebox[0.4\columnwidth]{ }\\
\includegraphics[width=0.42\columnwidth]{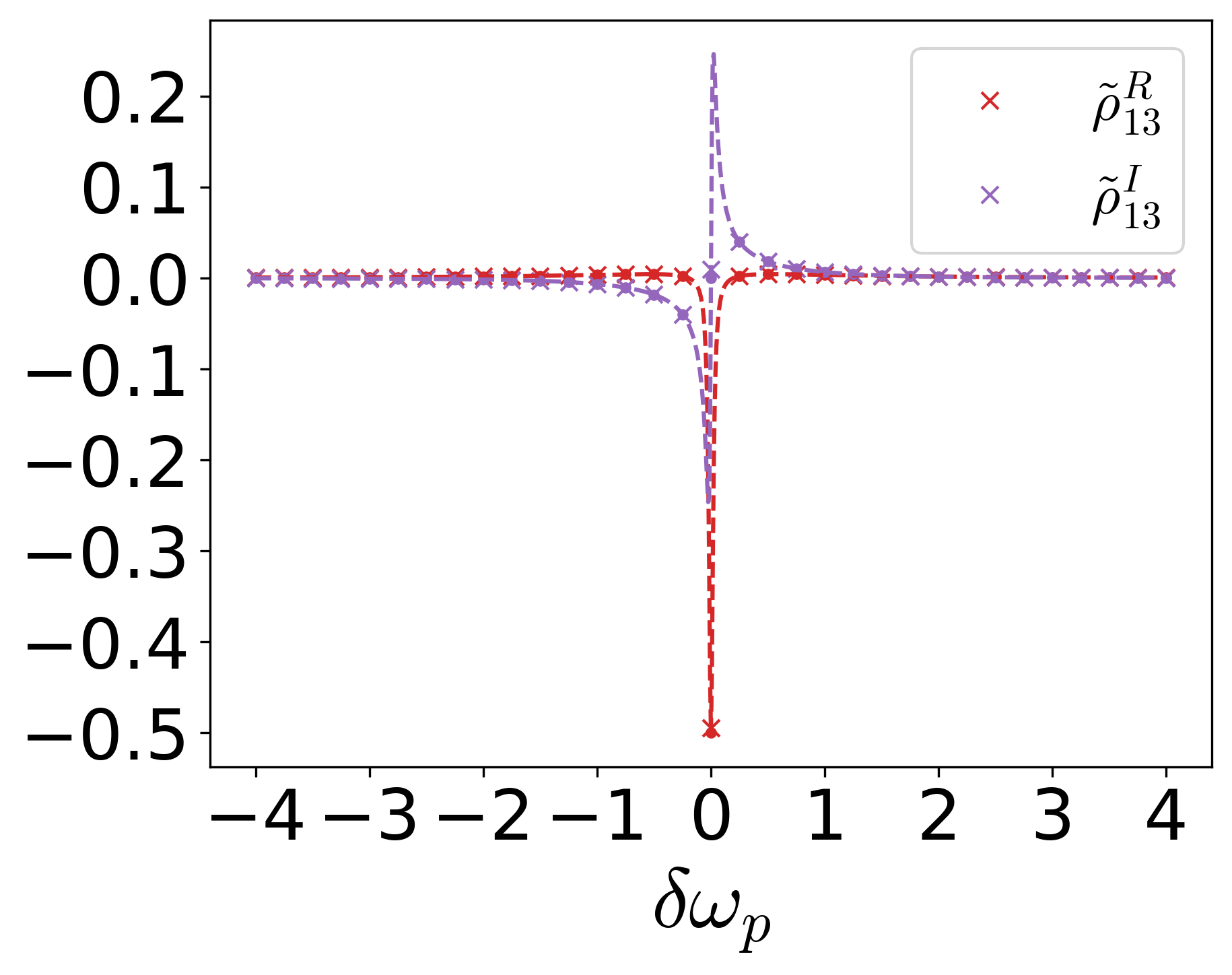} \hspace{.2in}\includegraphics[width=0.4\columnwidth]{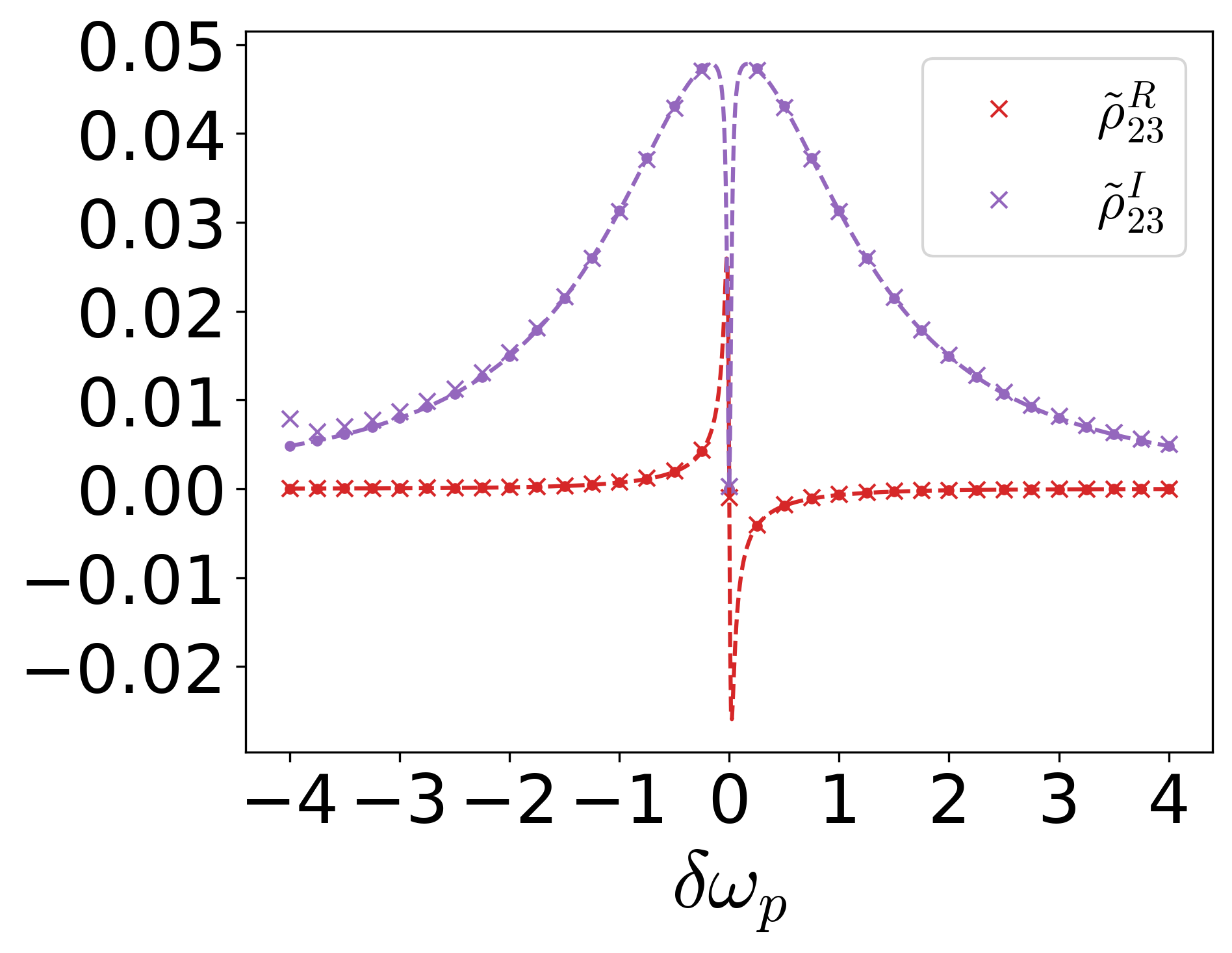}
\caption{Numerically calculated steady state average values of the open system case A-II for the full Hamiltonian, which are shown as data with ``$\times$" symbols.  The numerical results for the RWA Hamiltonian are shown with circle symbols, and dashed lines represent the analytical solution based on the RWA. }
\label{fig:steady_state_w_p_pops_B}
\end{figure}

\begin{figure}
(a)\makebox[0.4\columnwidth]{ }(b)\makebox[0.4\columnwidth]{ }\\
\includegraphics[width=0.44\columnwidth]{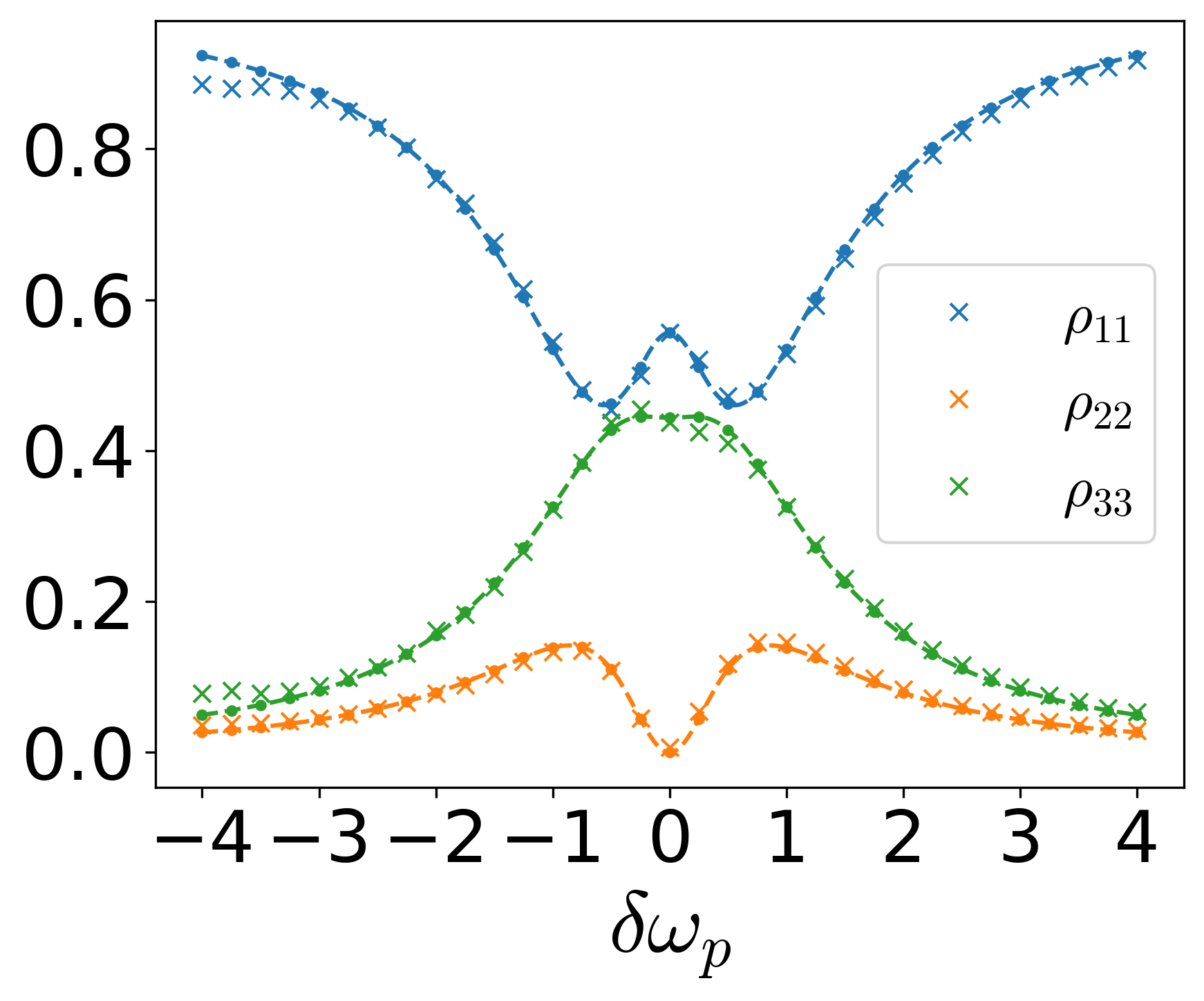} \hspace{.1in}\includegraphics[width=0.48\columnwidth]{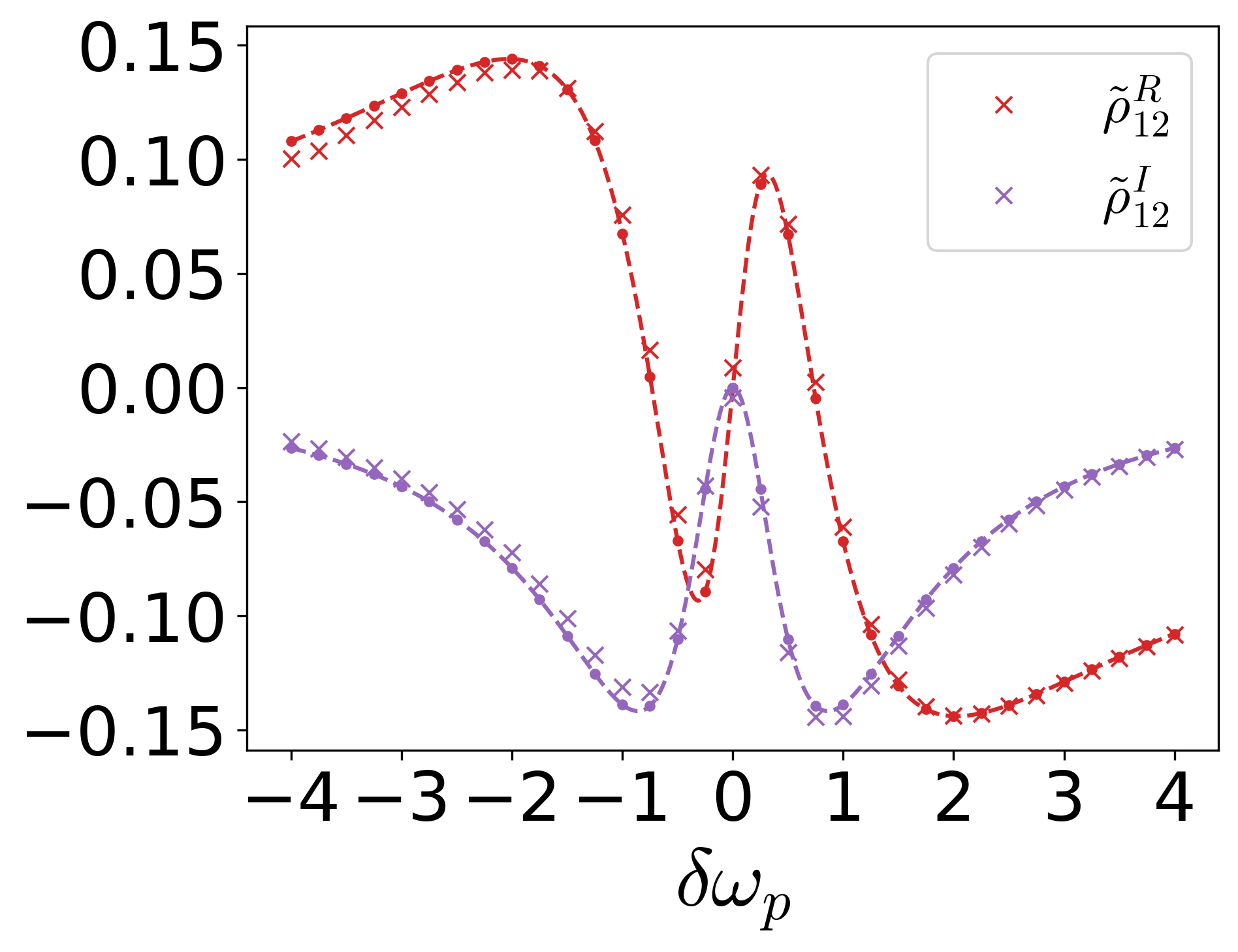} \\
(c)\makebox[0.4\columnwidth]{ }(d)\makebox[0.4\columnwidth]{ }\\
\includegraphics[width=0.46\columnwidth]{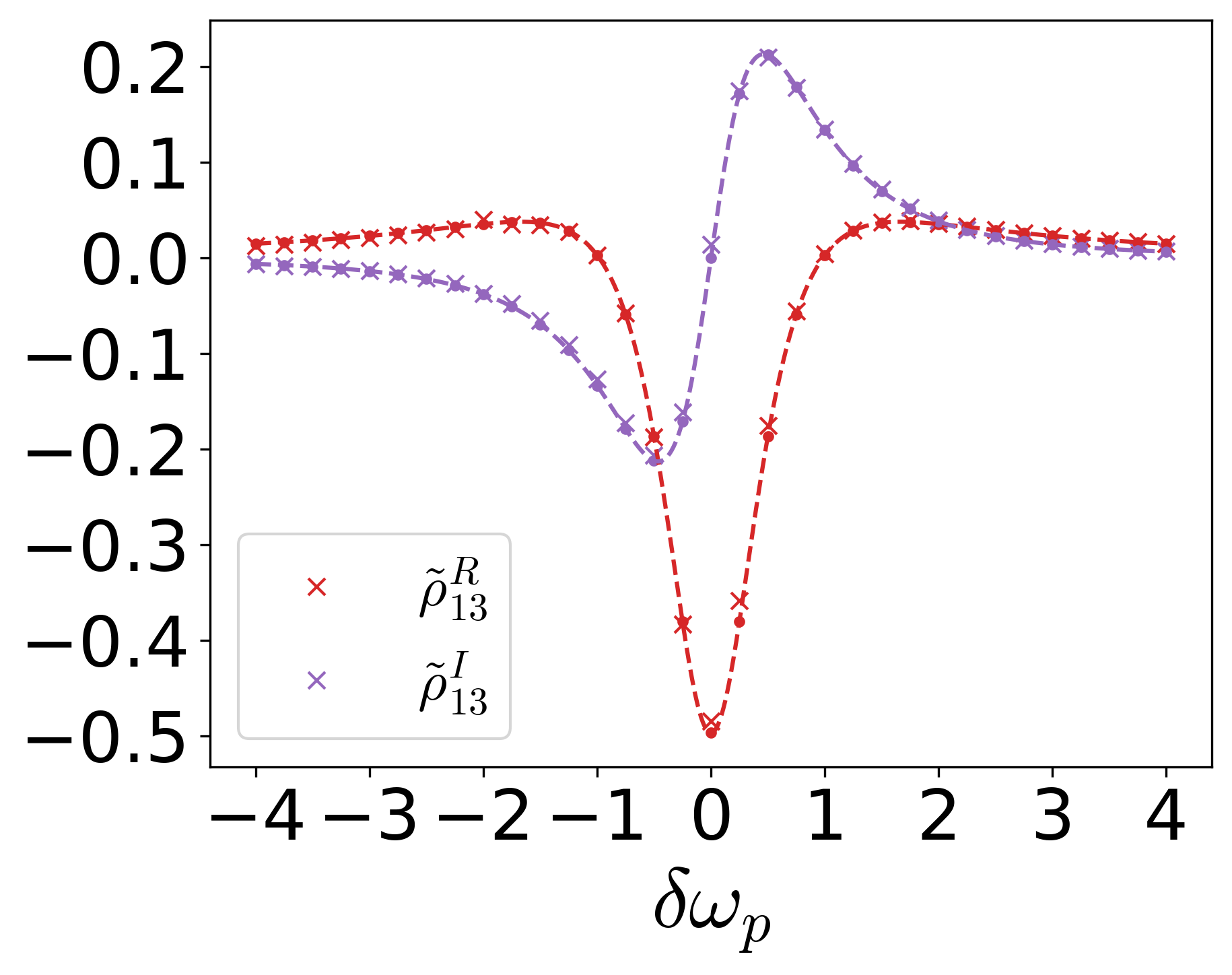} \hspace{.2in}\includegraphics[width=0.45\columnwidth]{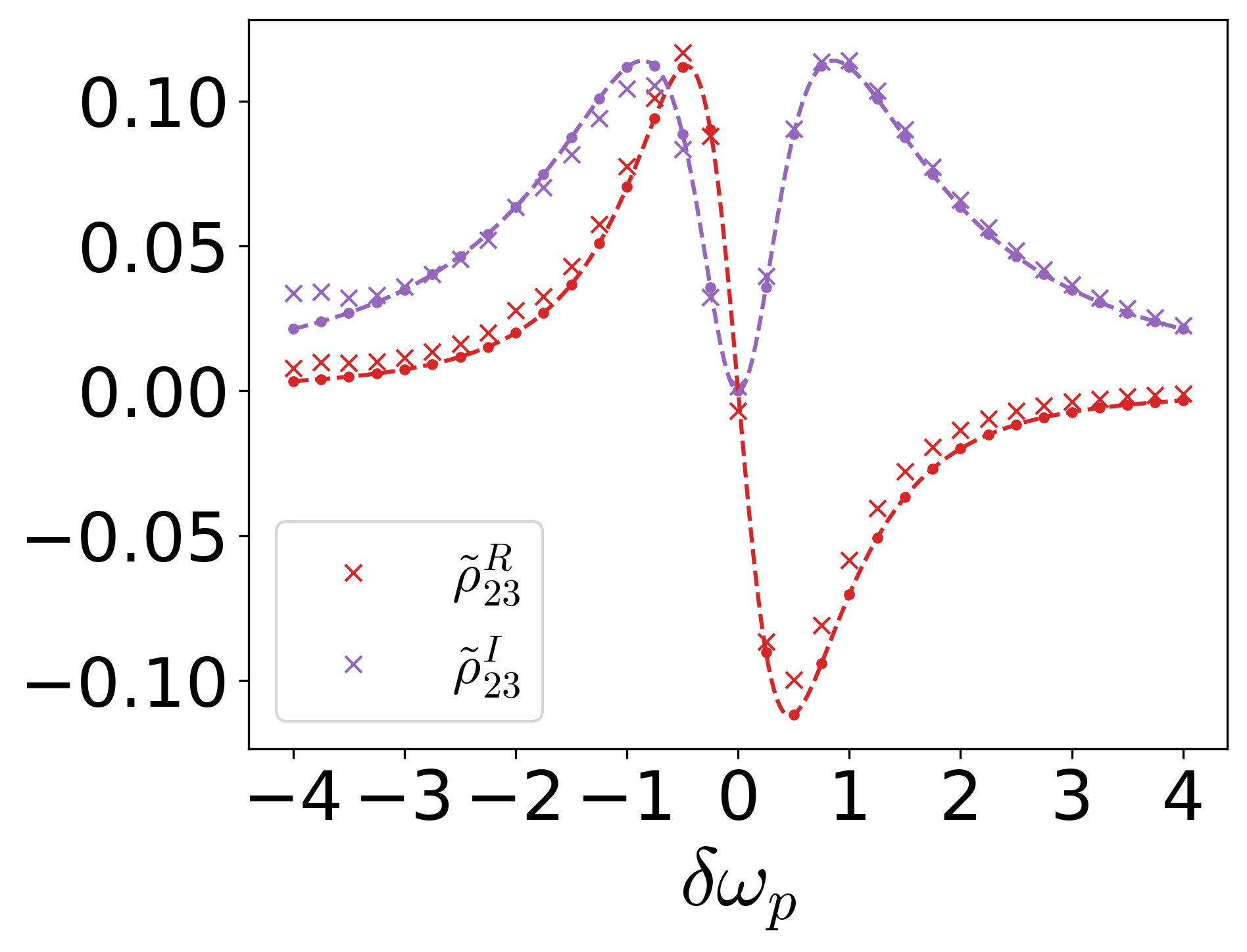}
\caption{Numerically calculated steady state average values of the open system case B-II for the full Hamiltonian, which are shown as data with ``x" symbols.  The numerical results for the RWA Hamiltonian are shown with circle symbols, and dashed lines represent the analytical solution based on the RWA.  }
\label{fig:steady_state_w_p_pops}
\end{figure}

\begin{figure}
(a)\makebox[0.4\columnwidth]{ }(b)\makebox[0.4\columnwidth]{ }\\
\includegraphics[width=0.4\columnwidth]{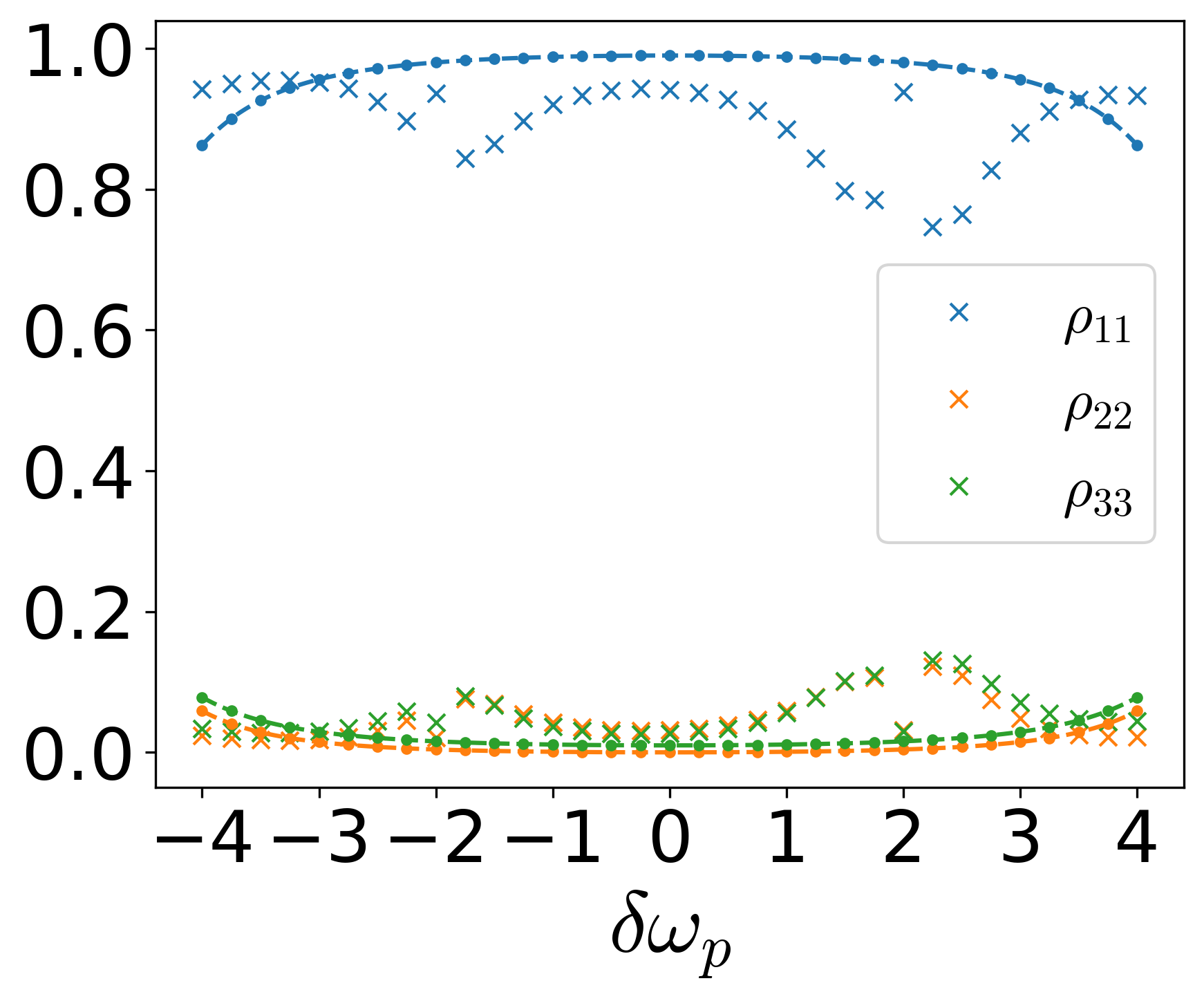} \hspace{.1in}\includegraphics[width=0.45\columnwidth]{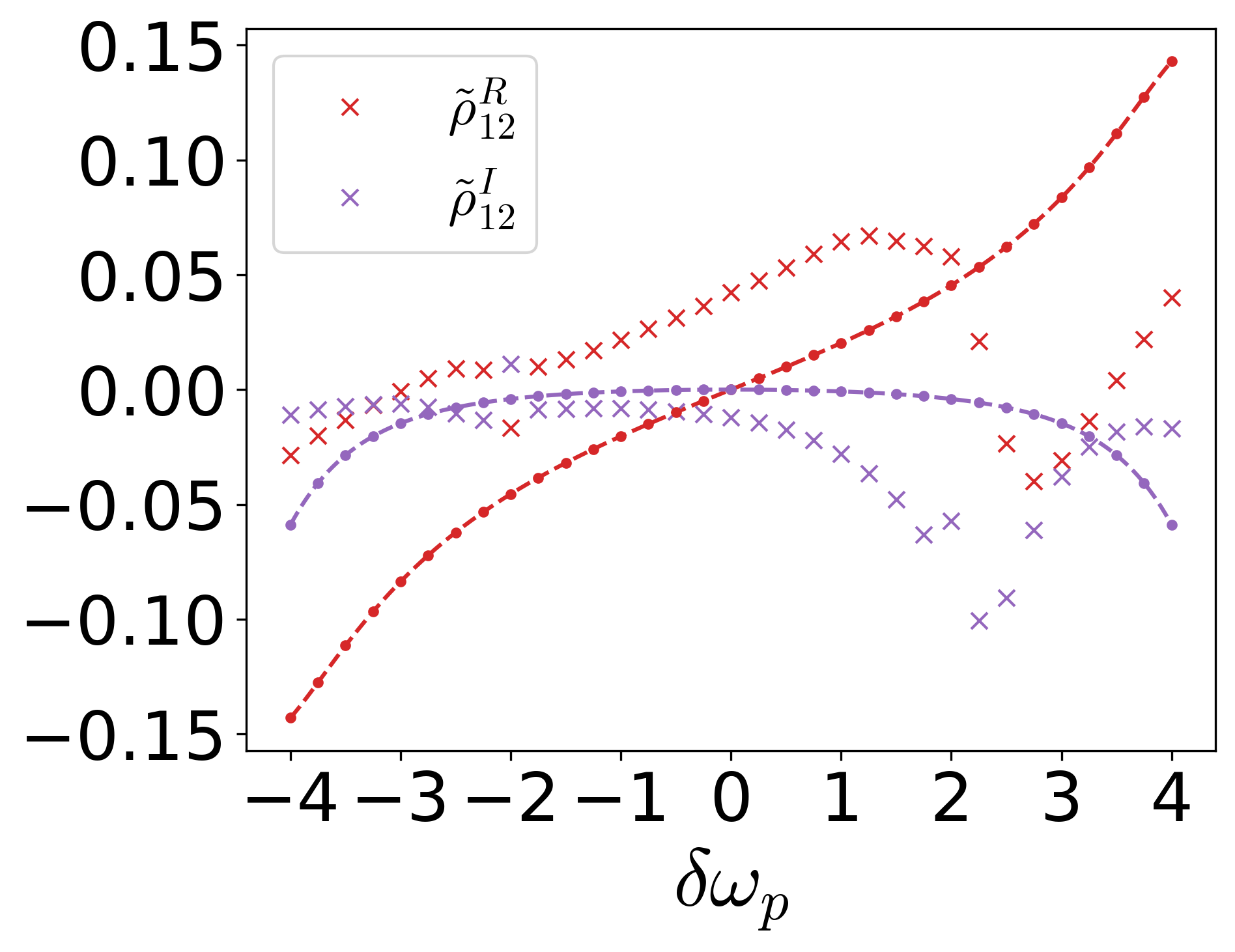} \\
(c)\makebox[0.4\columnwidth]{ }(d)\makebox[0.4\columnwidth]{ }\\
\includegraphics[width=0.4\columnwidth]{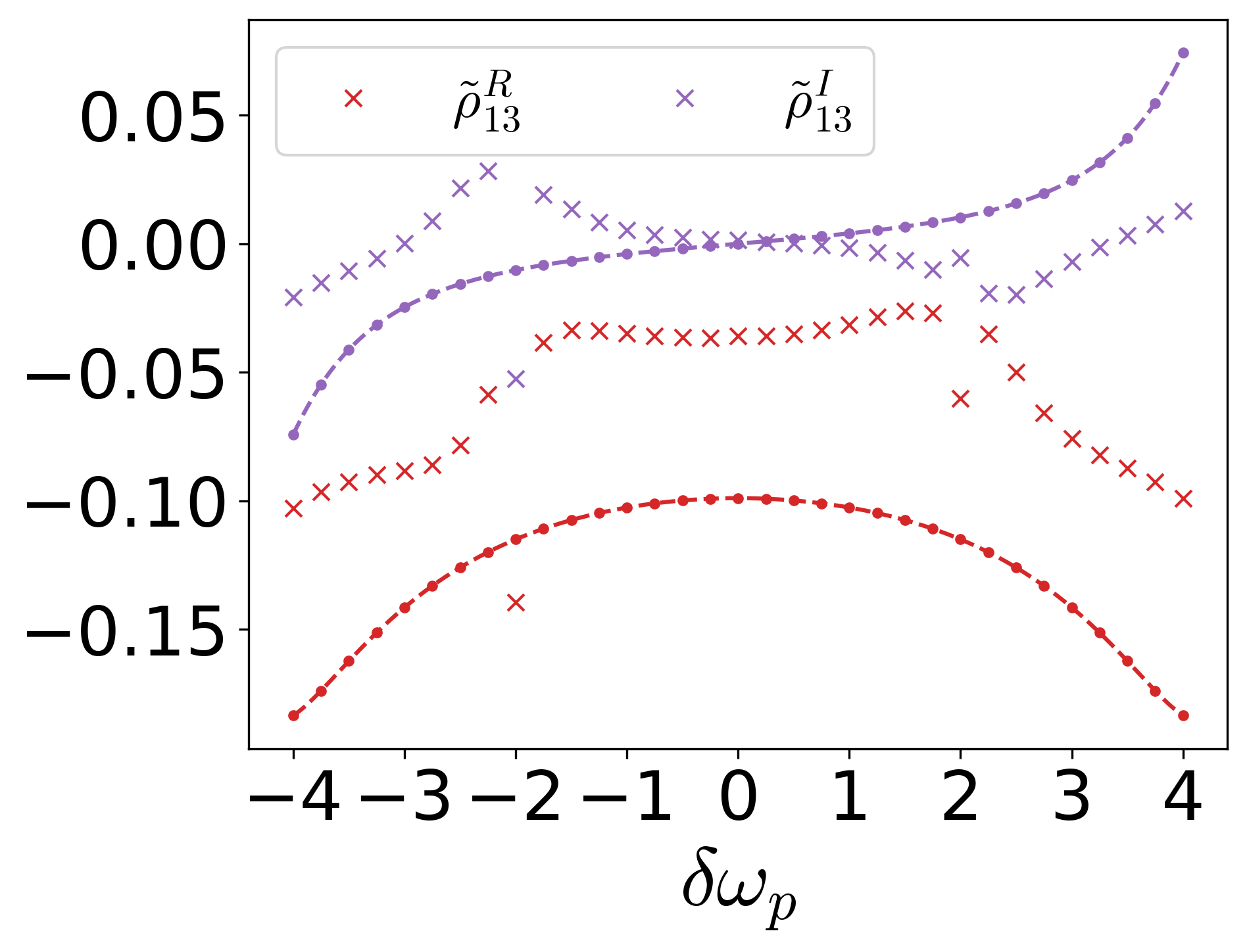} \hspace{.2in}\includegraphics[width=0.4\columnwidth]{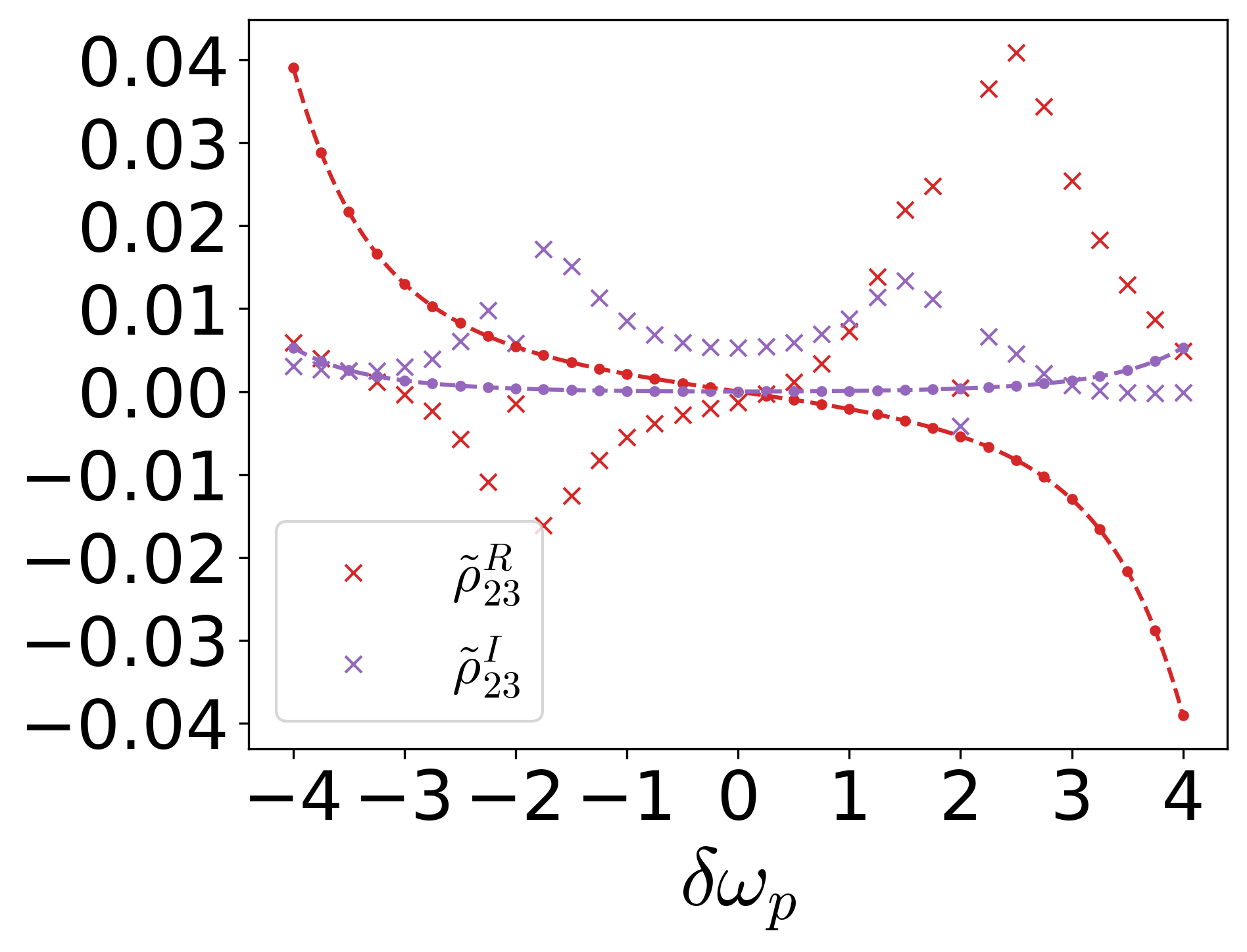}
\caption{Numerically calculated steady state average values of the open system case C-II for the full Hamiltonian, which are shown as data with ``x" symbols.  The numerical results for the RWA Hamiltonian are shown with circle symbols, and dashed lines represent the analytical solution based on the RWA.  }
\label{fig:steady_state_w_p_pops_C}
\end{figure}

In order to understand implications of the errors due to the RWA for the EIT behavior, we conducted calculations for the long time limit for values of  $\omega_p$ from 2 to 10 in increments of 0.25.  \rc{For all of these cases, the Hamiltonian remain periodic in time with a common period,  $T=8\pi$.  Thus, the approach of App. \ref{periodic} can still be used to effectively calculate the long time propagation.} Once we have conducted the long time dynamics, we then calculated the average steady-state values of the system density operator matrix elements by integrating their time dependent values over an interval of $8\pi$ with the trapezoid rule.  For the case A-II, Fig.  \ref{fig:steady_state_w_p_pops_B}  shows that most results for RWA agree very well with those for the full Hamiltonian. 
The results for the case B-II are shown in Fig. \ref{fig:steady_state_w_p_pops}.    All the diagonal and off-diagonal elements based on the RWA are in reasonable agreement with exact values over the range of  $\delta \omega_p$ tested. On the other hand, Fig.  \ref{fig:steady_state_w_p_pops_C} shows \rc{that there is }significant qualitative difference between the two density operators for the case C-II, which confirms the unreliability of the RWA for the case of strong  control fields. 	
	
\begin{figure}
\centering
(a)\makebox[0.4\columnwidth]{ }(b)\makebox[0.4\columnwidth]{ }\\
\includegraphics[width=0.45\columnwidth]{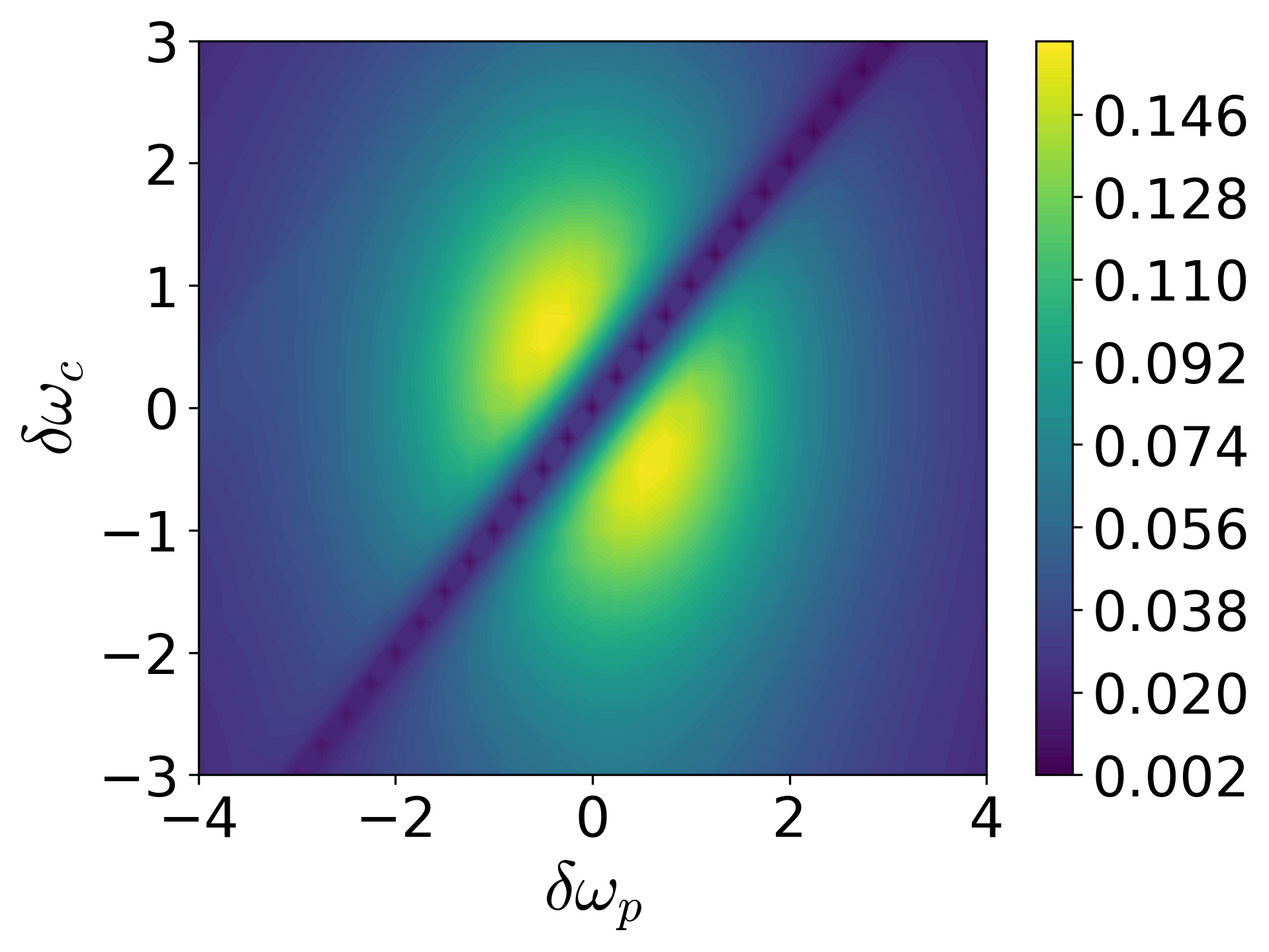}\hspace{.1in}\includegraphics[width=0.45\columnwidth]{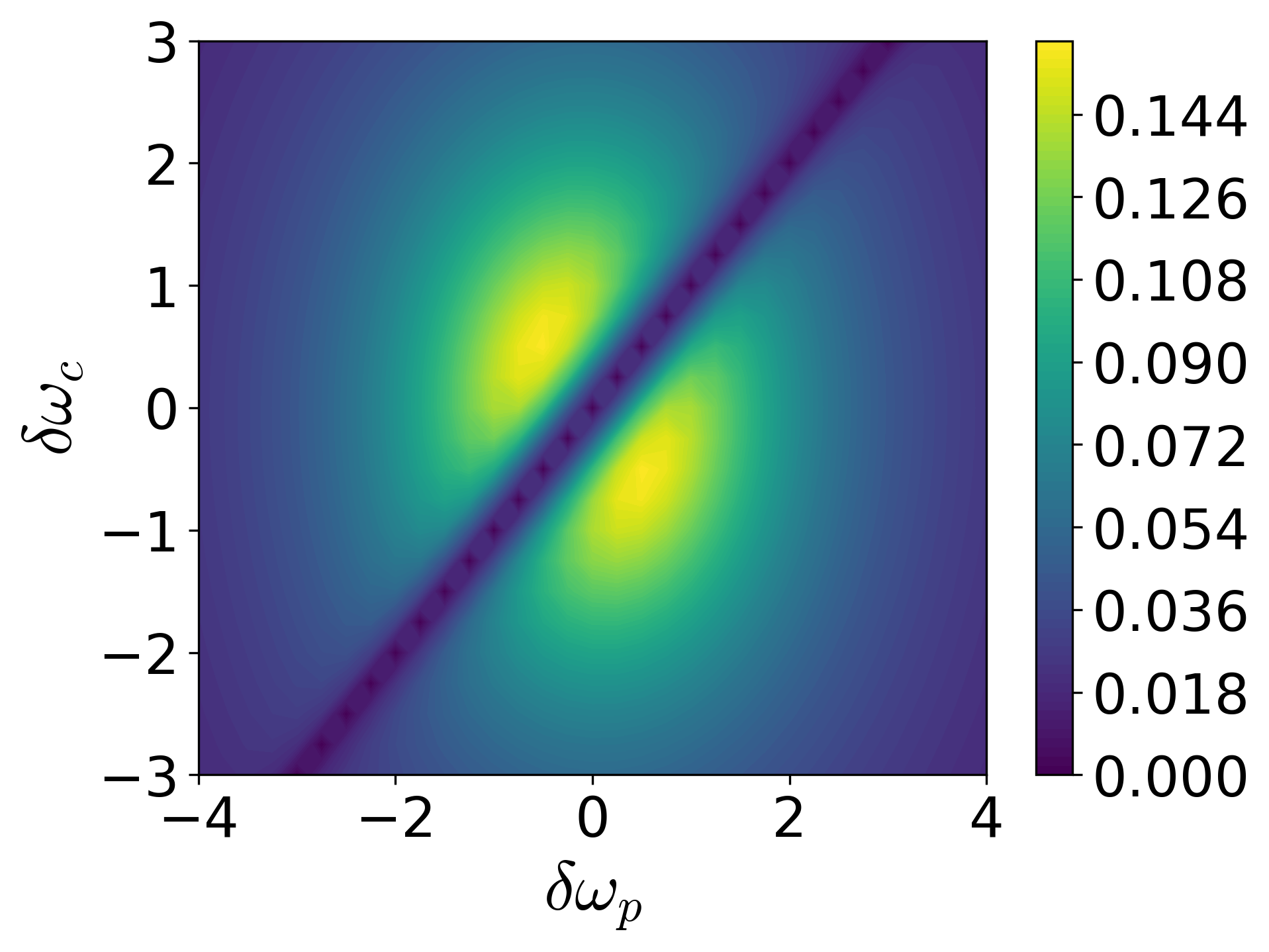}  \\
(c)\makebox[0.4\columnwidth]{ }(d)\makebox[0.4\columnwidth]{ }\\
\includegraphics[width=0.45\columnwidth]{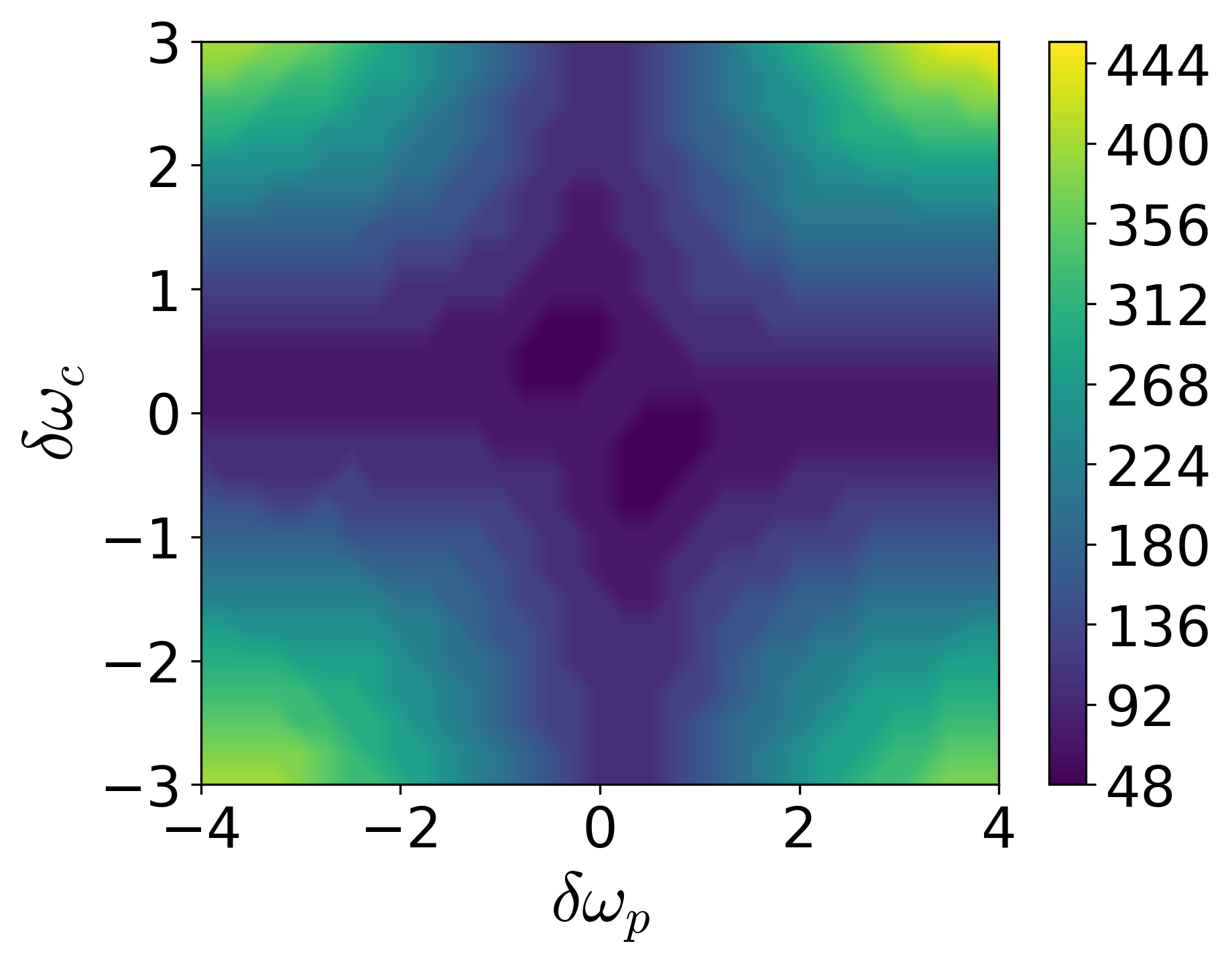}\hspace{.1in}\includegraphics[width=0.45\columnwidth]{Figures/BII_3D_time.png} 
\caption{Steady state limits of populations at state 2 ($\rho_{22}$) for the open system of case B-II with full Hamiltonian (a) and RWA (b), for which  times required for convergence are respectively plotted in (c) and (d). }
\label{fig:3D_A}
\end{figure}

\begin{figure}
\centering
(a)\makebox[0.4\columnwidth]{ }(b)\makebox[0.4\columnwidth]{ }\\
\includegraphics[width=0.45\columnwidth]{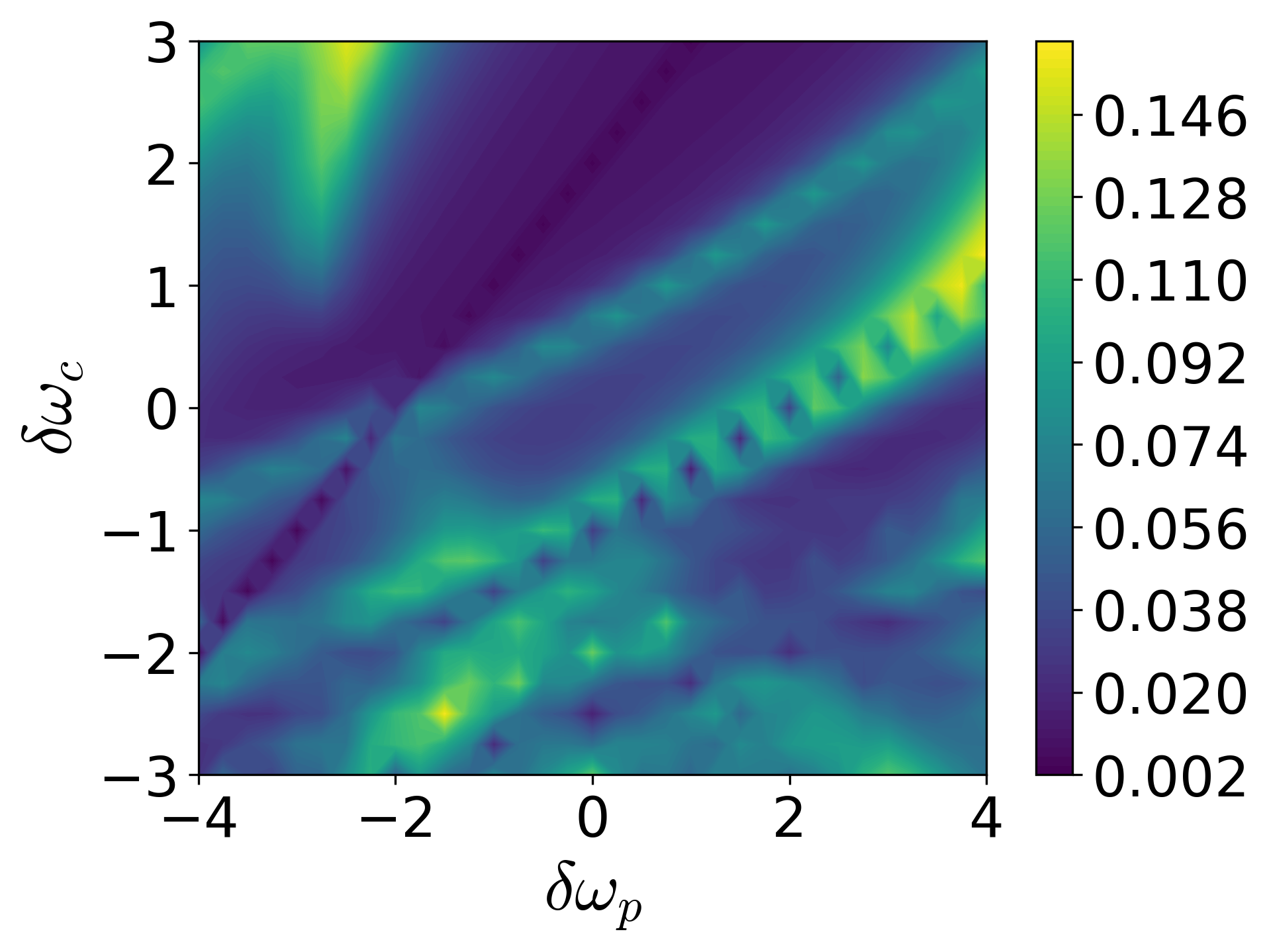}\hspace{.1in} \includegraphics[width=0.45\columnwidth]{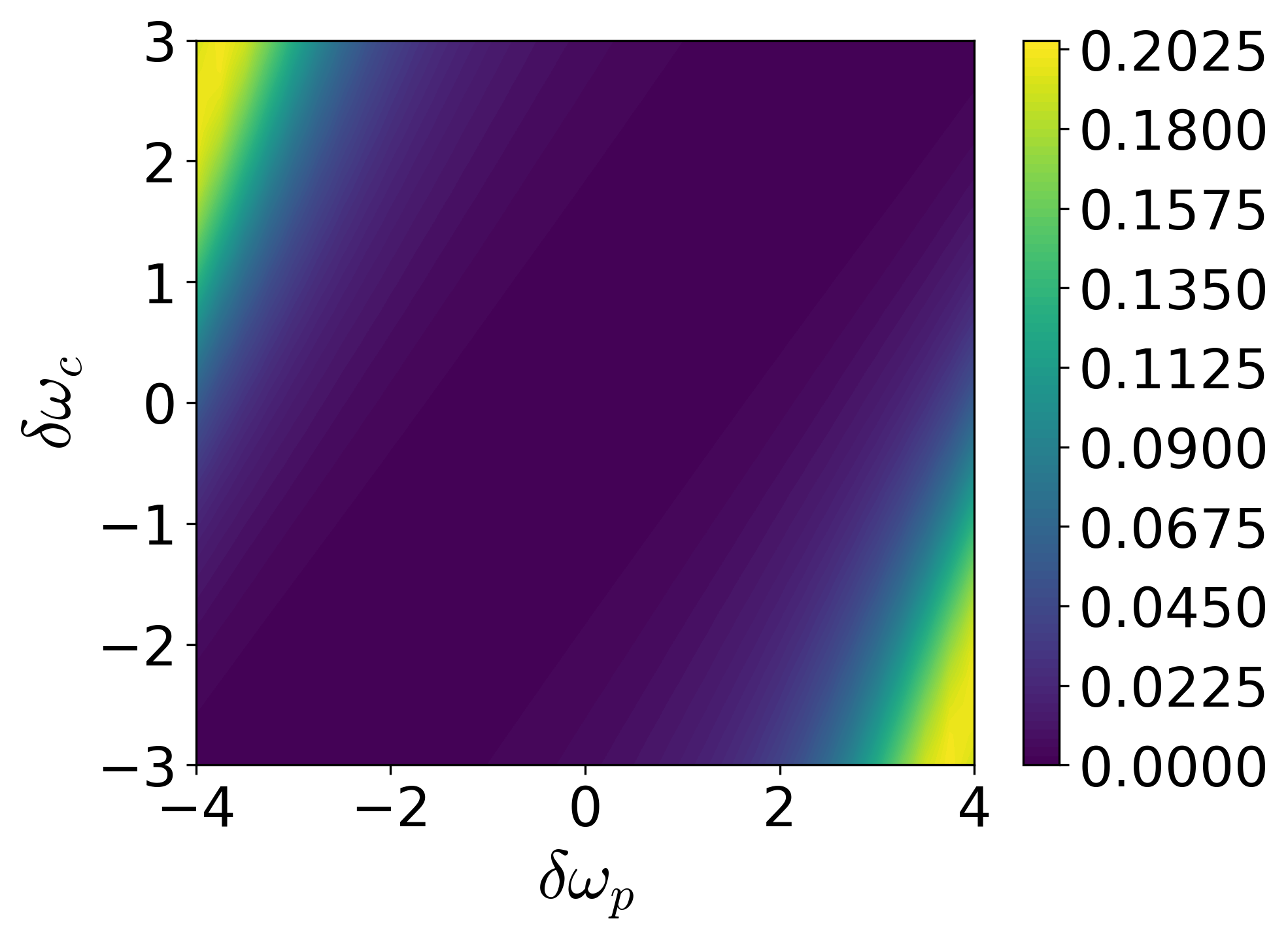} \\
(c)\makebox[0.4\columnwidth]{ }(d)\makebox[0.4\columnwidth]{ }\\
\includegraphics[width=0.45\columnwidth]{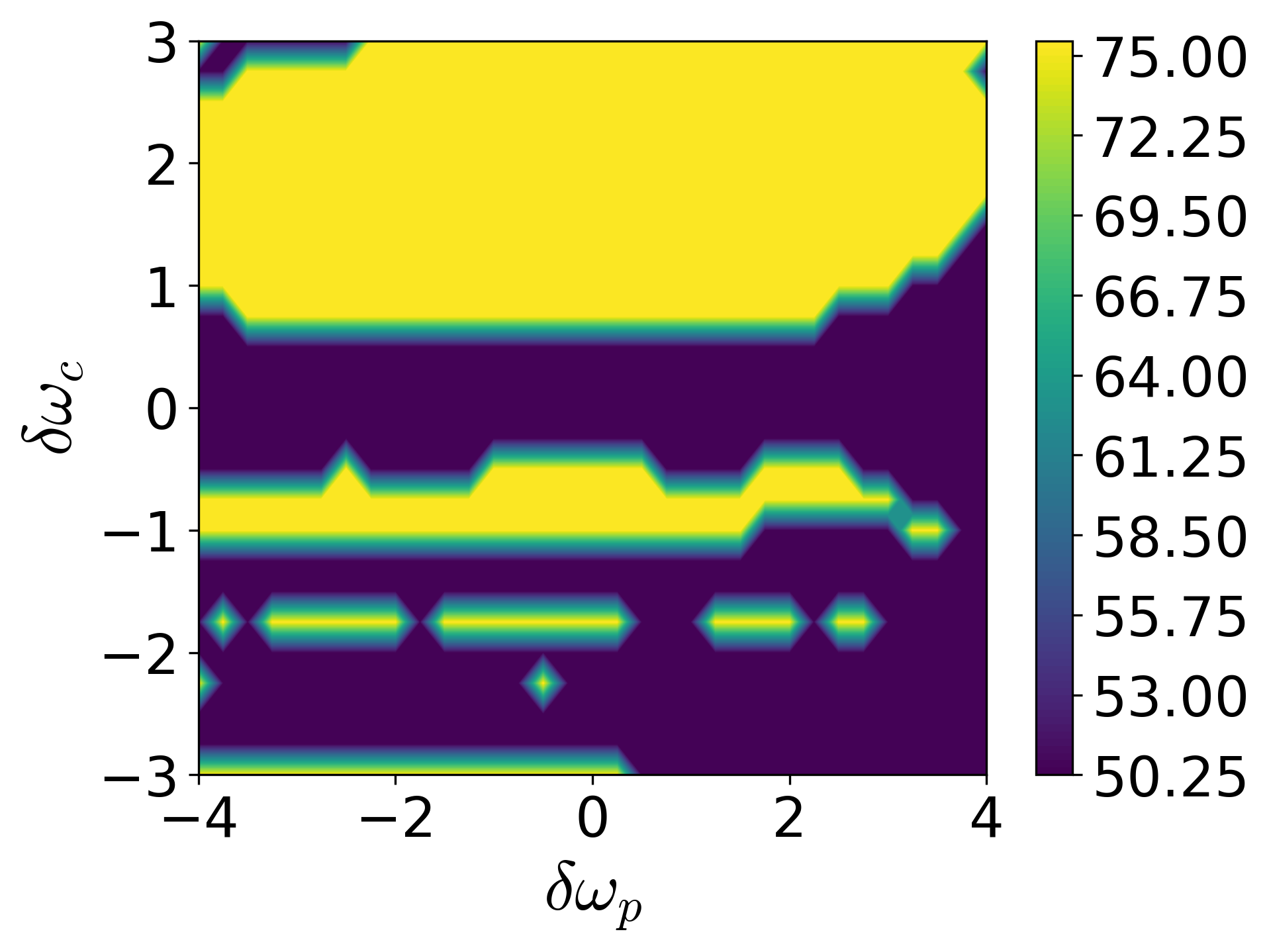}\hspace{.1in}\includegraphics[width=0.45\columnwidth]{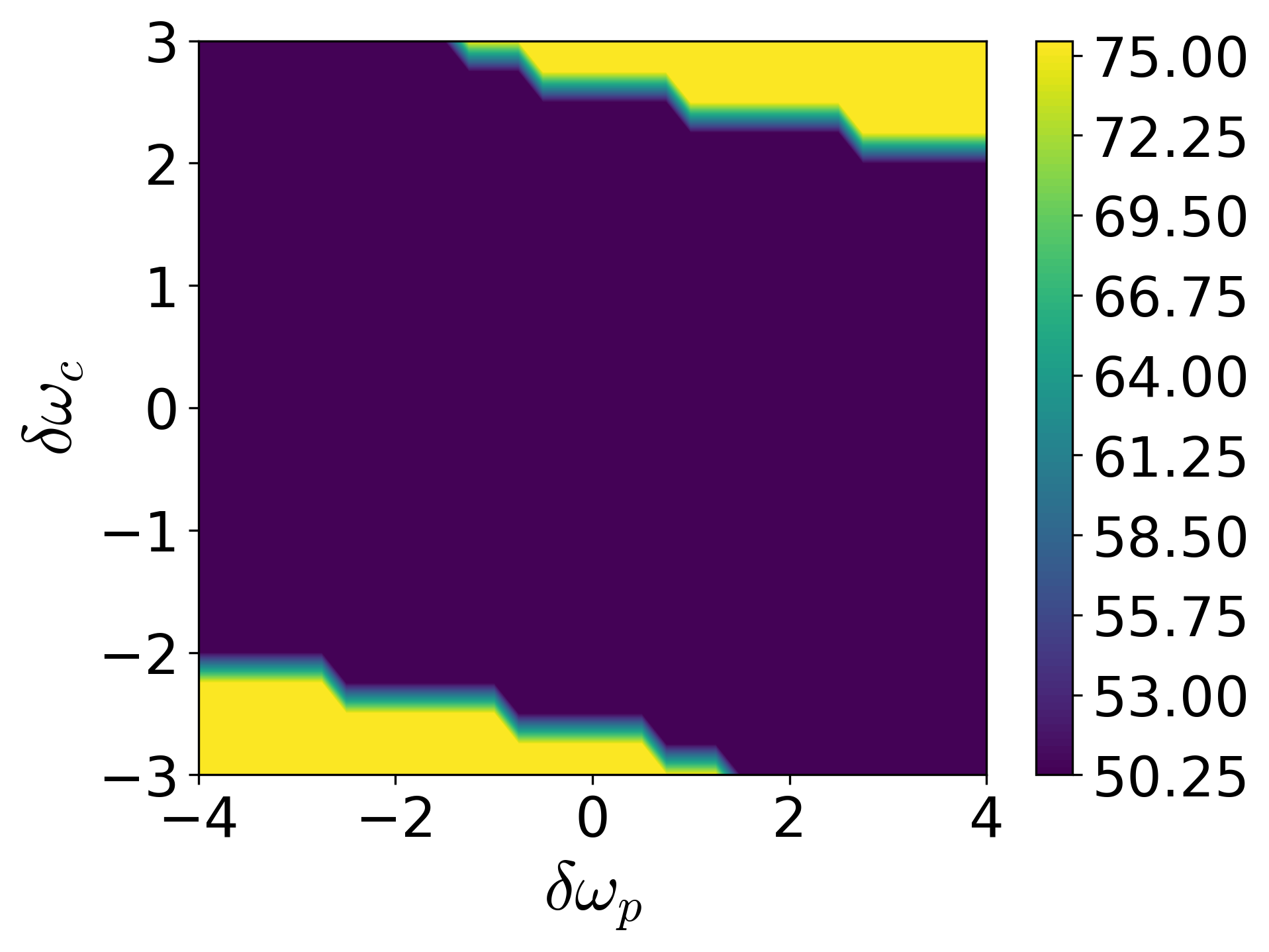} 
\caption{Steady state limits of populations at state 2 ($\rho_{22}$) for the open system of case C-II with full Hamiltonian (a) and RWA (b), for which  times required for convergence are respectively plotted in (c) and (d).  }
\label{fig:3D_C}
\end{figure}

Note that results presented so far \rc{do not cover the parameter regime where the TPR with a non-zero detuning ($\delta\omega_{p}=\delta\omega_{c} \neq 0$) occurs}, for which the phenomenon of CPT can still be observed. In recent works,\cite{vezvaee-prxq4,genov-jpb56}  these modified detunings were used in order to avoid transitions to an unwanted level, \cite{vezvaee-prxq4} and to swap populations of the two lower-lying states.\cite{genov-jpb56} Figures \ref{fig:3D_A} and \ref{fig:3D_C} provide \rc{such }results for cases B-II and C-II, respectively, 
for $\omega_c\ \in [1,1.25,...,6.75,7]$ and $\omega_p\ \in [2,2.25...,9.75,10]$.  A similar figure for case A-II is provided in the SM.

For weak and moderate field strengths, it is seen that there is no population in the excited state in the steady state limit under TPR conditions (the main diagonal in panels (a) and (b)). This explains why TPR is a requirement for stimulated Raman adiabatic passage (STIRAP),\cite{vitanov-rmp89} which aims to transfer population from state $|1\rangle$ to state $|3\rangle$ without significantly populating the excited state. However, for the strong control field case of C-II in Fig. 9, we see that the RWA population in panel (b) is a poor predictor of
the true population in panel (a). In particular, the population is a much more complicated function of detunings, and the intermediate state 2 has non-negligible population
even when the TPR condition under the assumption of the RWA is satisfied. This suggests  \rc{that} principles of STIRAP need to be examined more carefully beyond the RWA.

\subsection{Comparison of different fourth-order methods}
\rc{In order }to assess the accuracy and stability of our simple 4th order propagators, 
we compared ours with two widely used 4th order methods,\cite{butcher,griffiths-nm} explicit 4th order Runge-Kutta (RK4) method\cite{butcher} and implicit 4th order Adams-Moulton (AM4) method,\cite{griffiths-nm}  We also implemented an optimized 4th order CFME (O4-CFME)\cite{alvermann-jcp230,alvermann-njp14}  that utilizes three exponentials and Gauss-Legendre points.  Appendix \ref{sec:AM4_RK4} \rc{provides summaries of these three methods.}  
The error of each method was \rc{quantified by the following definition of error: }
\be	
\mbox{error}=\frac{|\mathcal{U}(t)-\mathcal{U}_{ref} (t)|_F}{|\mathcal{U}_{ref}(t)|_F}, \label{eq:error_frob_ref}
\ee \\
where  \rc{$\mathcal{U}_{ref}$ represents calculation results based on our 6th order ME propagator.}\cite{ture-jang-jpca124}  Errors for both short time ($t = 2\pi$) and long time ($t=16\pi$) \rc{dynamics were calculated for each algorithm}.

\begin{figure}
	\centering
	\includegraphics[width=0.85\columnwidth]{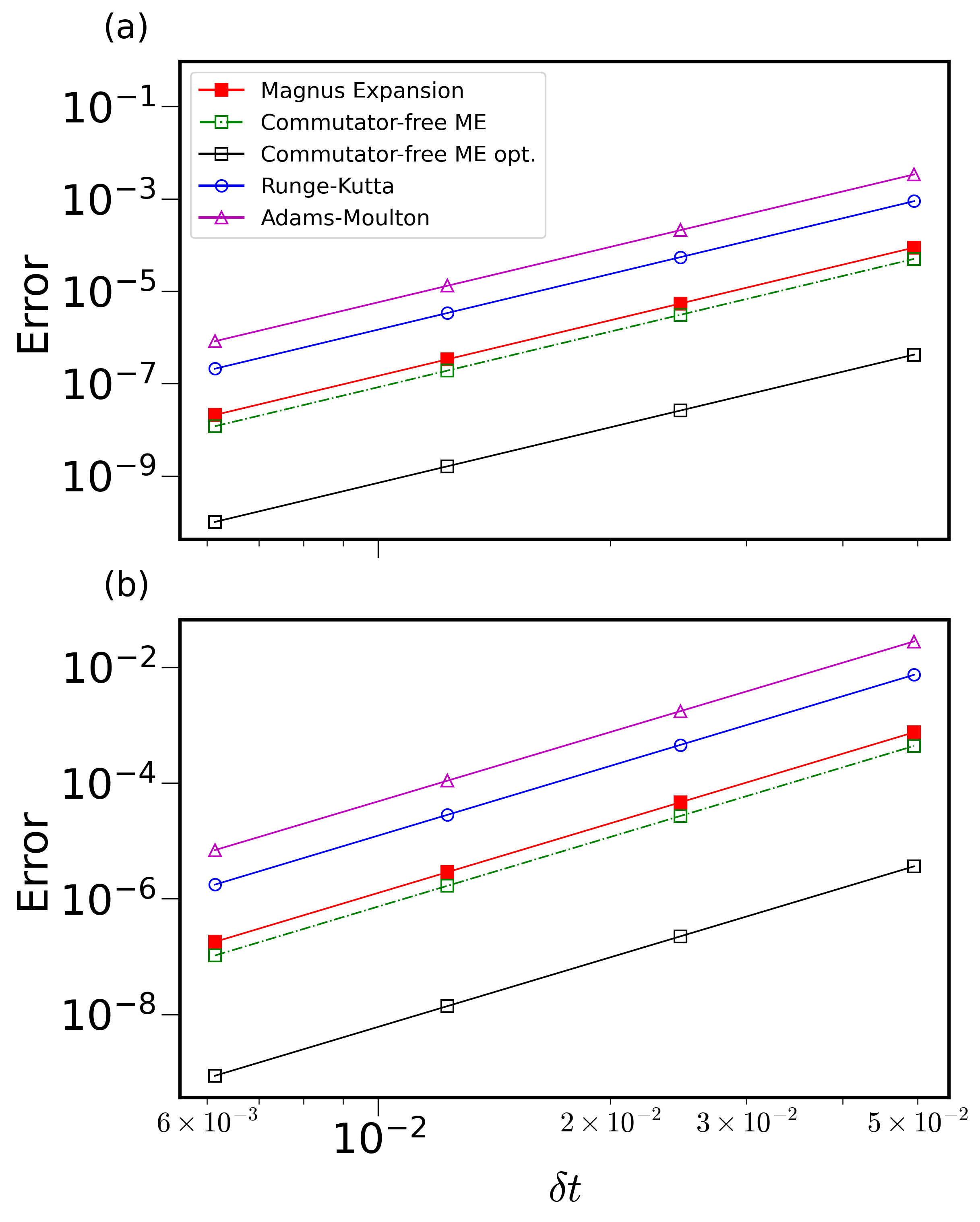}\\
	\caption{Comparison of 4th order ME-based methods with the RK4 and AM4 methods for the closed-system dynamics of case B-I (a) for short time evolution ($t = 2\pi$) and (b) for long time evolution ($t=16\pi$). The error was calculated using Eq. (\ref{eq:error_frob_ref}). } \label{fig:B_closed_error}
\end{figure}

\rc{Figure \ref{fig:B_closed_error} shows results} for the closed-system dynamics of case B-I. We see that our 4th order ME and CFME propagators have about an order of better accuracy than the RK4 and AM4 methods, both in the short-time and the long-time limits.  However, our results are significantly less accurate than those of the O4-CFME method.\cite{alvermann-jcp230,alvermann-njp14} This is understandable 
\rc{considering }that the latter utilizes 
one more \rc{value of time for each step of integration} and one more exponential.  In addition, the superior performance of O4-CFME suggests the benefit of using Gauss-Legendre points and at least three exponentials if possible. 

\begin{figure}
	\centering
	\includegraphics[width=0.85\columnwidth]{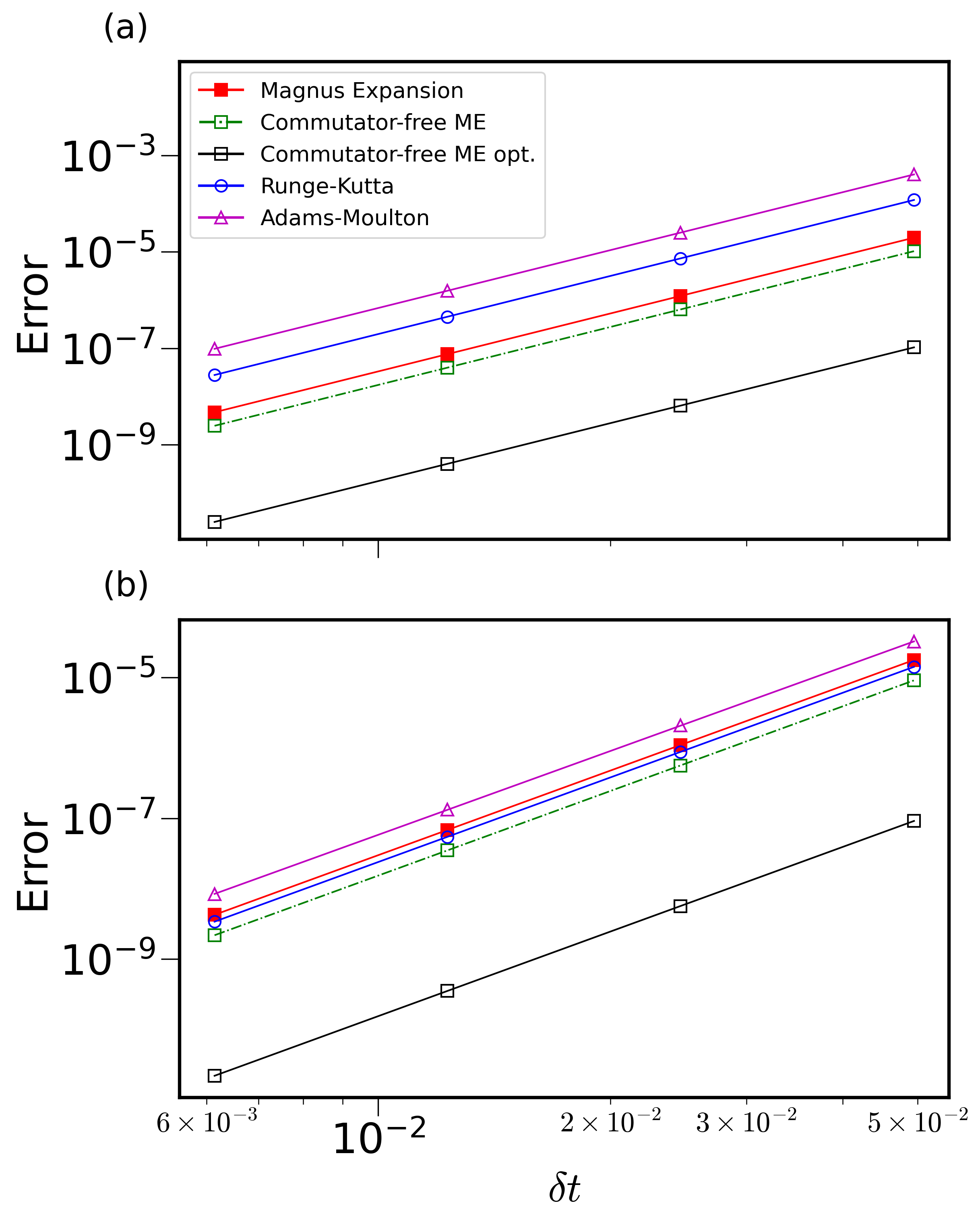}\\

	\caption{Comparison of  4th order ME-based methods with the RK4 and AM4 methods for the open-system dynamics of case B-II (a) for short time evolution ($t = 2\pi$) and (b) for long time evolution ($t=16\pi$). The error was calculated using Eq. (\ref{eq:error_frob_ref}) } \label{fig:B_open_error}
\end{figure}

\rc{Figure \ref{fig:B_open_error} shows }results for the open-system dynamics of case B-II. 
While the short-time dynamics show that our ME-based methods work better than the RK4 and AK4 methods, its benefit worsens for the long-time dynamics. As a result, the accuracy of our 4th order ME propagator with commutator becomes comparable to the RK4 method, whereas our CFME method remains more accurate than the latter.  On the other hand, the performance of the O4-CFME method\cite{alvermann-jcp230,alvermann-njp14} remains best even in this case. Similar behavior can be seen for all models of open-system dynamics considered in this work with some exceptions (see the SM).

The results presented here demonstrate the best performance of CFME and the benefit of using Gauss-Legendre points if possible.   
While our 4th order CFME method is less accurate than O4-CFME, its performance is consistently better than RK4 and AM4.  Thus, it is expected to serve as a good choice offering both reasonable accuracy despite its simplicity for a broad range of closed and open system dynamics problems.  

\section{Conclusion}
 The main result of \rc{this work} is the extension and test of our simple 4th order ME-based propagator to the Liouville space for both closed and open system dynamics, as represented by Eqs. (\ref{eq:cfme-h}), (\ref{eq:us-4th}), (\ref{eq:usl-4th}), and (\ref{eq:usl-4th-cf}).   These propagators are simple to use and can be naturally incorporated into broad range of quantum dynamics calculations while based on evaluation of Hamiltonians at a mid point in addition to end points of each time step.  In particular, our 4th order CFME, Eq. (\ref{eq:usl-4th-cf}) was confirmed to be consistently better than the RK4 and AM4 methods while being simple to implement.    
 
 Another important aspect of our work is detailed assessment of the RWA, which has been used widely due to its convenience and conceptual clarity, and is indeed well established as an accurate approximation for conventional spectroscopy measurements in the weak field limit.  However, beyond that limit and/or in the presence of multiple pulses, its reliability remains poorly understood. Considering recent experimental advances in nonlinear spectroscopy,\cite{mukamel} QC, and QS and the need for more accurate calculation methods of optical signals involved, better understanding of how the RWA breaks down is an important issue.  Results of our calculation for a prototypical $\Lambda$-system are consistent with the general notion that the RWA 
 \rc{is} reliable in the weak field and steady state limits.  However, for moderate control and probe field strengths, we find some of the results based on the RWA Hamiltonian exhibit  subtle deviations from  those of full Hamiltonian.   For a strong control field strength, there are substantial differences at the qualitative level even though the probe field remains moderate.  Although these have been expected, the actual quantitative comparison \rc{provided here} can help better assessment of the RWA in general.

Our numerical tests including the O4-CFME method\cite{alvermann-jcp230,alvermann-njp14} demonstrate superior performance of the ME-based approaches and the benefit of using more Gauss-Legendre points and more exponentials instead of commutators.  As yet, these results are limited to the $\Lambda$-model with periodic time dependences as considered here.  Thus, further tests are needed for other systems and more complex time dependences in order to establish a reliable general trends of performance.  \rc{In fact, all of our ME-based propagators are easily applicable to other types of systems\cite{gemmer,grifoni-pr304,jang-njp15,xu-njp18}  that are important in the context of quantum thermodynamics or energy/charge transport.  In addition, although we considered only the case of periodic driving, our propagators can easily be applied to non-periodic driving equally well.  In this sense, our approaches are complementary to those based on Floquet approaches.\cite{casas-jpa34,grifoni-pr304,mori-arcmp14,eckardt-njp17,eckardt-rmp89,kuwahara-ap367,engelhardt-prl123} } 

It is also noteworthy to mention the stability of our ME based numerical method for the non-Hermitian dynamics (e.g. Figs. \ref{fig:open_both}).  We found that 
the trace of unity is maintained with only negligible deviations (on the order of $10^{-12}$ in the worst case) due to numerical error, even in the long-time limit.   This suggests that, for the time-dependent system Hamiltonian $\hat{H}(t)$ with time-independent Lindblad dissipative terms as in  Eq. (\ref{eq:lindblad}), the ME truncated at any order produces a trace preserving map. In fact, we find numerically that it produces CPTP maps, which not only guarantee that the trace is maintained but ensure that the resulting density operator is physical even if it only describes a subsystem. An analytical proof for the closed system case is shown in Appendix \ref{sec:positive_definite}. For the open system \rc{dynamics}, \rc{the CPTP map is produced by our CFME expression, Eq. (\ref{eq:usl-4th-cf}), at any time step.  However,   for our expression with commutator,  Eq. (\ref{eq:usl-4th}), we confirmed numerically that the CPTP map is preserved only for small enough time step sizes, }which ensure that the first order term of the ME is larger than the subsequent terms.  This can be understood from the fact that  the first order term of the ME defines a CPTP map, while it has been shown that higher order terms contribute to breaking the complete positivity.\cite{mizuta-pra103}      

As a further extension of this work, one can imagine using the nearly exact numerical time evolution provided here to produce large quantities of data for a deep learning or reinforcement learning algorithm, with the goal of finding improved pathways (beyond well-known shortcuts to adiabaticity methods) to produce the desired population transfer in a short amount of time, which is also relatively insensitive to experimental parameters.  \vspace{.1in}\\

\noindent
{\bf SUPPLEMENTARY MATERIAL}\vspace{.1in}\\
The Supplementary Material provides additional calculation results for cases in Table \ref{table-1} and provides a link to the site where python codes can be accessed.

\acknowledgments
S.J.J. acknowledges major support of this research from the US Department of Energy, Office of Sciences, Office of Basic Energy Sciences (DE-SC0021413, DE-SC0026114) and partial support during the initial stage of this research from the National Science Foundation (CHE-1900170).  S.J.J. also acknowledges travel support from Korea Institute for Advanced Study through its KIAS Scholar program. \\

\noindent
{\bf  AUTHOR DECLARATIONS} \vspace{.1in}\\
{\bf Conflict of Interest} \vspace{.1in}\\
The authors have no conflicts to disclose. \vspace{.2in}\\
\noindent
{\bf  DATA AVAILABILITY} \vspace{.1in}\\
Most data that support the findings of this article are contained in this article.  Additional data and programs for all computation are available in supplementary materials, and further data and information are available from the corresponding author upon request.

\appendix
\rc{
\section{Derivation of simple commutator free Magnus Expansion propagators}
\label{sec:cfme-der}
We here provide the derivations of the commutator free propagators defined in Eqs. (\ref{eq:cfme-h}) and (\ref{eq:usl-4th-cf}). For this, let us consider the following generic first order differential equation:
\be
\frac{dy(t)}{dt} = f(t,y(t)) = A(t)y(t) ,\label{eq:diffeq-gen}
\ee
where $y(t)$, $f(t,y(t))$, and $A(t)$ are all operators and we omit explicit operator notation in this section.
For the case of dynamics in the Hilbert space, $f(t,y(t))=(-i\hat H(t)/\hbar) y(t)$, with $y(t)$ the time evolution operator.  For the case of the dynamics in the Liouville space,   $f(t,y(t))=-i{\mathcal L}(t) y(t)$, with $y(t)$ the density operator.  For the above form of the differential equation, application of our 4th order ME propagator\cite{ture-jang-jpca124} results in
\be
y(t+\delta t) = e^{\delta t c_0 + \delta t^3 \frac{1}{12}c_2 - \delta t^3\frac{1}{12}[c_0, c_1] }y(t) + \mathcal{O}(\delta t^5), \label{eq:ME_taylor_coeffs}
\ee
where $c_n$ is the $n$th Taylor series coefficient of $A(t)$ and is given by $c_n = \frac{1}{n!} \frac{d^n}{d\tau^n}A(\tau)|_{\tau = t +\delta t/2}$. To develop a commutator free version, we look for a propagation of the following form
\begin{widetext}
\be
y(t+\delta t) = e^{\delta t(\text a_{20} c_0 + \text a_{21} \delta t c_1 + \delta_t^2 \text a_{22} c_2)}e^{\delta t(\text a_{10} c_0 + \delta t\text a_{11} c_1 + \delta t^2\text a_{12} c_2)}y(t) + \mathcal{O}(\delta t^5), \label{eq:CFME_taylor_coeffs}
\ee
\end{widetext}
where $\text a_{ij}$ are undetermined coefficients. The stability requirement constrains the coefficients to satisfy $\text a_{10} = \text a_{20}$, $\text a_{11} = - \text a_{21}$, $\text a_{12} = \text a_{22}$. The remaining free coefficients can be determined by applying the BCH formula to Eq. (\ref{eq:CFME_taylor_coeffs}), equating the resulting expression with the exponent in Eq. (\ref{eq:ME_taylor_coeffs}), and separating terms by their Taylor series coefficients.  This procedure results in the following equations:
\ben
2\delta t \text a_{10}c_0 &=&
\delta t c_0 \nonumber \\
2\delta t^3 \text a_{12}c_2 
&=&
\delta t^3 \frac{1}{12}c_2 \nonumber \\ 
\delta t^3  \text a_{10}\text a_{11} [c_0, c_1]
&=&
- \delta t^3 \frac{1}{12}[c_0, c_1]\re{,}
\een
where equations with $\mathcal{O}(\delta t^5)$ have been discarded.  Solving the above equations gives $\text a_{10} =\frac{1}{2}$, $\text a_{11} = -\frac{1}{6}$, $\text a_{12} = \frac{1}{24}$, which determines all the coefficients. The expression can now be converted to one involving three equally spaced points by using the corresponding Legendre interpolating polynomial for $A(t)$, and substituting in the definition of $c_n$. Combining like terms leads to our final commutator free expression:
\begin{widetext}
\be
y(t+\delta t) = e^{\frac{\delta t}{12}(-A_k +4A_{k + 1/2} + 3A_{k + 1})} e^{\frac{\delta t}{12}(3A_k + 4A_{k + 1/2} - A_{k+1})}y(t) + \mathcal{O}(\delta t^5) ,
\ee
\end{widetext}
where $A_k = A(t_k)$.
}

\section{Liouville-space transformation}
\label{sec:diff_propagators}
Liouville space representation\cite{fano-1964,mukamel,gyamfi-ejp41} offers a convenient way to treat time evolution of density operators by representing them as vectors (or superkets) and have long been used for open system quantum dynamics and spectroscopy.    It is important to note that the mapping from an operator in the Hilbert space to a superket in the Liouville space is not unique and thus care should be taken to maintain a consistent definition.  We here employ the approach described recently by Gyamfi.\cite{gyamfi-ejp41} According to this prescription, the mapping from a Hilbert space to the corresponding Liouville space is implemented by the following mapping that flips the bra side of an arbitrary outer product and creates a direct product as follows:
\be
|a\rangle\langle b| \rightarrow |a\rangle \otimes |b\rangle^*  ,  \label{eq:flipper}
\ee   
where $|b\rangle^*$ is the complex conjugate of $|b\rangle$.  According to this mapping, the density operator is mapped to a superket as follows:
\be
\hat{\rho} = \sum_i \sum_j\rho_{ij}\ket{i}\bra{j} \rightarrow   \sum_{ij}\rho_{ij}\ket{i}\otimes\ket{j}^* \equiv \ket{\rho}\rangle .
\ee
Then, all the operator identities in the Hilbert space can be mapped\cite{gyamfi-ejp41} uniquely into the Liouville space employing the mapping defined by Eq. (\ref{eq:flipper}). 
For the description of the time evolution of the density operator,  the main relationship needed is the following mapping of triple product of operators (in the Hilbert space):\cite{gyamfi-ejp41}
\be
\hat{A}\hat{\rho}\hat{C} \rightarrow  (\hat{A} \otimes \hat{C}^T)\ket{\rho}\rangle =|A\rho C\rangle\rangle,
\ee
where $(\hat{A} \otimes \hat{C}^T)$ is a super-operator acting on the superket $|\rho\rangle\rangle$ (in the Liouville space) from the lefthand side. 

\section{Proof that truncated Magnus expansion produces a completely positive trace preserving (CPTP) map}
\label{sec:positive_definite}
Using the same Liouville space notation as in Appendix \ref{sec:diff_propagators}, a map is CPTP if and only if it can be written in the following form:\cite{choi-laa10}
\ben
\label{eq:CP_map}
\mathcal{U}(t,0) \ket{\rho}\rangle = \sum_{k}(\hat S_k \otimes \hat S_k^*) \ket{\rho}\rangle,
\een
where $S_k$ is called the Krauss operator and sum of these must be the identity operator in the system Hilbert space $\hat I$ as follows:
\ben
\label{eq:TP_map}
\sum_{k} \hat S_k^\dagger \hat S_k = \hat I .
\een

Application of the Magnus expansion truncated at the $n$th order to the closed system dynamics in Eq.  (\ref{eq:closed_Liouville}) yields  
\begin{align}
\label{eq:ME_superoperator}
&\mathcal{U}^{(N)}(t,0)\ket{\rho}\rangle \nonumber\\
&= \exp\left(	\sum_{n=1}^{N}\frac{1}{n!}\left(-\frac{i}{\hbar}\right)^n\mathcal{M}_n[\hat{H}(t)\otimes \hat I
-\hat I\otimes \hat{H}(t)^T]\vphantom{\sum_{n=1}^{N}}\right)		\ket{\rho}\rangle,  	
\end{align}
where $\mathcal{M}_n$ is a functional of superoperators in the exponent of Eq. (\ref{eq:closed_Liouville}), consisting of iterated integrals and commutators (for precise definitions, see Ref. \citenum{ture-jang-jpca124})  . However, note that because 	 $[\hat{H}(t)\otimes \hat I,\hat I\otimes \hat{H}(t)^T] = 0$, it becomes possible to separate the exponent in Eq. (\ref{eq:ME_superoperator}) into two terms:
\ben
&&\mathcal{U}^{(N)}(t,0)\ket{\rho}\rangle = \left(\exp\left(\sum_{n=1}^{N}\frac{1}{n!}\left(-\frac{i}{\hbar}\right)^nM_n[\hat{H}]\right) \otimes \hat I\right) \nonumber \\
&&\hspace{.3in}\times \left(\hat I \otimes \exp\left(\sum_{n=1}^{N}\frac{1}{n!}\left(-\frac{i}{\hbar}\right)^nM_n[-\hat{H}^T]\right)\right)\ket{\rho}\rangle ,\nonumber \\ \label{eq:un-rho-2}
\een
where $e^{\hat{H} \otimes \hat I} =e^{\hat{H}} \otimes \hat I  $ and $e^{\hat I \otimes \hat{H}^T} = \hat I \otimes e^{\hat{H}^T} $ was used. Now, multiplying and using the fact that $M_n[-\hat{H}^T] = (-1)^nM_n[\hat{H}^*]$, Eq. (\ref{eq:un-rho-2}) can be expressed as 
\ben
\mathcal{U}^{(N)}(t,0)\ket{\rho}\rangle =&& \left(\exp\left(\sum_{n=1}^{N}\frac{1}{n!}\left(-\frac{i}{\hbar}\right)^nM_n[\hat{H}]\right)\right. \nonumber \\ \otimes &&\left.\exp\left(\sum_{n=1}^{N}\frac{1}{n!}\left(\frac{i}{\hbar}\right)^nM_n[\hat{H}^*]\right)\right)\ket{\rho}\rangle, \nonumber 
\een
which clearly has the form of Eq. (\ref{eq:CP_map}) with $\hat S=\exp\left(\sum_{n=1}^{N}\frac{1}{n!}\left(-\frac{i}{\hbar}\right)^nM_n[\hat{H}]\right)$. 

Since $\hat S$ is guaranteed to be unitary at any finite order,  Eq. (\ref{eq:TP_map}) is also satisfied for any finite order approximation for the ME. This proof shows that the Hilbert space property of the ME (guaranteed unitary operator at any order of truncation) translates into the corresponding Liouville space property of CPTP map for Hermitian dynamics. Also in analogy with the Hilbert space situation, propagators based on time-dependent perturbation theory fail to be CPTP maps; they clearly violate Eq. (\ref{eq:TP_map}).  Note that the above proof for closed system dynamics remains true even with RWA.

\section{Closed system dynamics in the rotating wave frame (RWF)}
\label{sec:rotating_frame}
The RWF is defined by the following time dependent unitary operator:
\be
\hat{W}(t) = \ket{1}\bra{1} + e^{-i\omega_pt}\ket{2}\bra{2} + e^{-i(\omega_p-\omega_c)t}\ket{3}\bra{3}  .\label{eq:rotating_frame}
\ee
Thus, for a state $|\psi(t)\rangle$,  the corresponding state in the RWF $|\tilde \psi(t)\rangle$ is defined as
\be
\ket{\psi (t)} = \hat W(t)\ket{\tilde \psi(t)} .
\ee
Given that the time evolution of $\ket{\psi (t)}$ is governed by $\hat H^{\rm RWA} (t)$, which is the sum of Eqs. (\ref{eq:hs0}) and (\ref{eq:hsr-rwa}), the time evolution equation for $\ket{\tilde \psi(t)}$  is 
\ben
i\hbar\frac{\partial}{\partial t} \ket {\tilde \psi(t)} &=&\left (i\hbar \frac{\partial}{\partial t} \hat W^\dagger(t) \right)|\psi (t)\rangle\nonumber \\
&&+\hat W^\dagger(t)\hat H^{\rm RWA} (t) |\psi (t)\rangle \nonumber \\
&=&\hat {\tilde H}_{I}^{\rm RWA} |\tilde \psi (t)\rangle ,
\een 
where 
\ben
\hat {\tilde H}_{I}^{\rm RWA} &=&\left (i\hbar \frac{\partial}{\partial t} \hat W^\dagger(t) \right) \hat W(t)+\hat W^\dagger(t)\hat H^{\rm RWA} (t)\hat W(t) \nonumber \\
&=& \hat H_0-\hbar\omega_p|2\rangle\langle 2|-\hbar (\omega_p-\omega_c)|3\rangle\langle 3| \nonumber \\
&&-\frac{\hbar\Omega_p}{2} \left (|1\rangle\langle 2|+|2\rangle\langle 1| \right ) \nonumber \\
&&-\frac{\hbar\Omega_c}{2} \left (|2\rangle\langle 3|+|3\rangle\langle 2| \right ) .
\een
For a density operator in the original frame, we can also define the density operator in the RWF as follows:
\be
\hat {\tilde \rho}(t)=\hat W^\dagger(t)\hat \rho(t) \hat W(t) .
\ee
The diagonal components of $\hat {\tilde \rho}(t)$ are the same as those of $\hat \rho_s(t)$.  On the other hand, the off-diagonal components are related by
\ben
\tilde{\rho}_{12}(t) &=& e^{-i\omega_pt}\rho_{12} , \label{eq:rho_tilde_def-12}\\ 
\tilde{\rho}_{13}(t) &=& e^{-i(\omega_p-\omega_c)t} \rho_{13}(t) , \label{eq:rho_tilde_def-13}\\ 
\tilde{\rho}_{23}(t) &=& e^{i\omega_ct}\rho_{23}(t)  , \label{eq:rho_tilde_def-23}
\een
and complex conjugates of the above relations.

The time evolution operator for $\ket{\tilde \psi (t)}$ can be found  easily by diagonalizing $\hat {\tilde H}_{I}^{\rm \small RWA}$, which is time independent.
Assuming that $E_1=0$, $E_2=\hbar\omega_2$, and $E_3=\hbar\omega_3$,  the matrix representation of $\hat {\tilde H}_{I}^{\rm RWA}$ in the basis of $|1\rangle$, $|2\rangle$, and $|3\rangle$ is given by
\be
\hat{\tilde H}_{I}^{\rm \small RWA}=
\frac{\hbar}{2}\begin{pmatrix}
0 &&-\Omega_p && 0 \\
-\Omega_p && 2(\omega_2-\omega_p) && -\Omega_c\\
0 && -\Omega_c && 2(\omega_3-(\omega_p-\omega_c))
\end{pmatrix} .
\label{eq:time_independent}
\ee

\section{Solution of closed system dynamics using RWA at the TPR conditions} 	
\label{RWA-solution}
Diagonalizing Eq. (\ref{eq:time_independent}) involves solving a cubic equation for the eigenvalues, which can be quite cumbersome. However, under TPR conditions with zero detuning, the resulting cubic equation is simpler to solve because zero is a root of the characteristic polynomial. The other two eigenvalues are given by $\lambda$ and $-\lambda$, where $\lambda =\frac{1}{2}\sqrt{\Omega_c^2+\Omega_p^2}$.\cite{shore-qc} Using these values and their associated eigenvectors, we find that the time evolution operator for 
$\ket {\tilde \psi(t)}$ is given by 
\begin{widetext}
\begin{align}
\hat{\tilde U}_{I}^{\rm \small RWA}(t) &=\frac{\Omega_c^2+\Omega_p^2\cos(\lambda t)}{\Omega_c^2+\Omega_p^2}\ket{1}\bra{1} + i\frac{\Omega_p\sin(\lambda t)}{2\lambda}\ket{1}\bra{2}\nonumber +\frac{\Omega_c\Omega_p(\cos(\lambda t)-1)}{\Omega_c^2+\Omega_p^2}\ket{1}\bra{3}\nonumber\\ 
&+
i\frac{\Omega_p\sin(\lambda	 t)}{2\lambda}\ket{2}\bra{1}\nonumber 
 + \cos(\lambda t) \ket{2}\bra{2} +i\frac{\Omega_c\sin(\lambda t)}{2\lambda}\ket{2}\bra{3}\nonumber \\
& + \frac{\Omega_c\Omega_p(\cos(\lambda t)-1)}{\Omega_c^2+\Omega_p^2}\ket{3}\bra{1}+i\frac{\Omega_c\sin(\lambda t)}{2\lambda}\ket{3}\bra{2}+ \frac{\Omega_c^2\cos(\lambda t)+\Omega_p^2}{\Omega_c^2+\Omega_p^2}\ket{3}\bra{3} .
\end{align}
Employing the above expression, $\hat {\tilde \rho}(t)=\hat{\tilde U}_{I}^{\rm \small RWA}(t) \hat {\tilde \rho}(0) \left (\hat{\tilde U}_{s,I}^{{\rm \small RWA}} (t) \right)^\dagger$ can be calculated for the initial condition $\hat {\tilde \rho}(0)=\ket{1}\bra{1}$.  The resulting matrix expression in the basis of $|1\rangle$, $|2\rangle$, and $|3\rangle$ is 
\be
\hat{\tilde \rho}(t) = \begin{pmatrix}\frac{2\Omega_c^4 + \Omega_p^4+ 4\Omega_c^2\Omega_p^2\cos(\lambda t) + \Omega_p^4\cos(2\lambda t) }{2\left(\Omega_c^2+\Omega_p^2\right)^2}&
-i\frac{2  \Omega_{c}^{2} \Omega_{p} \sin{\left(\lambda t  \right)} + \Omega_{p}^{3} \sin{\left(2\lambda t\right)  }}{2 \left(\Omega_{c}^{2} + \Omega_{p}^{2}\right)^{\frac{3}{2}}}&
\frac{2\Omega_c\Omega_p(\Omega_p^2-\Omega_c^2)(1-\cos(\lambda t))+\Omega_c\Omega_p^3(\cos(2\lambda t)-1)}{2\left(\Omega_c^2+\Omega_p^2\right)^2}  \\ 

i\frac{2 \Omega_{c}^{2} \Omega_{p} \sin{\left(\lambda t  \right)} + \Omega_{p}^{3} \sin{\left(2\lambda t\right)  }}{2 \left(\Omega_{c}^{2} + \Omega_{p}^{2}\right)^{\frac{3}{2}}} &
\frac{\Omega_p^2(1-\cos(2\lambda t))}{2\left(\Omega_c^2+\Omega_p^2\right)} &
-i\frac{\Omega_c\Omega_p^2(2\sin(\lambda t)-\sin(2\lambda t))}{2\left(\Omega_c^2+\Omega_p^2\right)^{\frac{3}{2}}} \\ 
\frac{2\Omega_c\Omega_p(\Omega_p^2-\Omega_c^2)(1-\cos(\lambda t))+\Omega_c\Omega_p^3(\cos(2\lambda t)-1)}{2\left(\Omega_c^2+\Omega_p^2\right)^2}  &
i\frac{\Omega_c\Omega_p^2(2\sin(\lambda t)-\sin(2\lambda t))}{2\left(\Omega_c^2+\Omega_p^2\right)^{\frac{3}{2}}}&
\frac{\Omega_c^2\Omega_p^2(-4\cos(\lambda t)+\cos(2\lambda t)+3)}{2\left(\Omega_c^2+\Omega_p^2\right)^2}
\end{pmatrix}.
\label{eq:rho_closed_analytical}
\ee
\end{widetext}
It is easily verified that the analytical solution defined above satisfies $\text{Tr}\left(\hat{\tilde \rho}(t)\right)=1$ and $\hat{\tilde \rho}^\dagger(t) =\hat{\tilde \rho}(t)$. In addition, taking 
$\Omega_c=\Omega_p$ produces a solution consistent with previous analytical results for the population.\cite{shore-qc} The exact result is indistinguishable from the numerical results in Fig. \ref{fig:closed_both} and Figs. S1 and S5 in the SM, and were therefore not plotted.  The off-diagonal elements of the above density operator elements in the rotating frame can be expressed in terms of those in the original frame by using Eqs. (\ref{eq:rho_tilde_def-12})-(\ref{eq:rho_tilde_def-23}) and their complex conjugates.

\section{Dynamics for periodic Hamiltonian \label{periodic}}
Given that the Hamiltonian is periodic in time with period $T$, $\hat H(t+nT) = \hat H(t)$ for $0< t\leq T$ and any positive value of integer $n$.  As a result, in practice, explicit calculation of $\hat U(t,0)$ for only $0< t\leq T$ suffices.  This is because for any value of $t$, the following relation holds:
\be
\hat U(nT+t,0) = \hat U(t,0)\hat U(T,0)^n  ,  \label{eq:periodic-dynamics}
\ee
where $n=1,2,3,\cdots$.

 For the calculation of $\hat U(t,0)$ for $0< t\leq T$, we employ an expression based on the ME for the differential operator, $\hat U(t_k+\delta t,t_k)$, where $t_k=k\delta t$, for a small enough value of $\delta t$.   Thus, we first determine $\hat U(t,0)$ for $0< t \leq T$ at each discretized $t_k$ by repeated multiplication of $\hat U(t_{k+1},t_k)$.  Once $\hat U(T,0)$ is determined this way,  it is possible to make large (stroboscopic) jumps corresponding to the size of the period. This results in orders of magnitude of savings in terms of calculating matrix exponentials and commutators, and makes the calculation of long-time dynamics feasible. The dynamics at intra-period points (micro-motion) can then be determined using Eq. (\ref{eq:periodic-dynamics}).  If the major focus is the steady state limit, even further acceleration is possible by using the approach of numerical matrix multiplication,\cite{thirumalai-jcp79-2} namely, using the fact that $\hat U(2^mT,0)=\hat U(2^{m-1}T,0)\hat U(2^{m-1} T,0)$ for any positive integer $m$.  

Conceptually, our approach is similar to the Floquet-ME.\cite{casas-jpa34,eckardt-njp17,kuwahara-ap367}  In fact, the time evolution operator for a single period, $\hat U(T,0)$,  corresponds to the unitary operator for the Floquet Hamiltonian\cite{eckardt-njp17} $\hat{H}_F$(sometimes referred to as the average or effective Hamiltonian\cite{brinkmann-cmra45}) defined as follows:
\ben
\label{eq:AHT}
e^{-\frac{i}{\hbar}\hat{H}_F} = e_{(+)}^{-\frac{i}{\hbar}\int_0^{T}dt\hat{H}(t)}=\hat U (T,0)  .
\een
For practical purposes, our numerical procedure for solving the Floquet problem has some advantages compared to the approach to determine $\hat H_F$. For instance, it has been noted that the Floquet-ME may require high order approximation for reasonable accuracy,\cite{kuwahara-ap367} which may only be possible for simple Hamiltonians. In addition, truncation of the Floquet-ME beyond the first order produces a dependence of the eigenvalues of $H_F$ on the initial point (referred to as the Floquet gauge), which makes them no longer related by a unitary transformation.\cite{eckardt-njp17} Finally, our procedure enables calculation of the intra-period motion as well, from which detailed time dependences in the long time limit can be identified (see Fig. \ref{fig:steady_state_w_p_pops}). The procedure above allows for calculation of the density-operator for the closed-system dynamics by using $\hat \rho(t)=\hat U(t,0)\hat \rho(0)\hat U^\dagger(t,0)$.

\section{Other 4th order methods}
\label{sec:AM4_RK4}
\subsection{Runge-Kutta and Adams-Moulton methods}
\rc{We here describe well-known 4th order algorithms, RK4\cite{butcher} and AM4,\cite{griffiths-nm} for the case of operator differential equation given by Eq. (\ref{eq:diffeq-gen}), by adopting those for ordinary differential equations.}\cite{butcher,griffiths-nm}	

The RK4  method evolves from $y_n$ to $y_{n+1}$ by using stages labeled $k_i$ given by:\cite{butcher}
\ben
k_1 &=& f(t_n,y_n) ,\nonumber \\
k_2 &=& f(t_{n+1/2},y_n+\frac{\delta t}{2}k_1) ,\nonumber \\
k_3 &=& f(t_{n+1/2},y_n+\frac{\delta t}{2}k_2)  , \nonumber \\
k_4 &=& f(t_{n+1},y_n+\delta tk_3)  .
\een 
The above stages are then used to calculate $y_{n+1}$ as follows:
\be
y_{n+1} = y_n + \frac{\delta t}{6}(k_1+2k_2+2k_3+k_4) .
\ee

The AM4 method uses the following implicit equation:\cite{butcher}
\ben
y_{n+3} = y_{n+2} + \frac{\delta t}{24} && \Bigl(9f(t_{n+3},y_{n+3})+19f(t_{n+2},y_{n+2}) \nonumber \\  &&-5f(t_{n+1},y_{n+1}) + f(t_n,y_n)\Bigr), \label{eq:Adams-Moulton}		
\een
which requires three previous values of $y$. To begin, one uses the initial condition $y_0$ and uses an explicit method (such as the RK4 method above) to find $y_1$ and $y_2$. For all subsequent time steps, Eq. (\ref{eq:Adams-Moulton}) is used to find $y_{n+3}$, which requires the solution of a system of equations.

\subsection{Optimized commutator free Magnus Expansion}
We here describe an optimized 4th order CFME (O4-CFME) that uses three exponentials.\cite{alvermann-jcp230,alvermann-njp14} For a differential equation given by the second equality of Eq. (\ref{eq:diffeq-gen}), this algorithm entails propagating $y(t)$ from time $t$ to $t +\delta t$ according to the following equation:
\be
y(t+\delta t) = e^{\delta t\sum_{i=1}^3 a_{3i}A_i}e^{\delta t\sum_{i=1}^3 a_{2i}A_i}e^{\delta t\sum_{i=1}^3 a_{1i}A_i}y(t).
\ee
In the above equation, $A_i$ is shorthand for $A(t + \delta tc_i)$ where $c_i$ is the ith Gauss-Legendre point defined as $c_1 = \frac{1 - \sqrt{3/5}}{2}$ , $c_2 = \frac{1}{2} $ , $c_3 = \frac{1 + \sqrt{3/5}}{2}$, and the coefficient $a_{ij}$ \rc{denotes an element of the following matrix $a$}:
\be 
a = \begin{pmatrix} 
      \frac{10 \sqrt{15}}{261} + \frac{37}{240} & -\frac{1}{30} &  - \frac{10\sqrt{15}}{261} + \frac{37}{240} \\ \\
      -\frac{11}{360} & \frac{23}{45} & -\frac{11}{360} \\ \\
      - \frac{10\sqrt{15}}{261} + \frac{37}{240} & -\frac{1}{30} & \frac{10 \sqrt{15}}{261} + \frac{37}{240}
      
    \end{pmatrix}
\ee
\ \vspace{.2in}\\
\noindent
{\bf REFERENCES}\vspace{.05in}\\

\newpage

\renewcommand{\theequation}{S\arabic{equation}}
\renewcommand{\thefigure}{S\arabic{figure}}
\renewcommand{\thetable}{S\arabic{table}}
\renewcommand\thepage{S\arabic{page}}
\setcounter{figure}{1}
\setcounter{page}{1}
\begin{widetext}
\section{Supplementary Material -  A simple fourth order propagator based on the Magnus expansion in the Liouville space: Application to a $\Lambda$-system and assessment of the rotating wave approximation}



This Supplementary Material provides additional results of calculation for cases in Table I of the main text, mostly for cases A-I, A-II, C-I, and C-II  and some additional details for cases B-I and B-II.    
All Python codes to reproduce data and figures can be found at \href{https://github.com/TanerTure/floquet_dynamics}{https://github.com/TanerTure/floquet\_dynamics}.

\begin{figure}
	(a)\makebox[0.8\columnwidth]{ }\\
	\includegraphics[width=0.4\columnwidth]{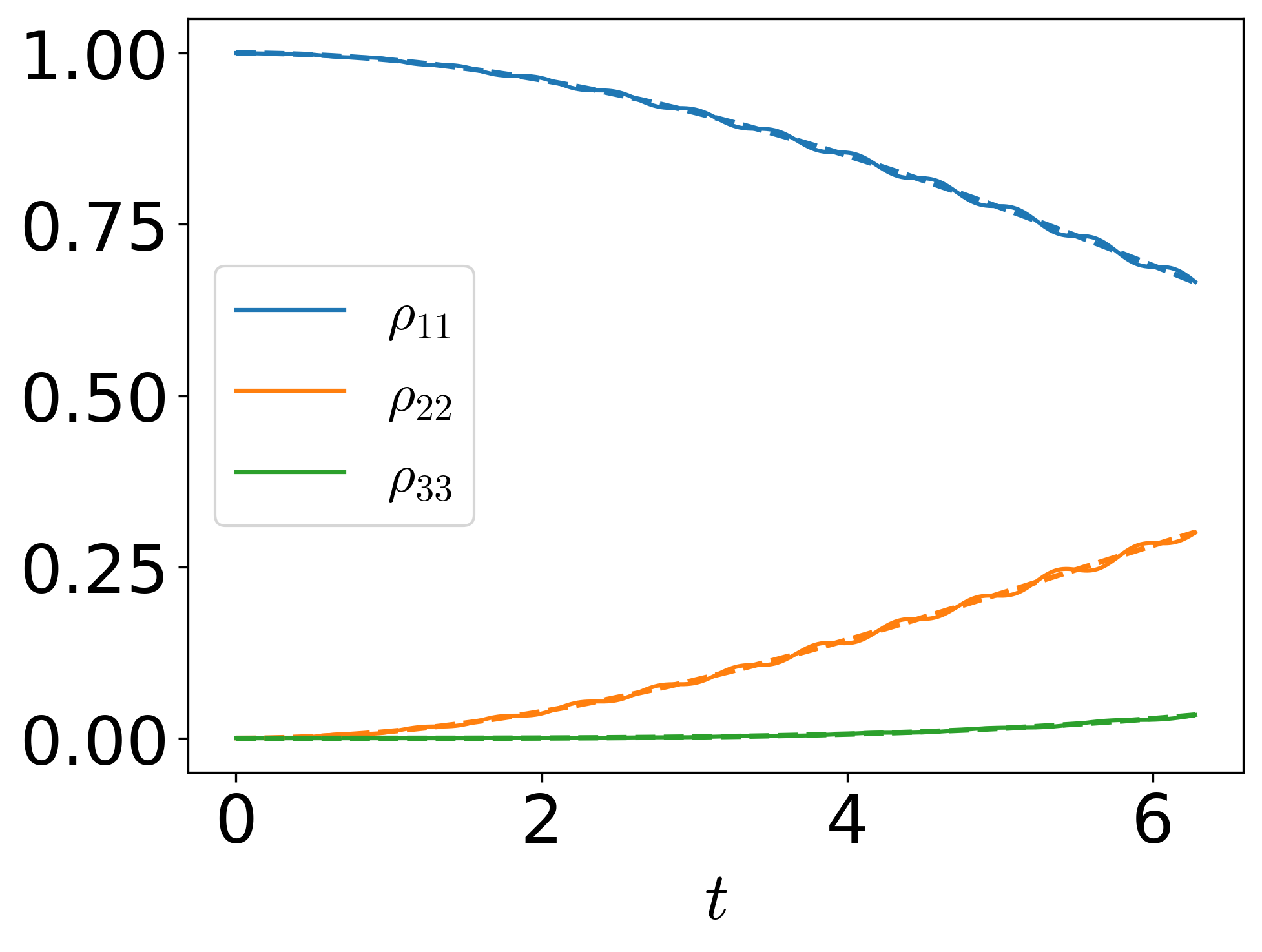}\\
	(b)\makebox[0.8\columnwidth]{ }\\
	\includegraphics[width=0.4\columnwidth]{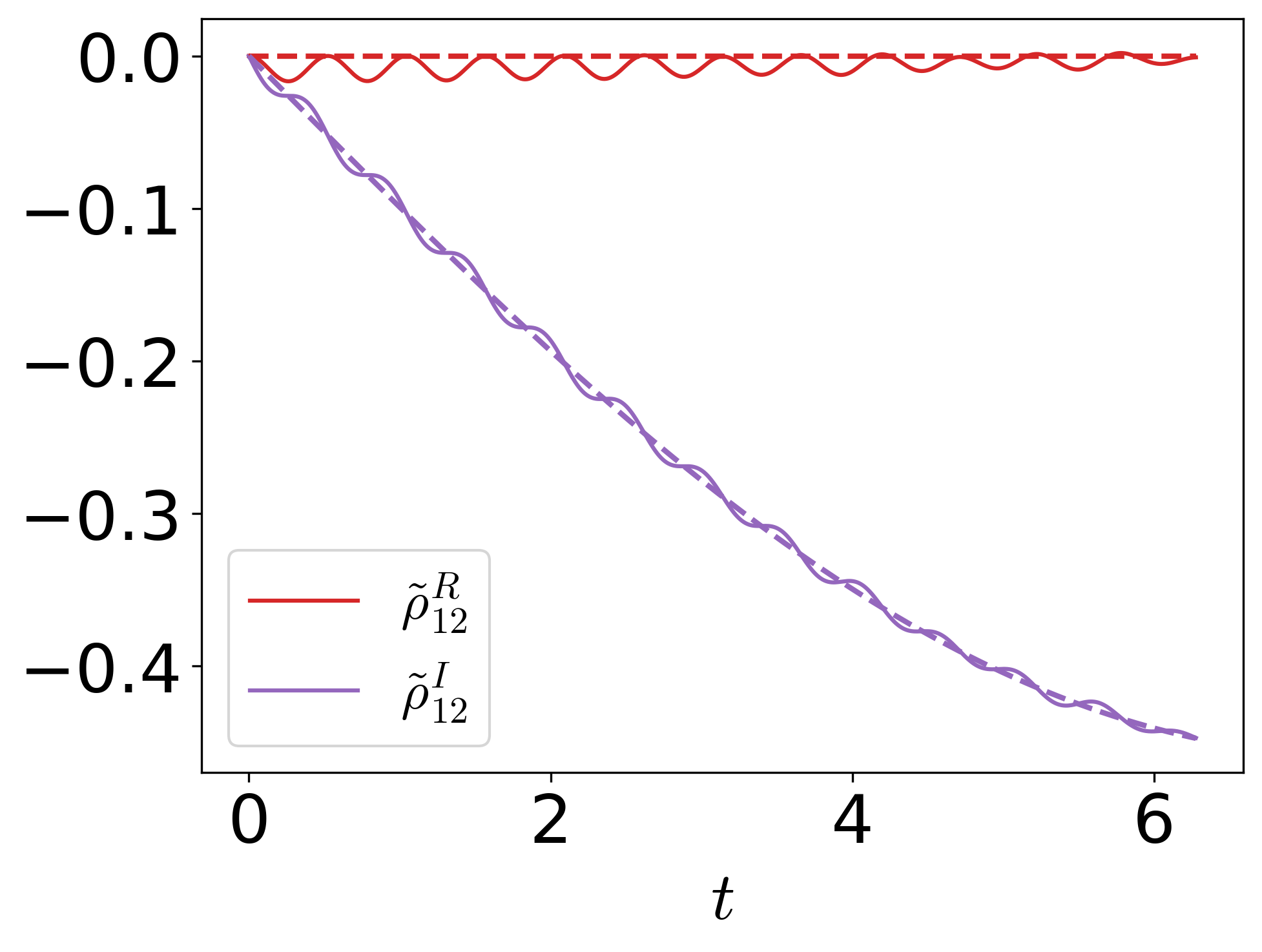} \\
	(c)\makebox[0.8\columnwidth]{ }\\
	\includegraphics[width=0.4\columnwidth]{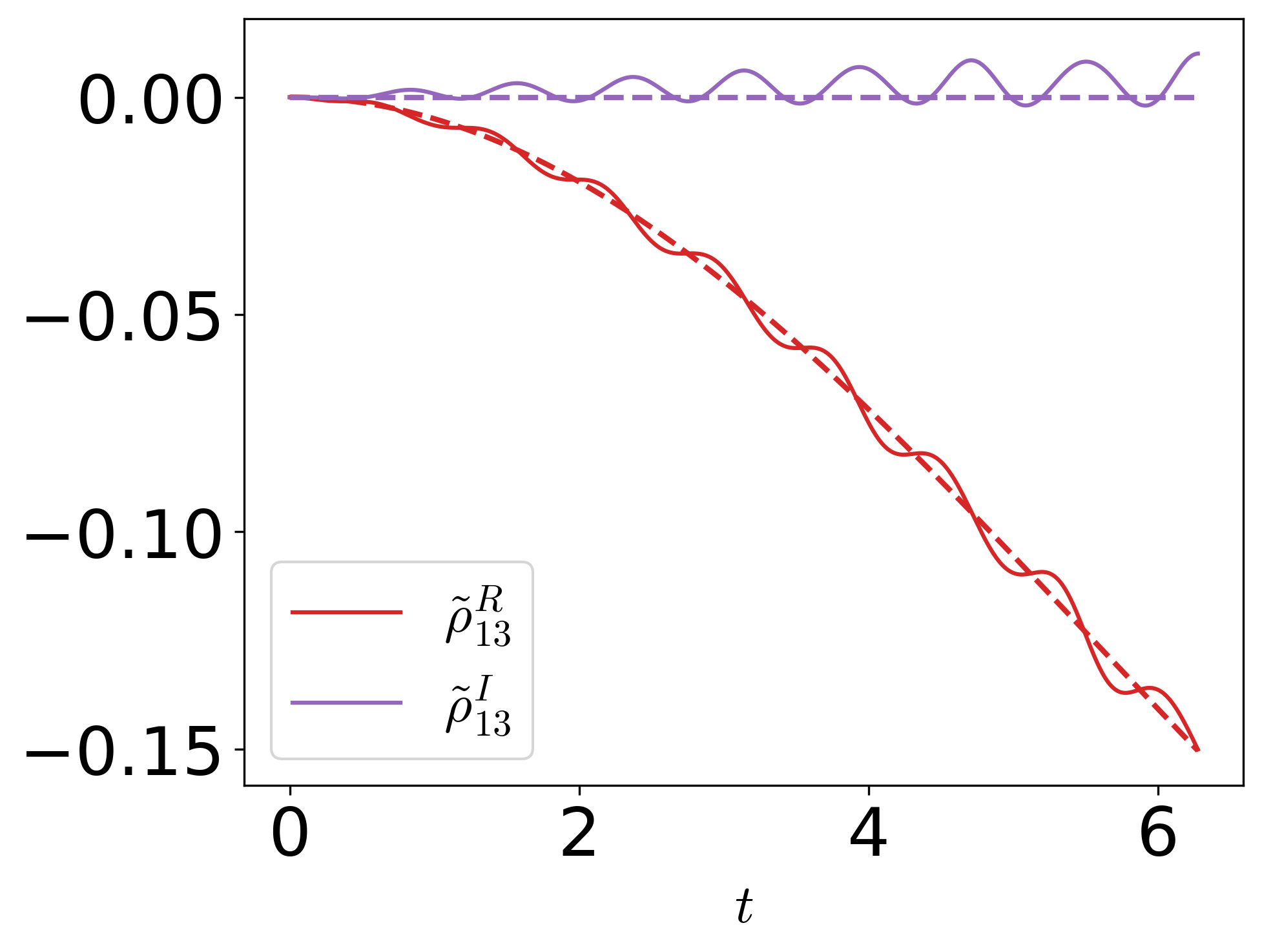}\\
	(d)\makebox[0.8\columnwidth]{ }\\
	\includegraphics[width=0.4\columnwidth]{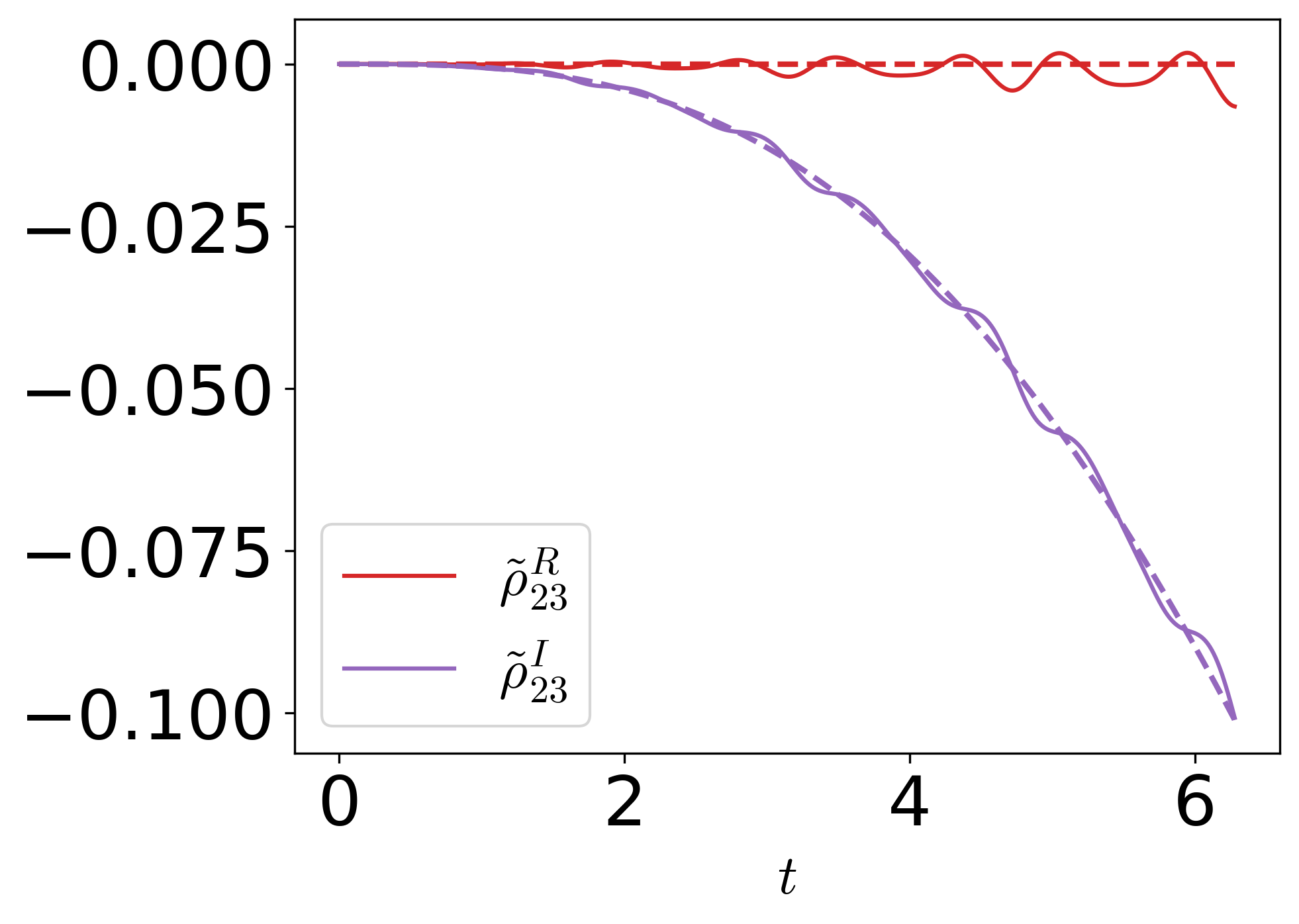}
	\caption{Elements of the time dependent system density operator for the case A-I in the Table I of the main text,  a closed system unitary dynamics, with full Hamiltonian (solid) and the RWA(dashed), for which the Hamiltonians are respectively $\hat H(t)$ and $\hat H^{\rm \small RWA}(t)$.  Both were calculated using the 4th order ME-propagator with commutator, Eq. (8) in the main text. }
	\label{fig:closed_both_B}
\end{figure}

\begin{figure}
	(a)\makebox[0.8\columnwidth]{ }\\
	\includegraphics[width=0.4\columnwidth]{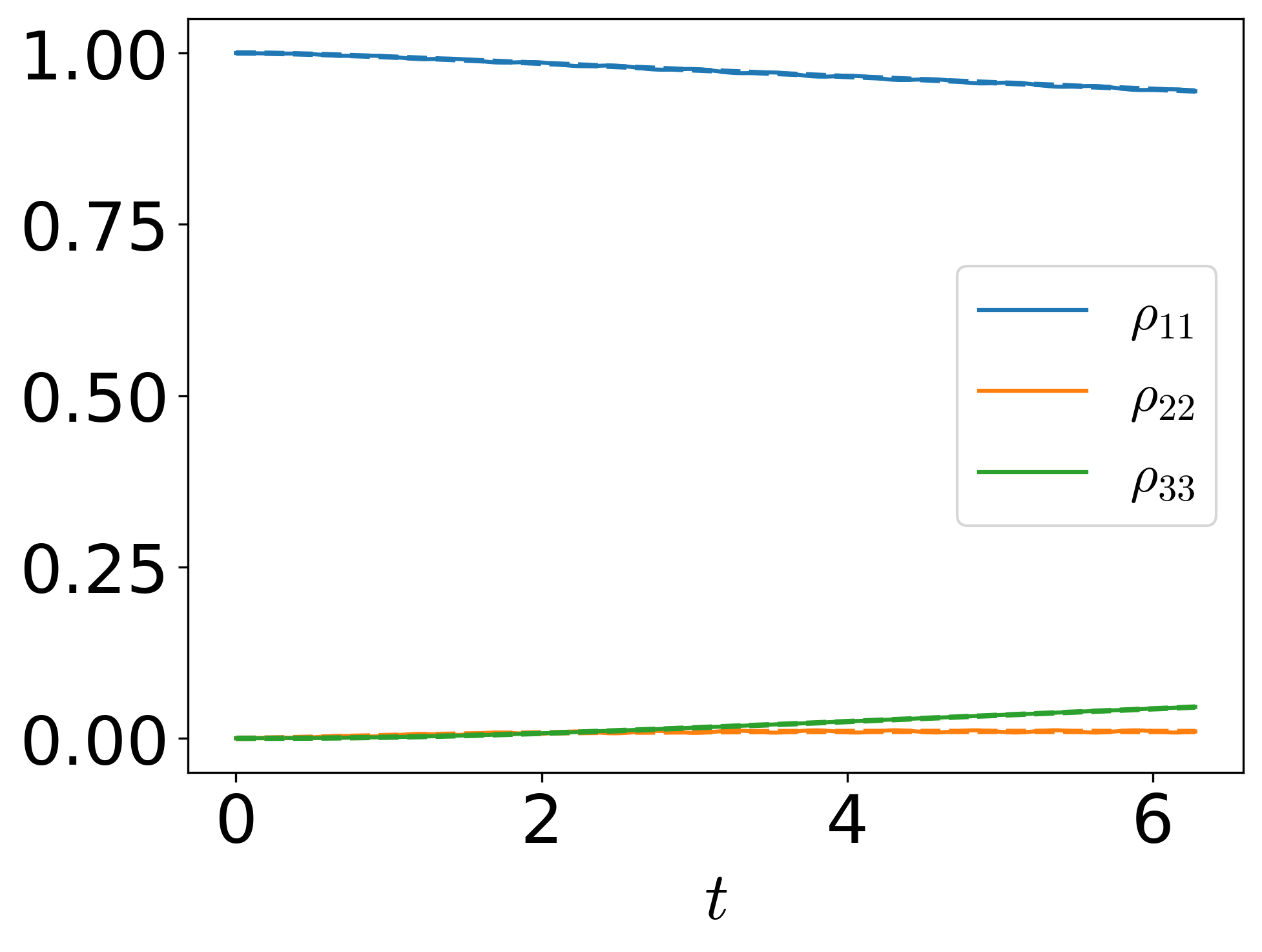}\\
	(b)\makebox[0.8\columnwidth]{ }\\
	\includegraphics[width=0.4\columnwidth]{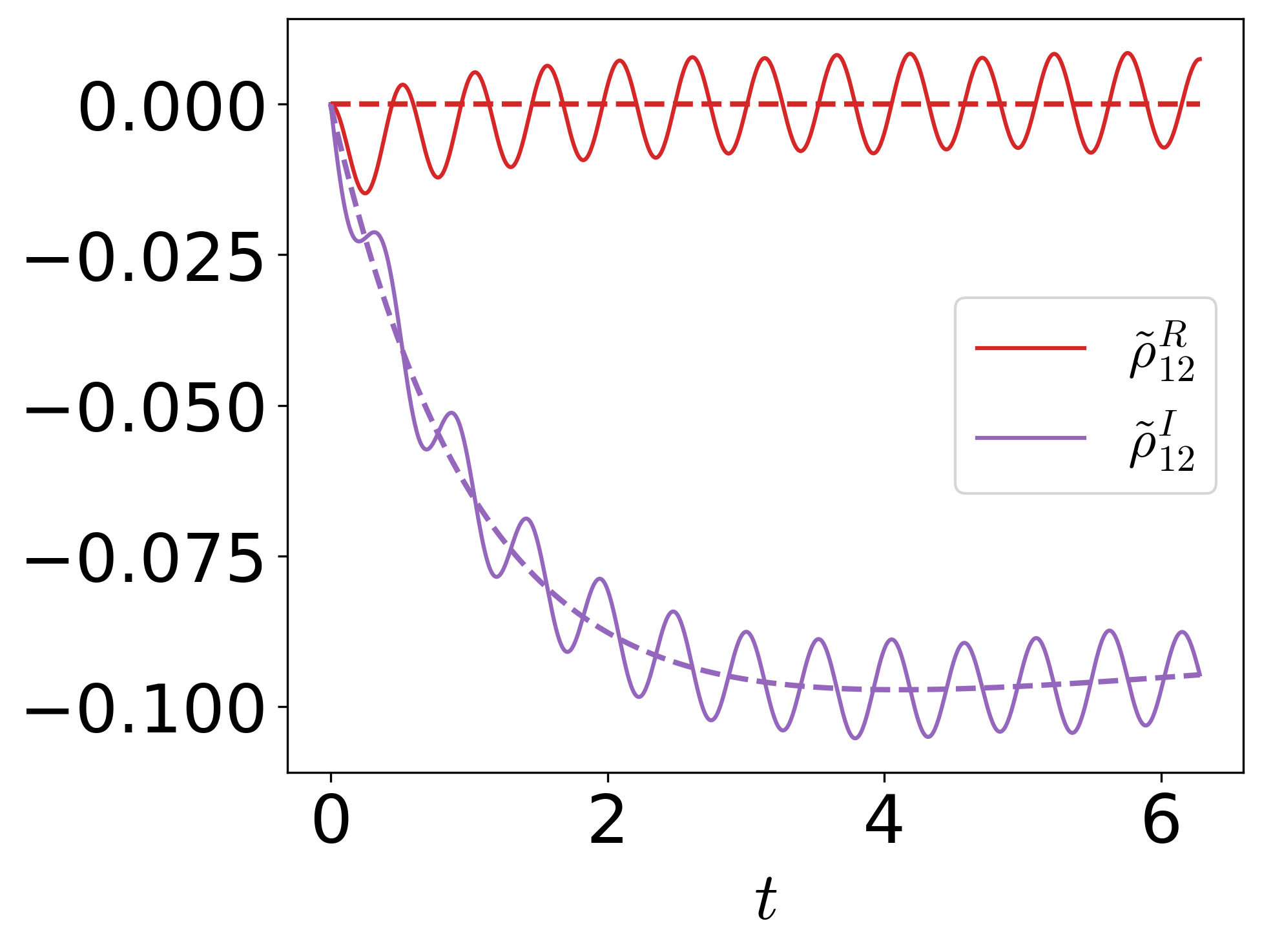} \\
	(c)\makebox[0.8\columnwidth]{ }\\
	\includegraphics[width=0.4\columnwidth]{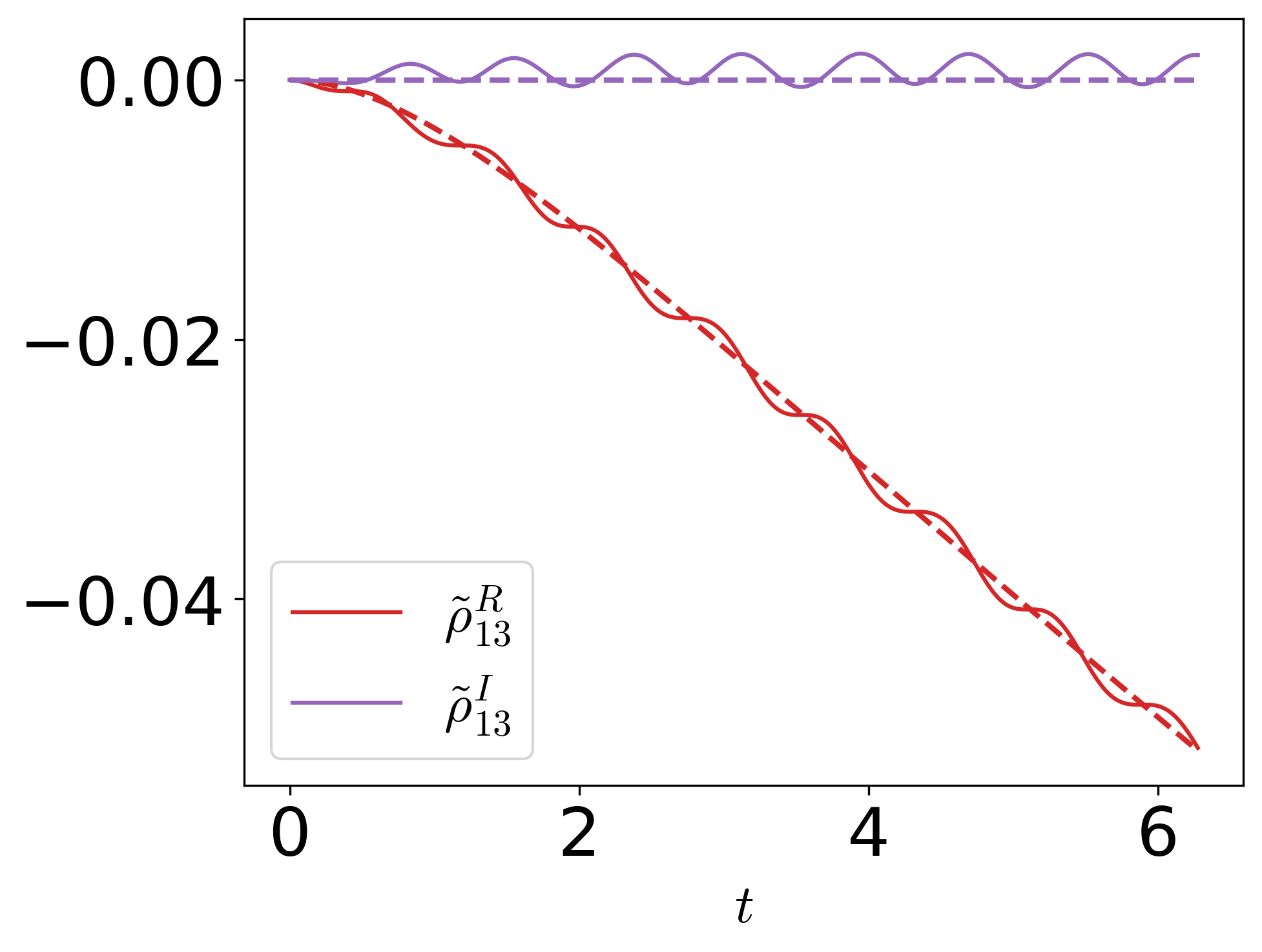}\\
	(d)\makebox[0.8\columnwidth]{ }\\
	\includegraphics[width=0.4\columnwidth]{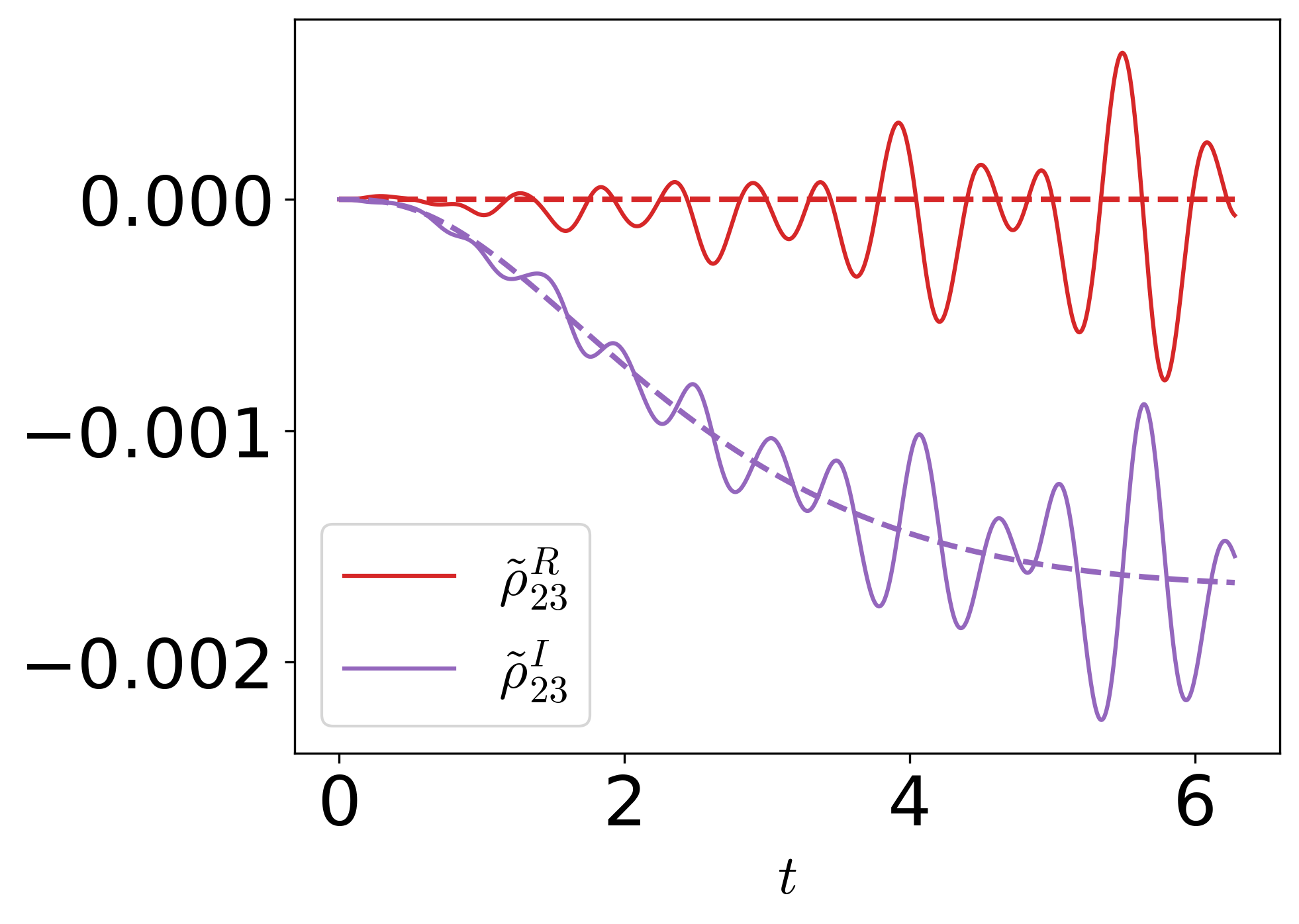}
	\caption{Elements of the time dependent system density operator for the case A-II in Table I of the main text, an open system non-unitary dynamics, with full Hamiltonian (solid) and the RWA(dashed), for which the Hamiltonians are respectively $\hat H(t)$ and $\hat H^{\rm \small RWA}(t)$.  Both were calculated using the 4th order ME-propagator with commutator, Eq. (11) in the main text.}
	\label{fig:open_both_B}
\end{figure}

\begin{figure}
	(a)\makebox[0.8\columnwidth]{ }\\
	\includegraphics[width=0.4\columnwidth]{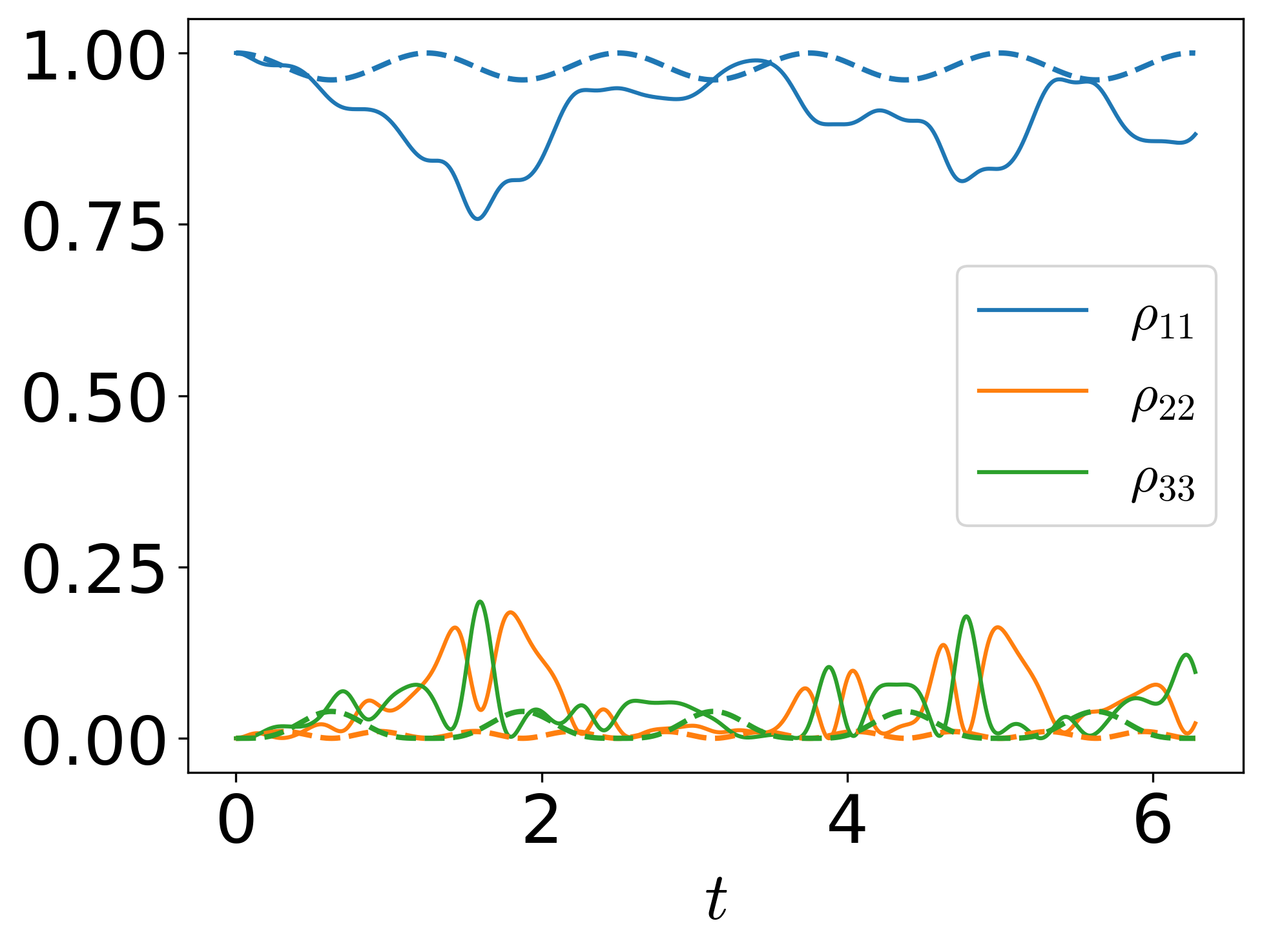}\\
	(b)\makebox[0.8\columnwidth]{ }\\
	\includegraphics[width=0.4\columnwidth]{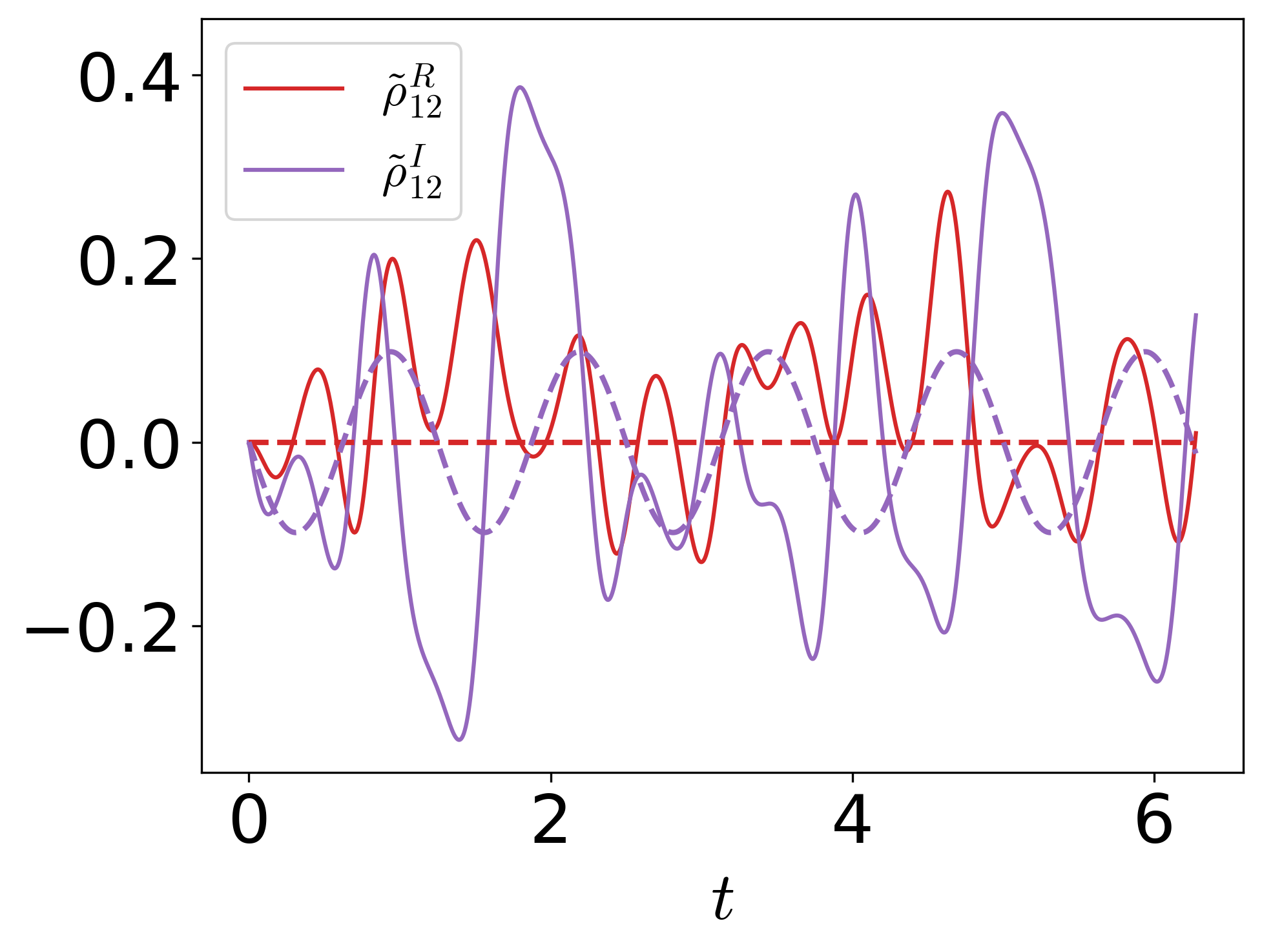} \\
	(c)\makebox[0.8\columnwidth]{ }\\
	\includegraphics[width=0.4\columnwidth]{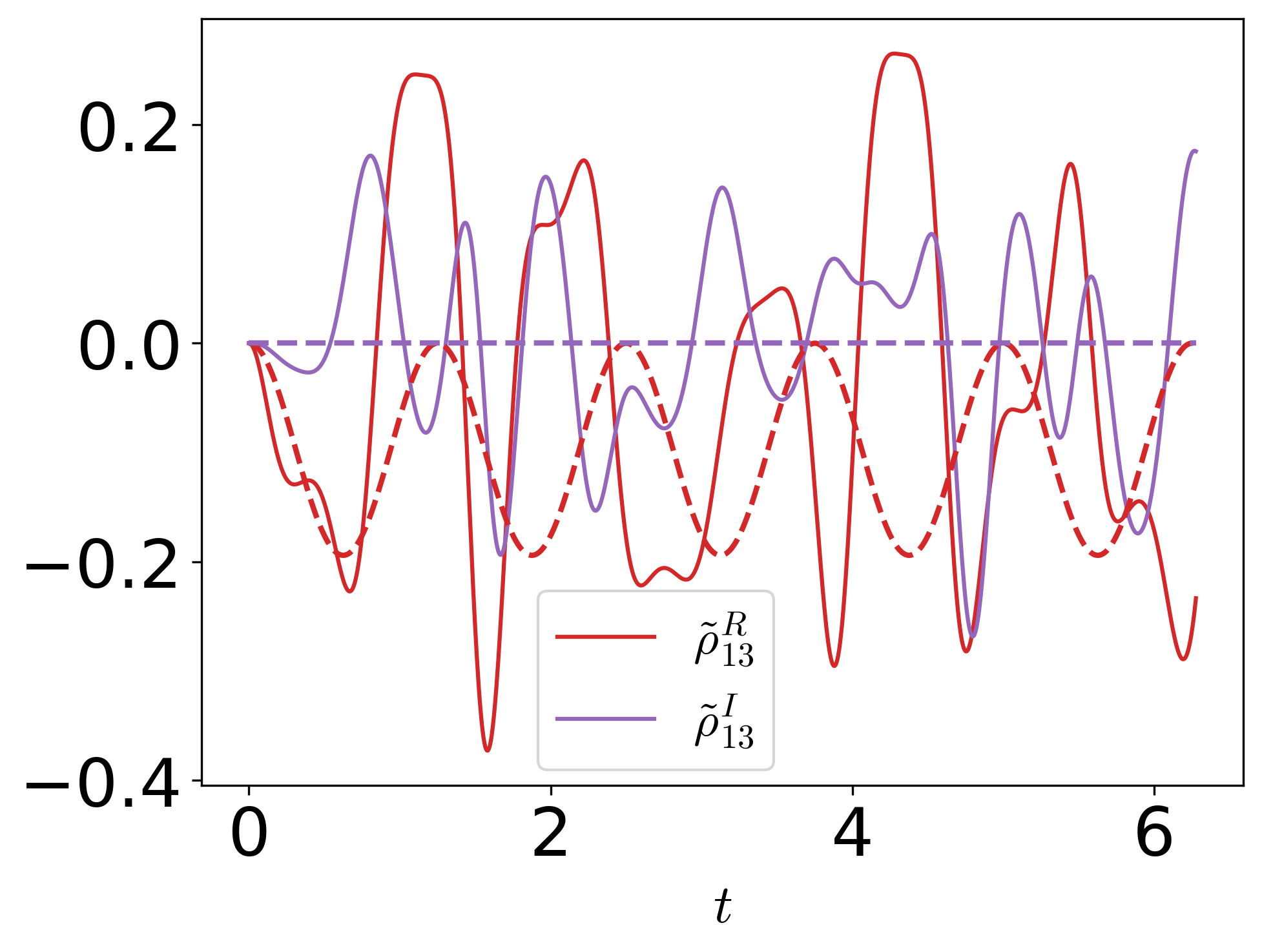}\\
	(d)\makebox[0.8\columnwidth]{ }\\
	\includegraphics[width=0.4\columnwidth]{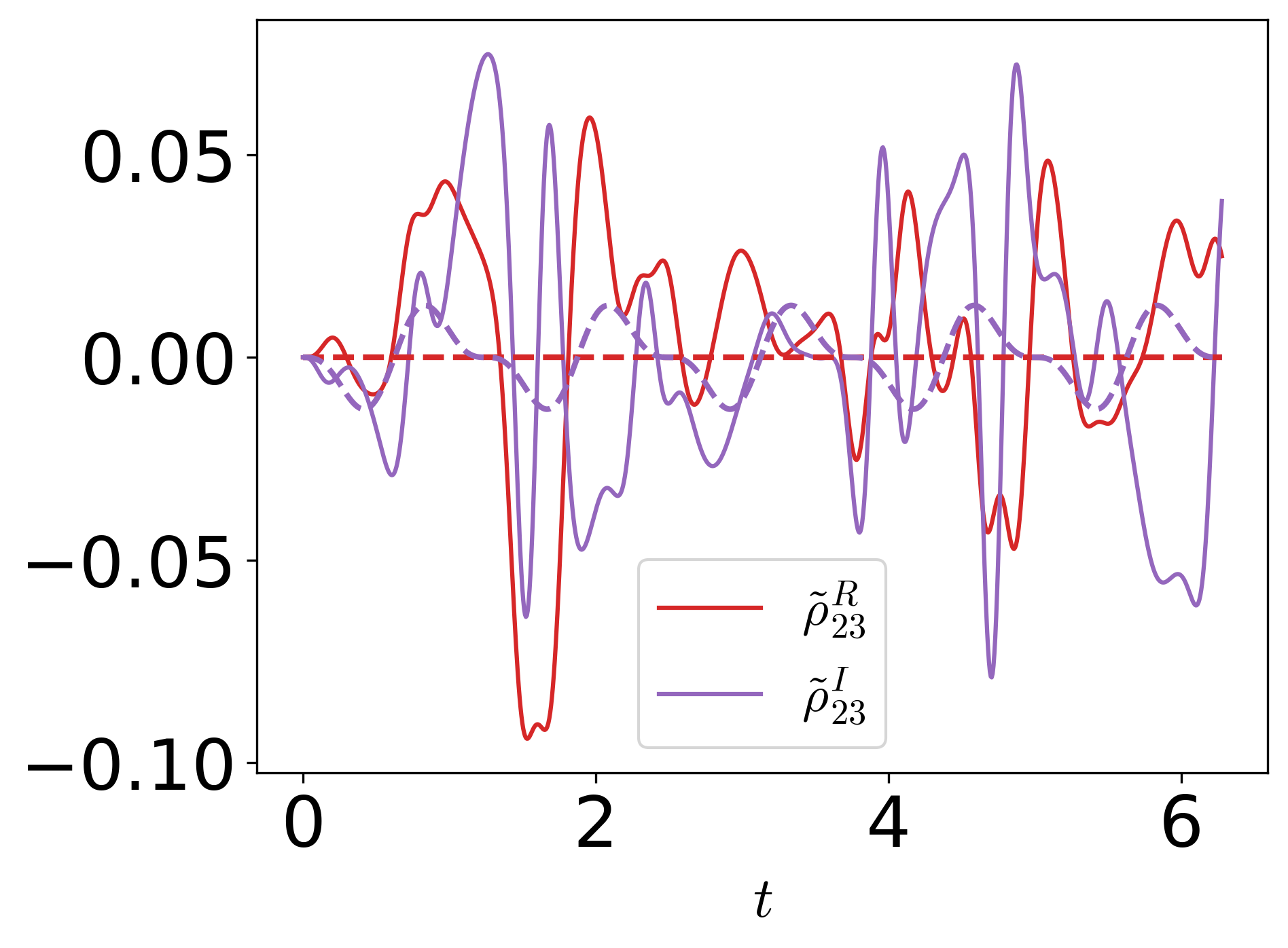}
	\caption{Elements of the time dependent system density operator for the case C-I in Table I of the main text, a closed system unitary dynamics, with full Hamiltonian (solid) and the RWA(dashed), for which the Hamiltonians are respectively $\hat H(t)$ and $\hat H^{\rm \small RWA}(t)$.  Both were calculated using the 4th order ME-propagator with commutator, Eq. (8) in the main text.}
	\label{fig:closed_both_C}
\end{figure}

\begin{figure}
	(a)\makebox[0.8\columnwidth]{ }\\
	\includegraphics[width=0.4\columnwidth]{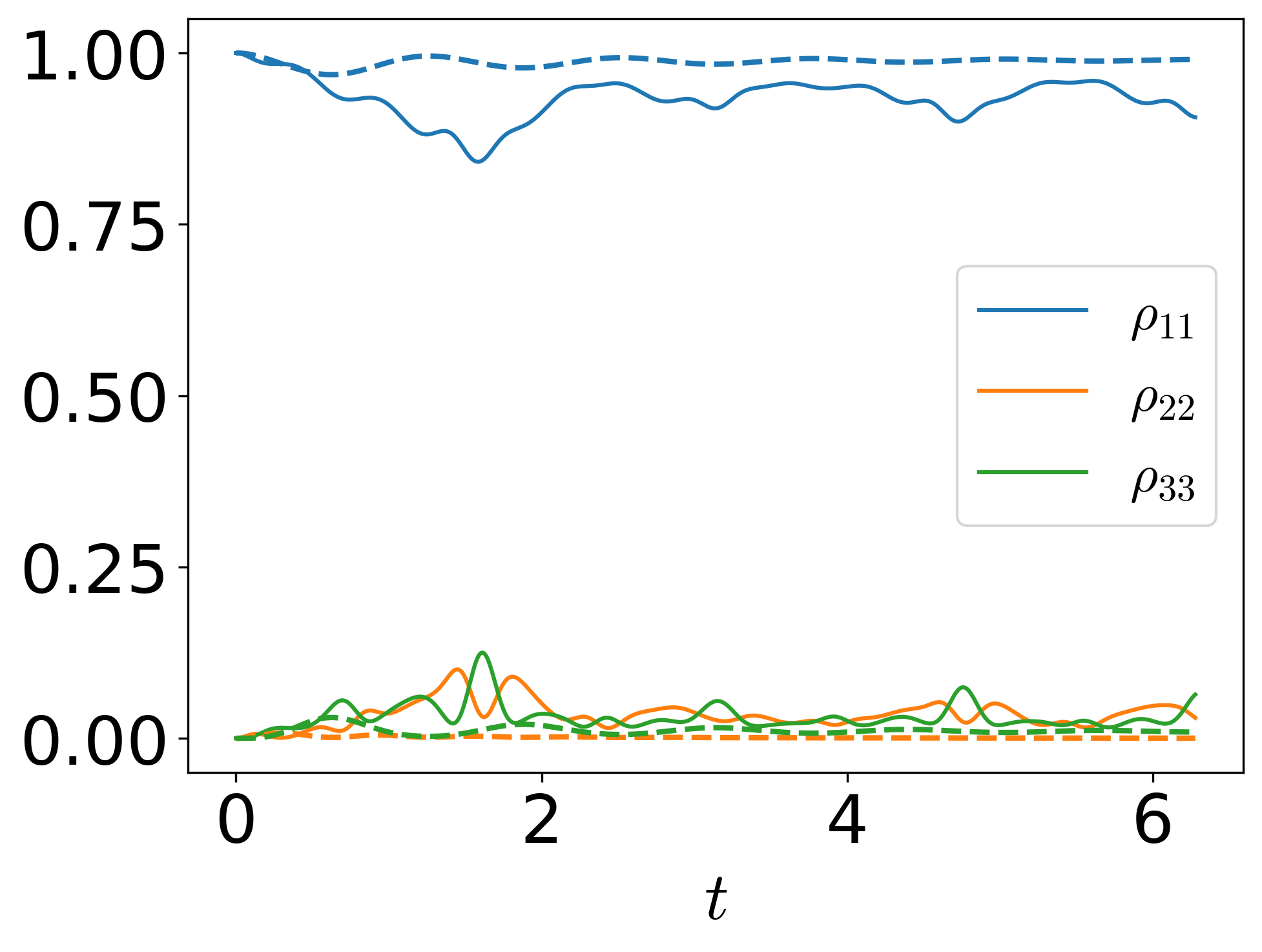}\\
	(b)\makebox[0.8\columnwidth]{ }\\
	\includegraphics[width=0.4\columnwidth]{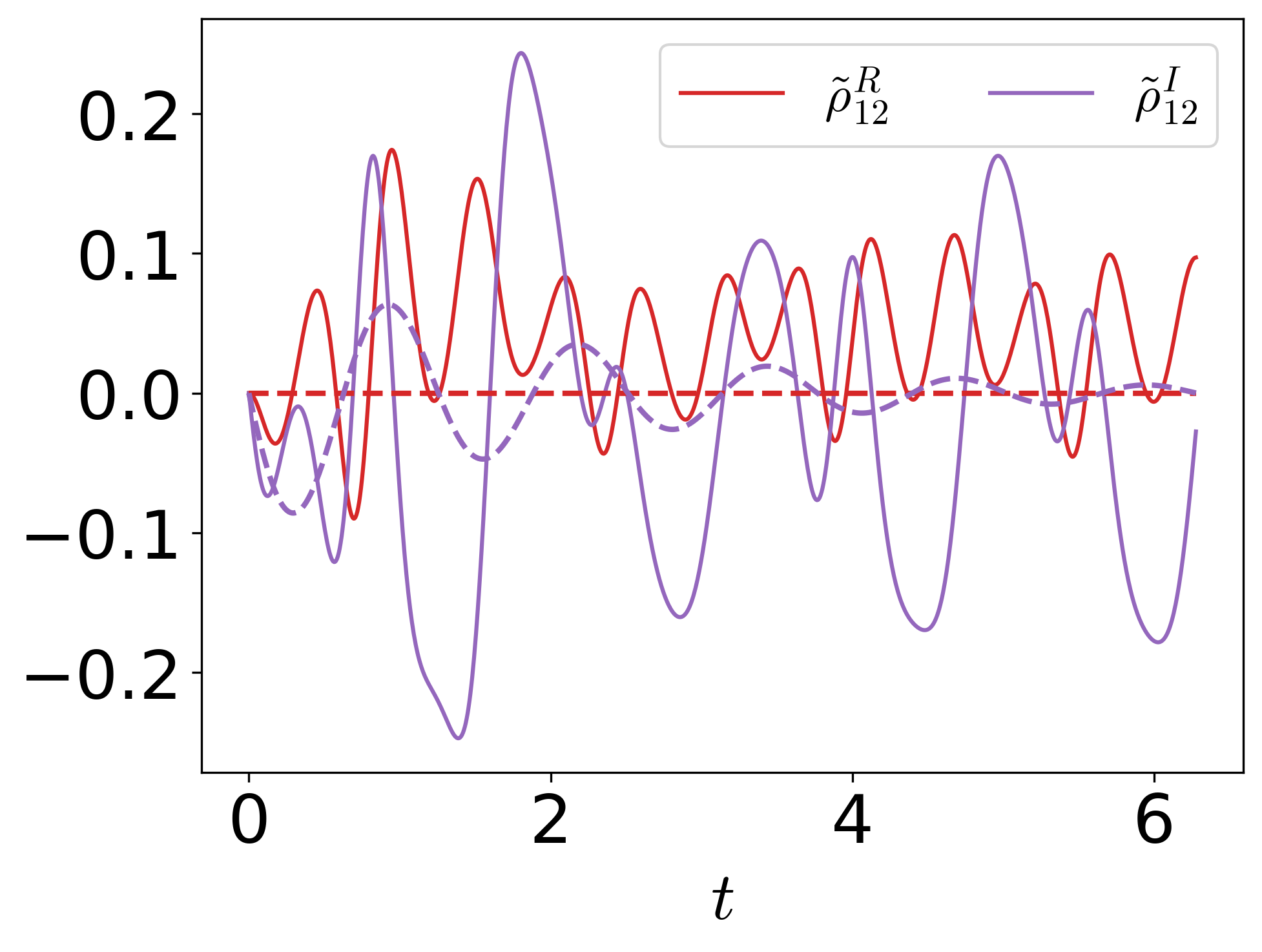} \\
	(c)\makebox[0.8\columnwidth]{ }\\
	\includegraphics[width=0.4\columnwidth]{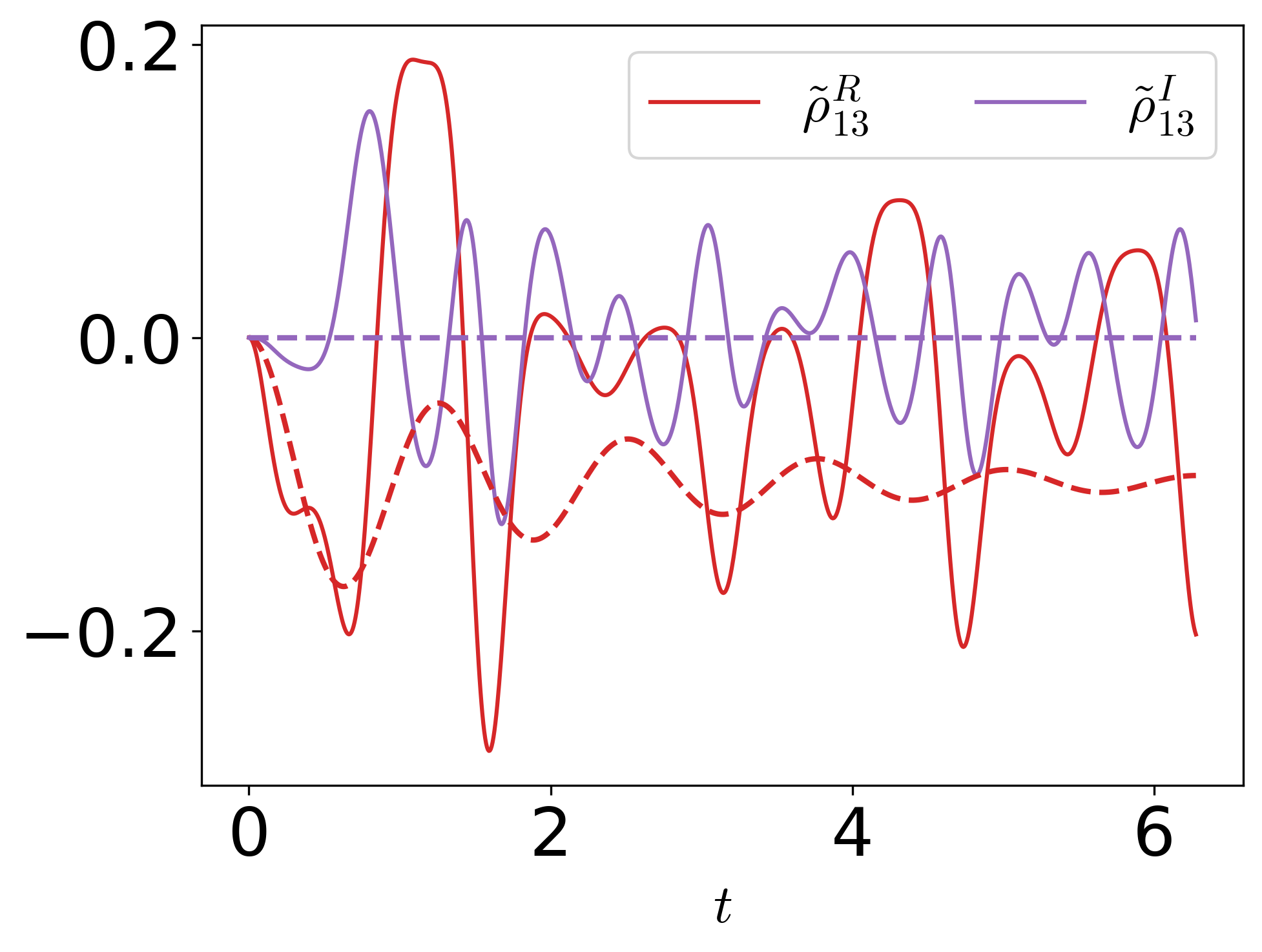}\\
	(d)\makebox[0.8\columnwidth]{ }\\
	\includegraphics[width=0.4\columnwidth]{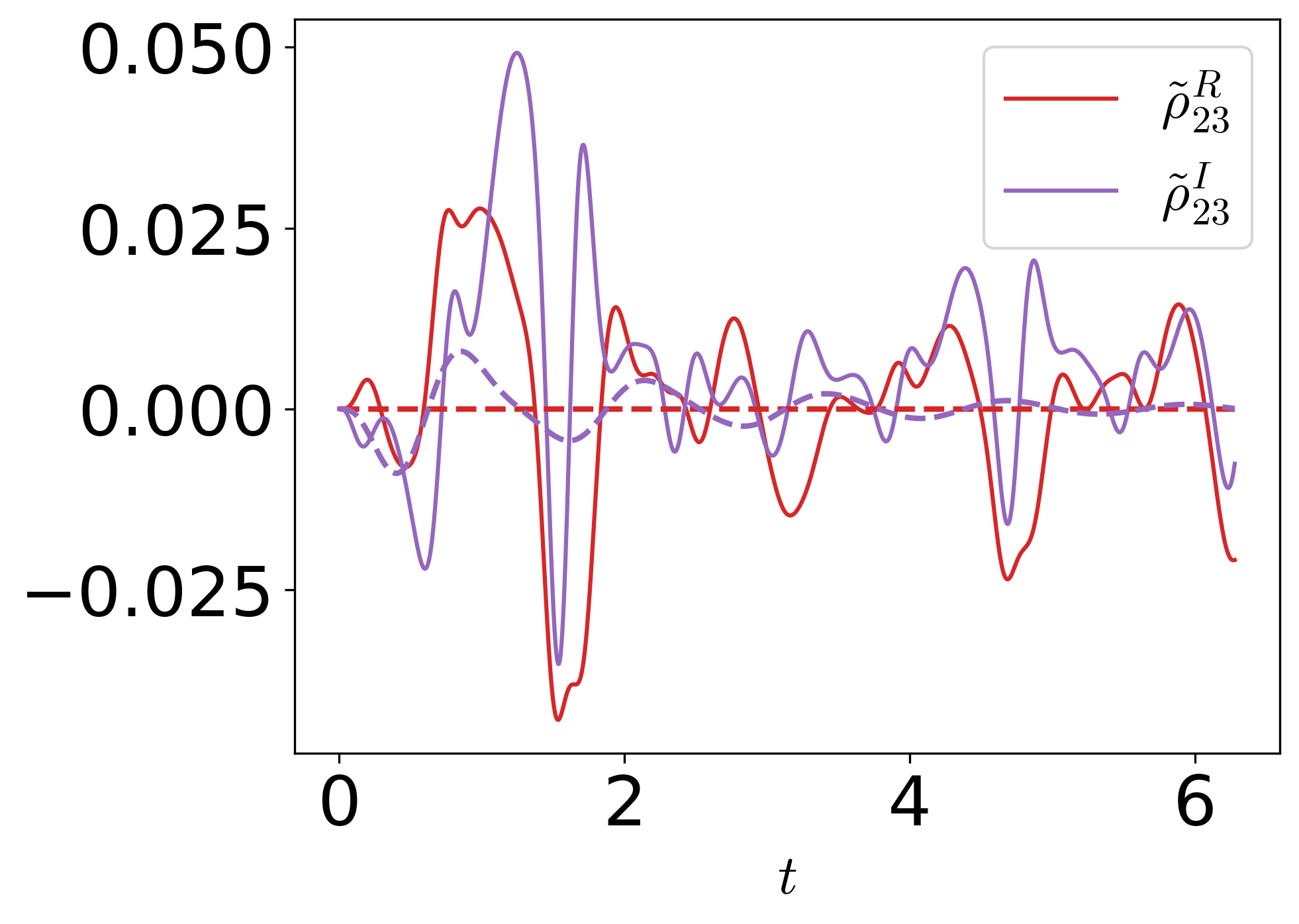}
	\caption{Elements of the time dependent system density operator for case C-II in Table I of the main text, an open system non-unitary dynamics, with full Hamiltonian (solid) and the RWA(dashed), for which the Hamiltonians are respectively $\hat H(t)$ and $\hat H^{\rm \small RWA}(t)$.  Both were calculated using the 4th order ME-propagator with commutator, Eq. (11) in the main text.}
	\label{fig:open_both_C}
\end{figure}

 \begin{figure}
 	
 	(a)\makebox[0.8\columnwidth]{ }\\
 	\includegraphics[width=0.4\columnwidth]{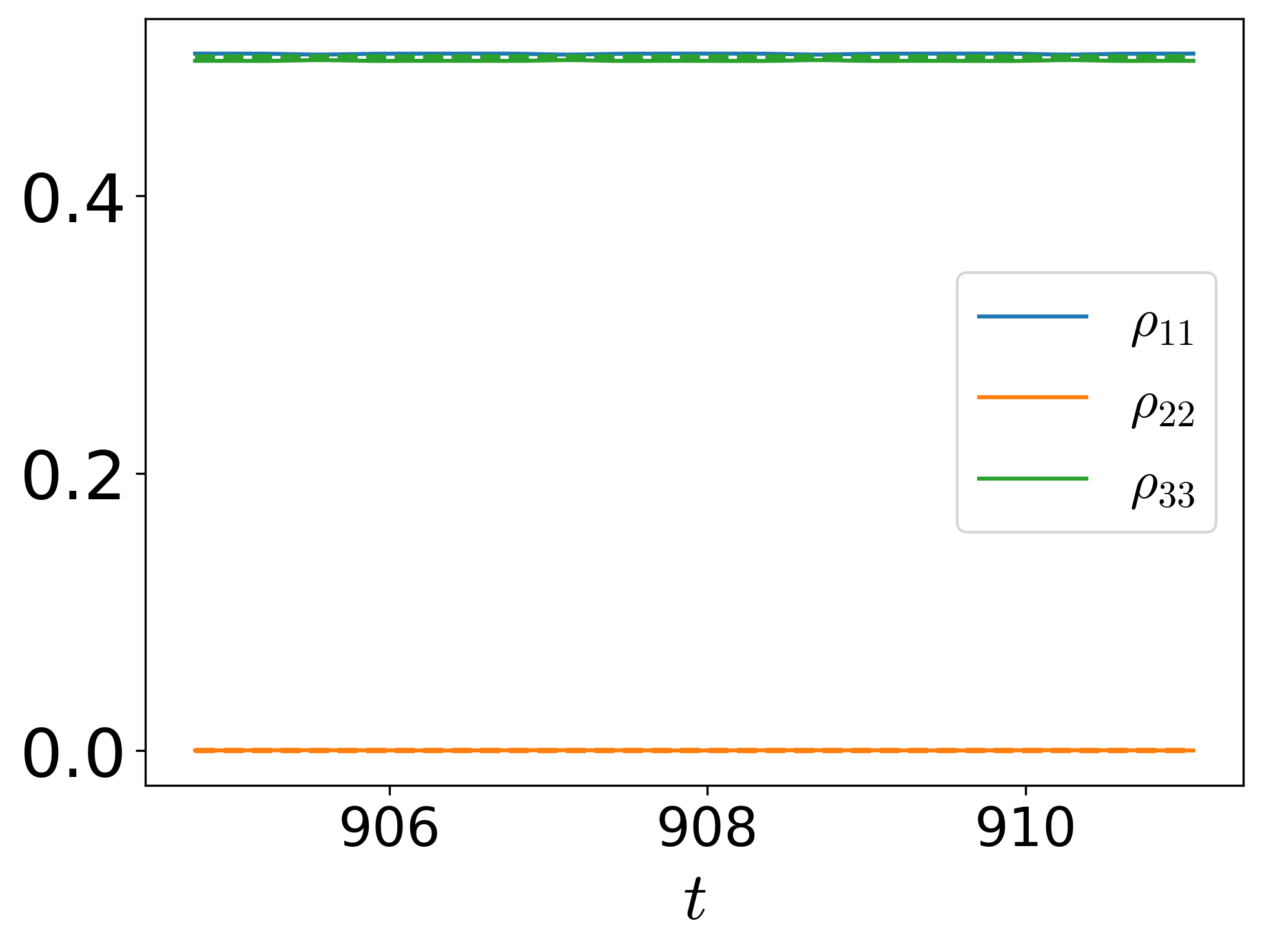} \\
 	(b)\makebox[0.8\columnwidth]{ }\\
 	\includegraphics[width=0.4\columnwidth]{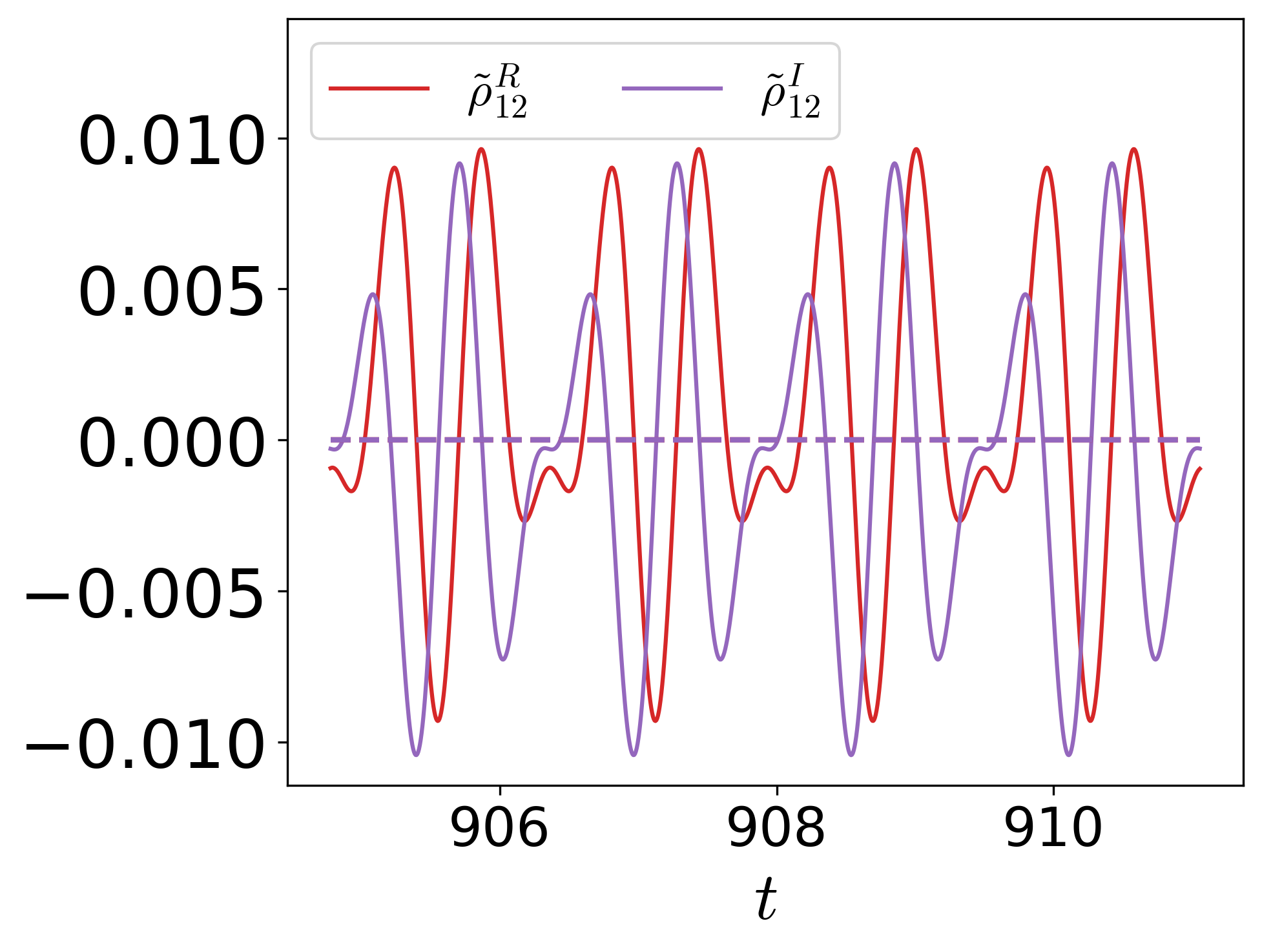} \\
 	(c)\makebox[0.8\columnwidth]{ }\\
 	\includegraphics[width=0.4\columnwidth]{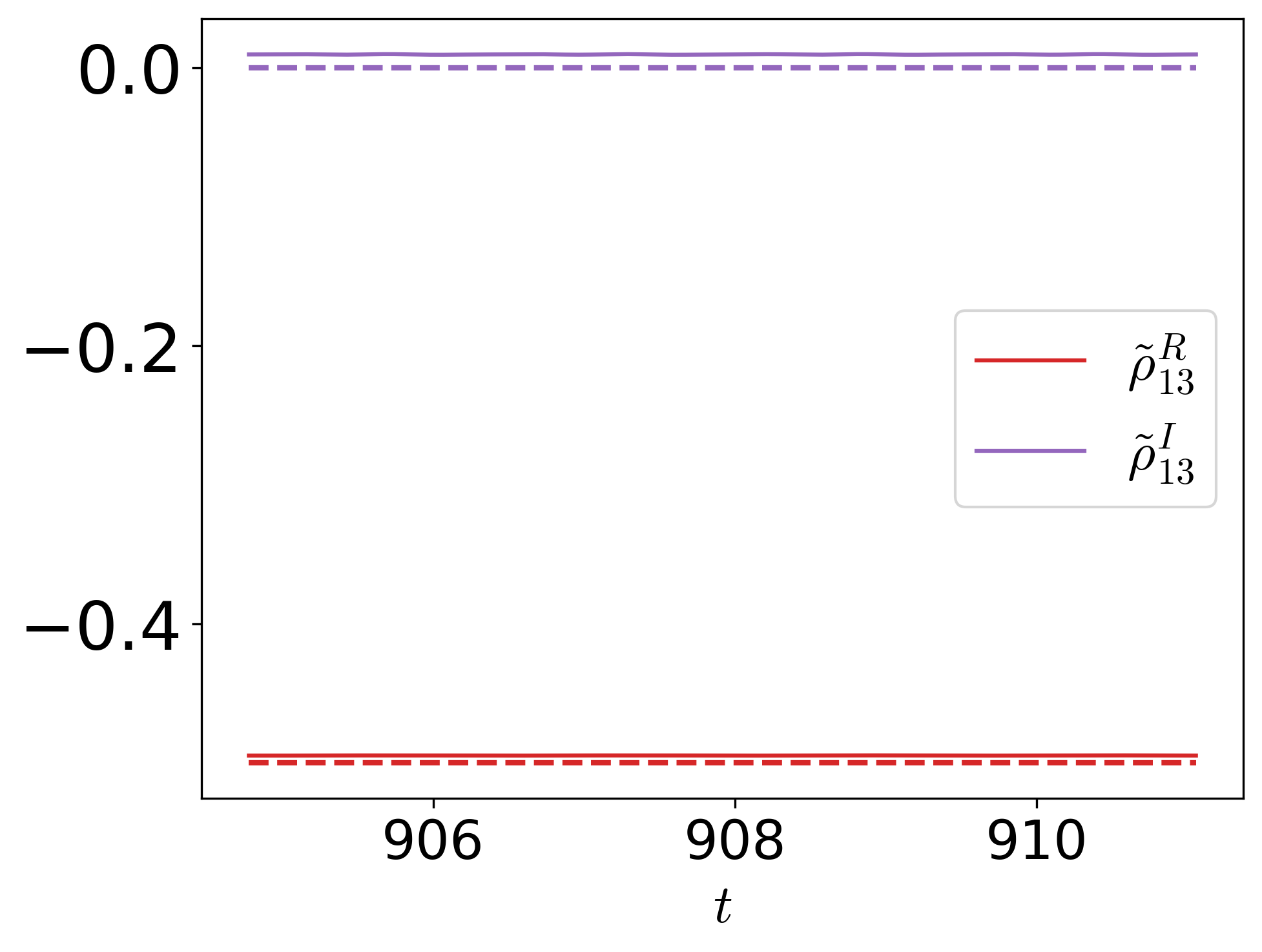}\\
 	(d)\makebox[0.8\columnwidth]{ }\\
 	\includegraphics[width=0.4\columnwidth]{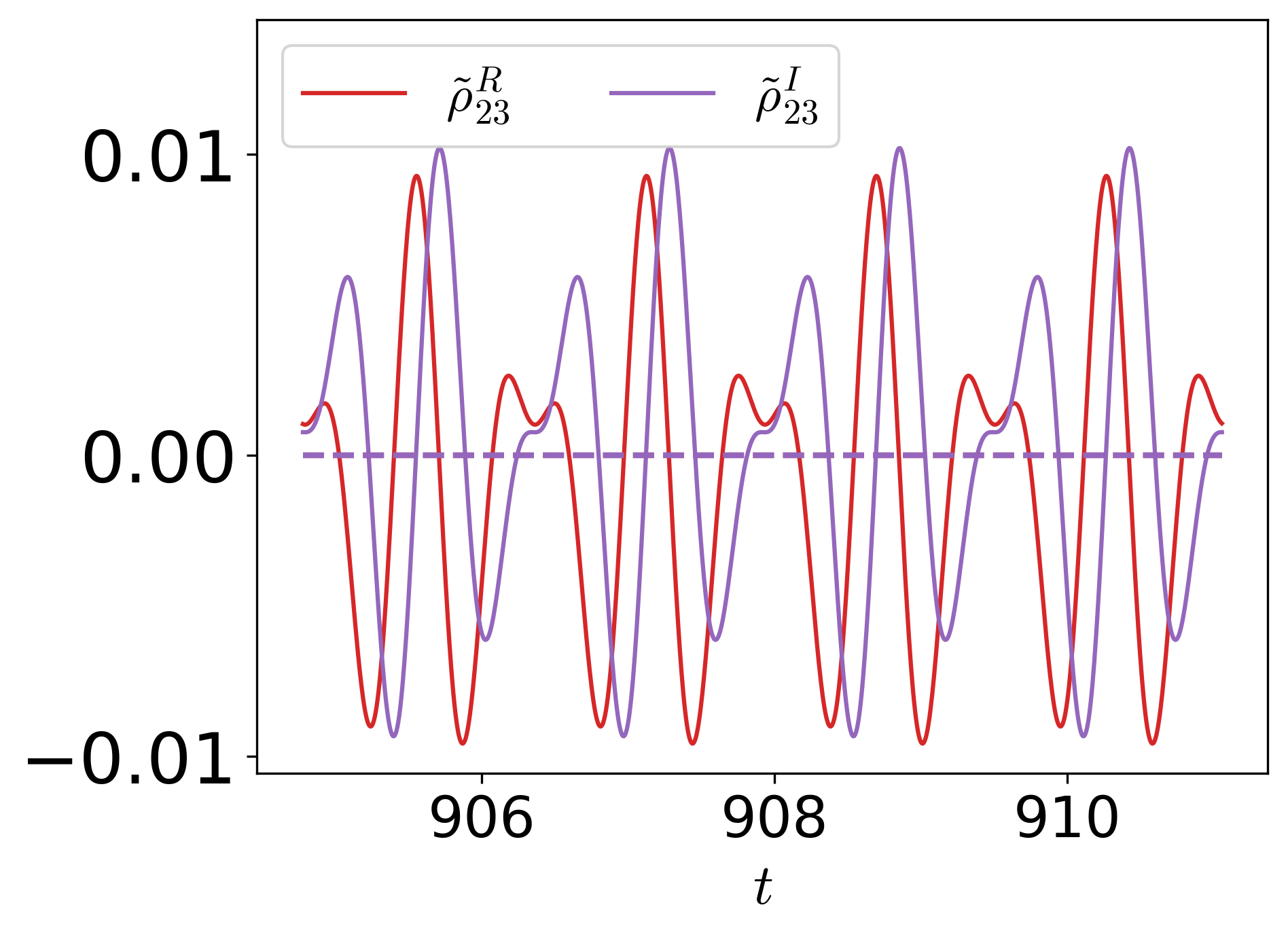}
 	\caption{Elements of the time dependent system density operator in the long time limit for the case A-II in Table I of the main text, an open system non-unitary dynamics, with full Hamiltonian (solid) and the RWA(dashed), for which the Hamiltonians are respectively $\hat H(t)$ and $\hat H^{\rm \small RWA}(t)$.  Both were calculated using the 4th order ME-propagator with commutator, Eq. (11) in the main text. }
 	\label{fig:open_both_steadystate_B}
 \end{figure}

 \begin{figure}
 	(a)\makebox[0.8\columnwidth]{ }\\
 	\includegraphics[width=0.4\columnwidth]{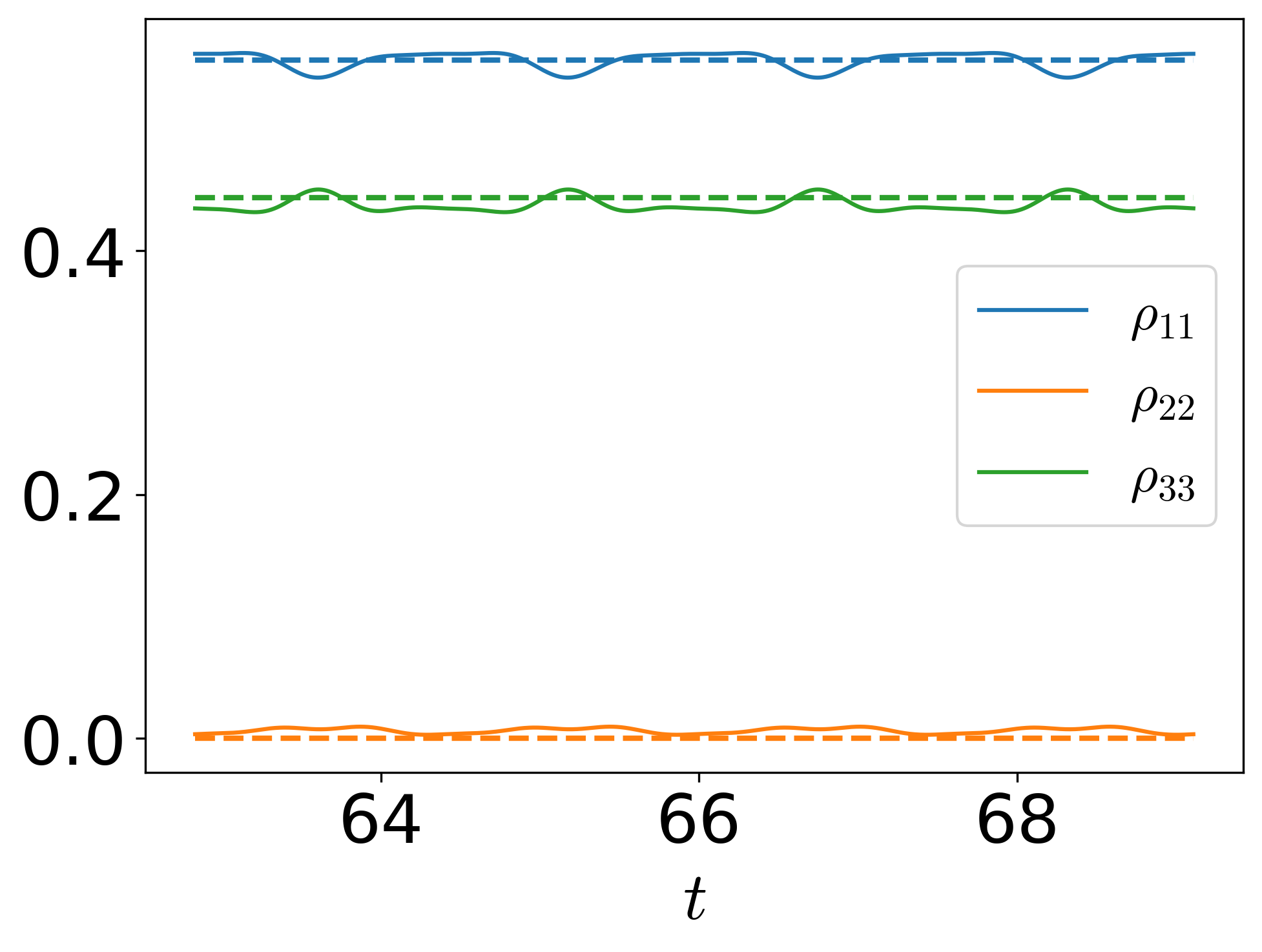} \\
 	(b)\makebox[0.8\columnwidth]{ }\\
 	\includegraphics[width=0.4\columnwidth]{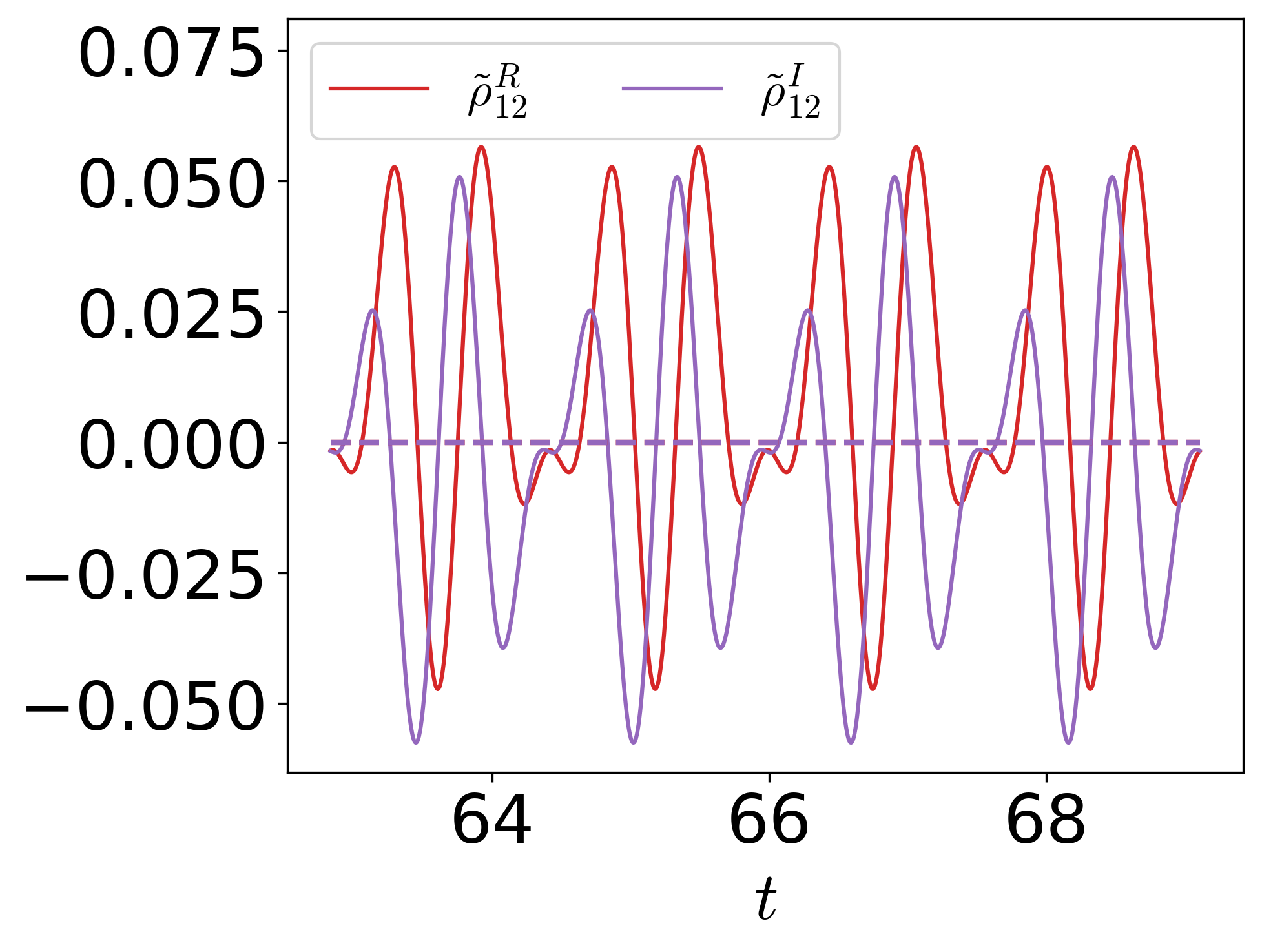} \\
 	(c)\makebox[0.8\columnwidth]{ }\\
 	\includegraphics[width=0.4\columnwidth]{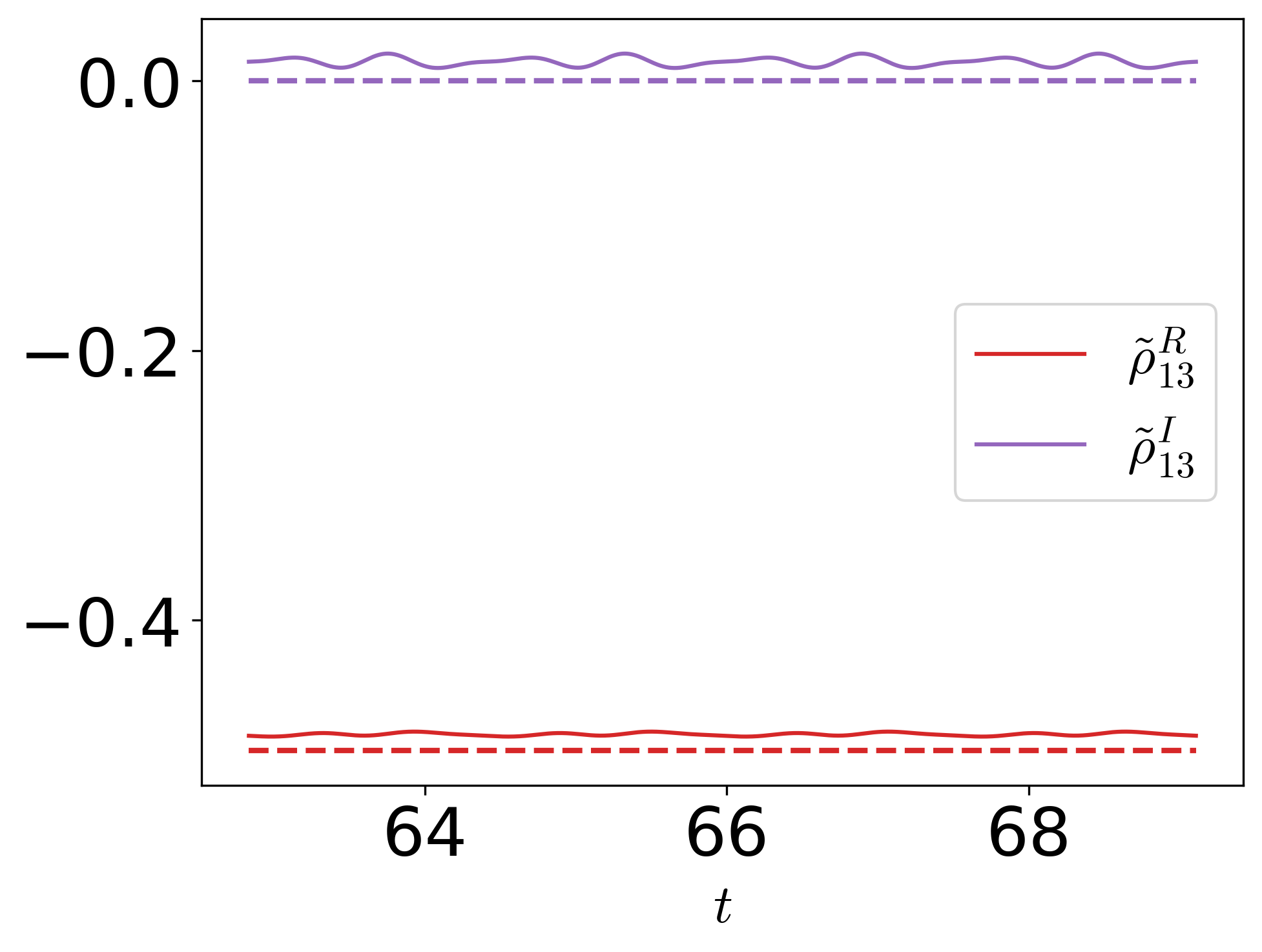}\\
 	(d)\makebox[0.8\columnwidth]{ }\\
 	\includegraphics[width=0.4\columnwidth]{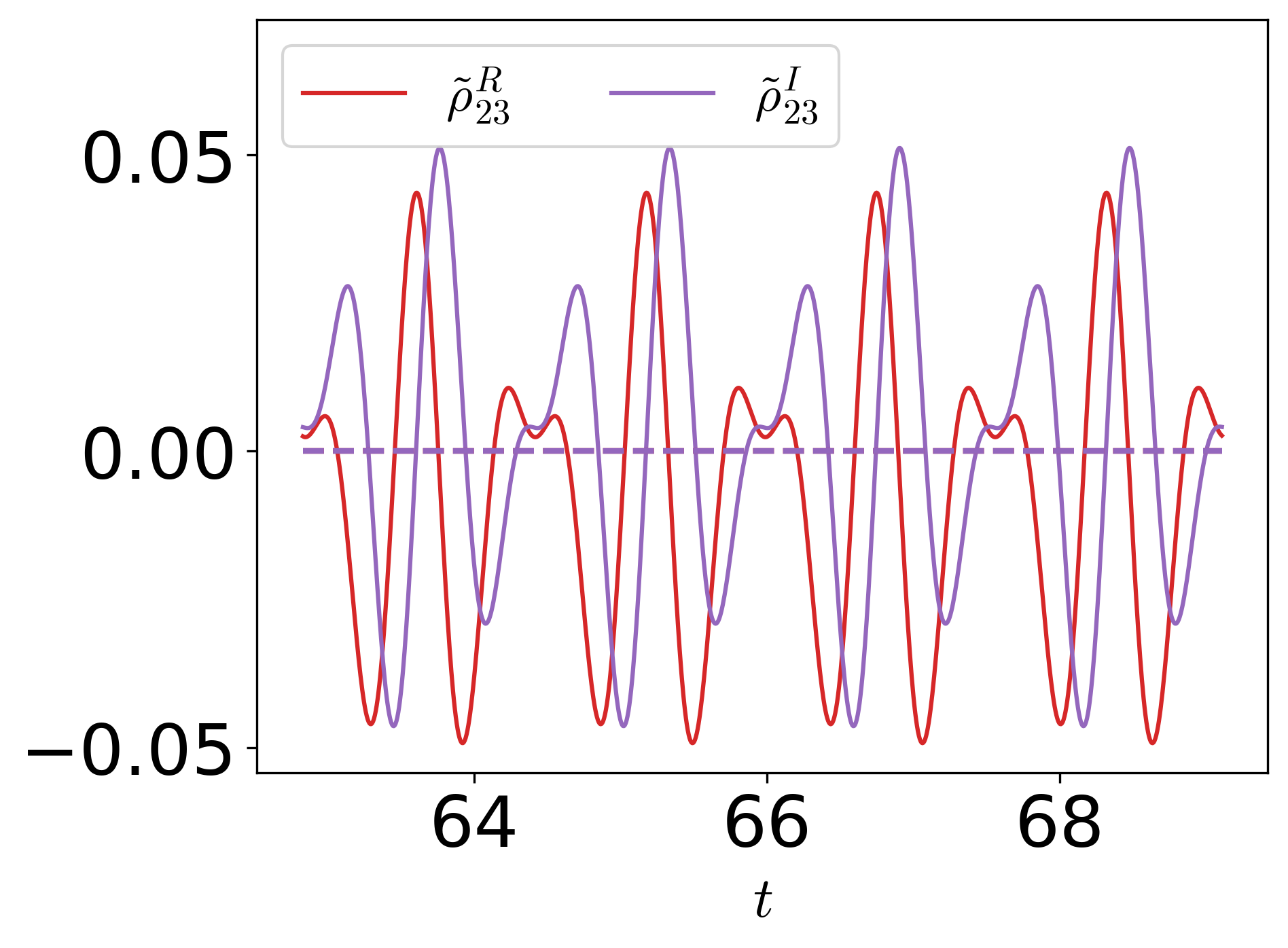}
 	\caption{Elements of the time dependent system density operator in the long time limit for the case B-II in Table I of the main text, an open system non-unitary dynamics, with full Hamiltonian (solid) and the RWA(dashed), for which the Hamiltonians are respectively $\hat H(t)$ and $\hat H^{\rm \small RWA}(t)$.  Both were calculated using the 4th order ME-propagator with commutator, Eq. (11) in the main text.}
 	\label{fig:open_both_steadystate}
 \end{figure}
 
   \begin{figure}
 	
 	(a)\makebox[0.8\columnwidth]{ }\\
 	\includegraphics[width=0.4\columnwidth]{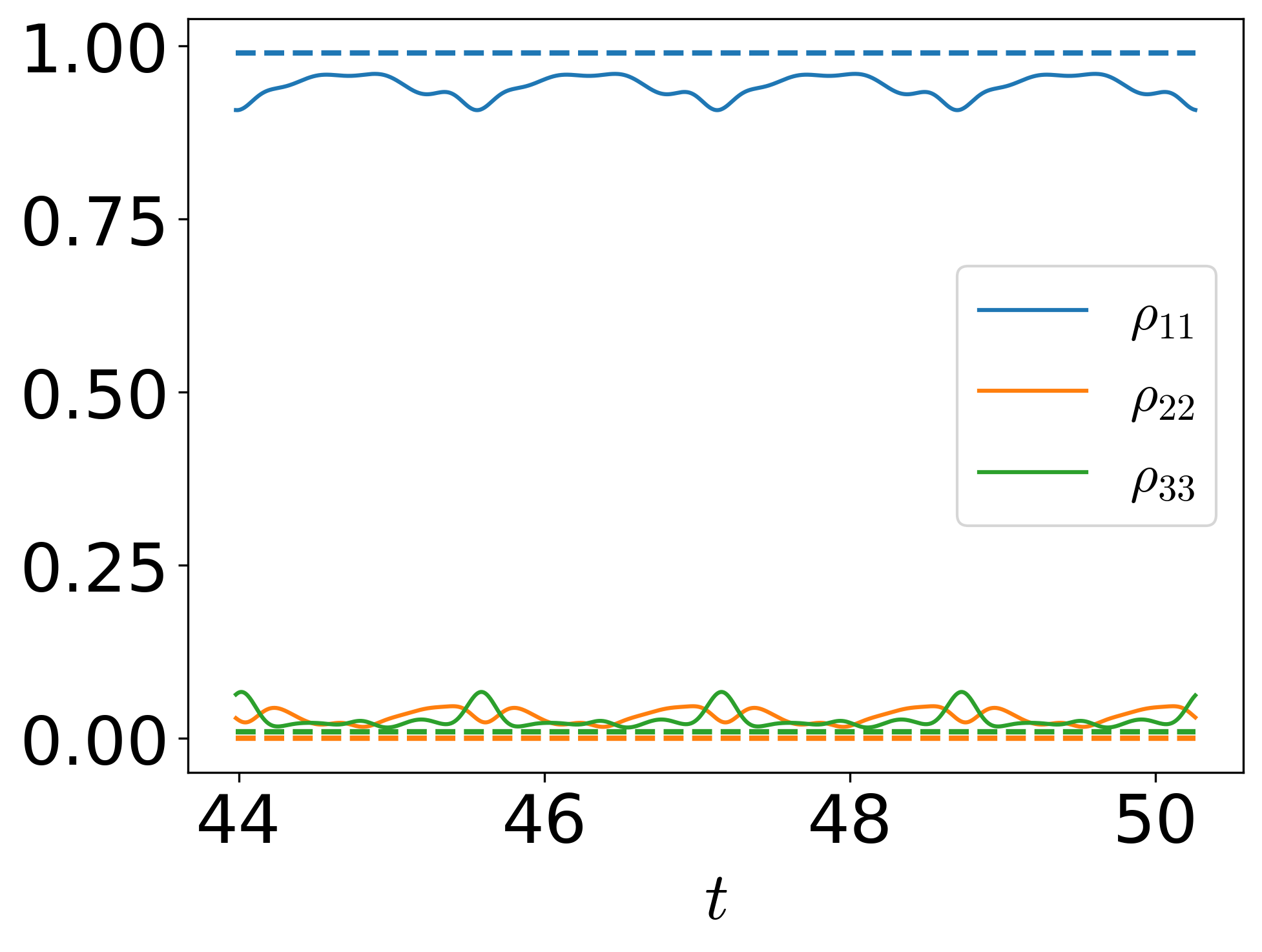} \\
 	(b)\makebox[0.8\columnwidth]{ }\\
 	\includegraphics[width=0.4\columnwidth]{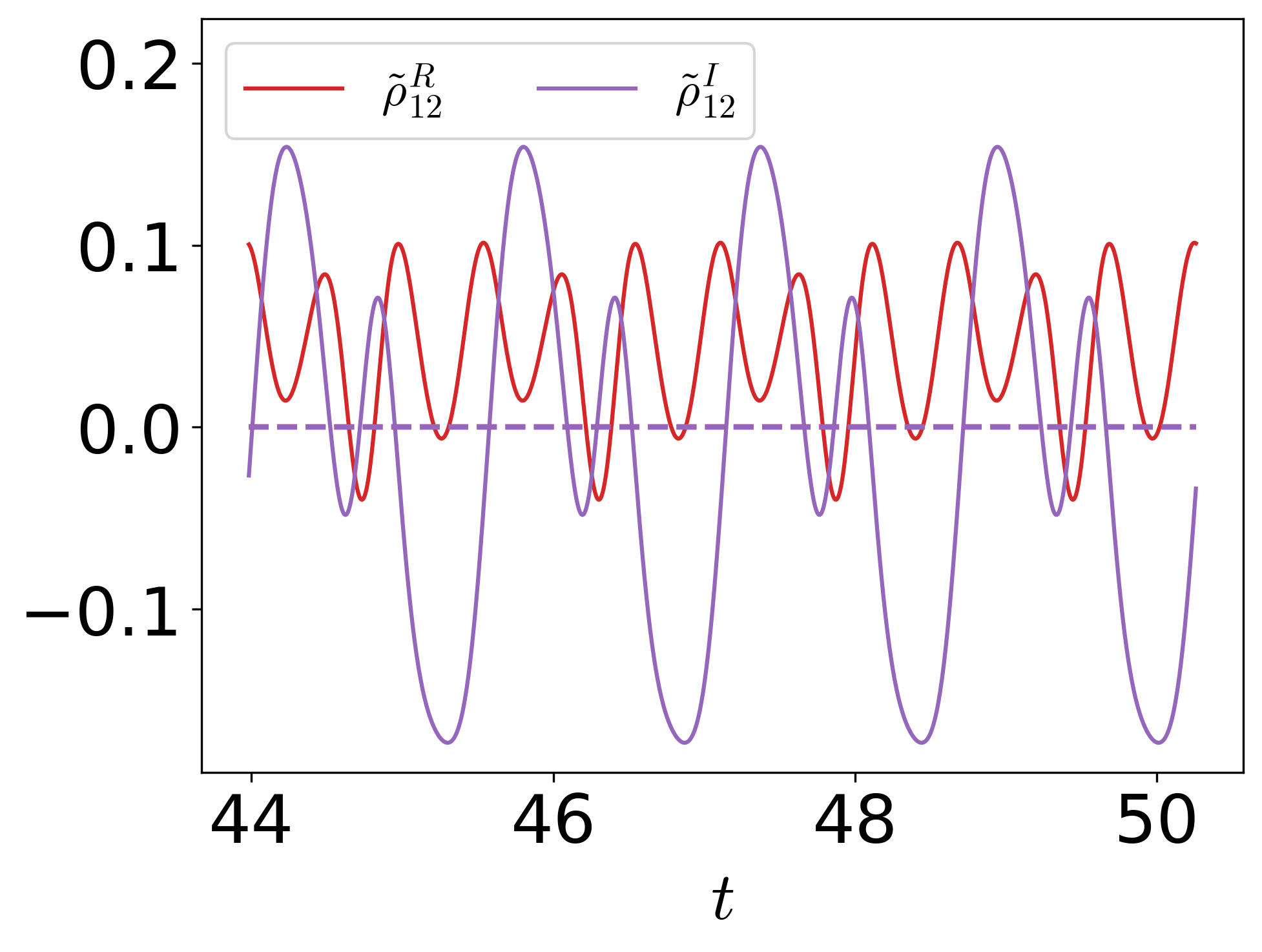} \\
 	(c)\makebox[0.8\columnwidth]{ }\\
 	\includegraphics[width=0.4\columnwidth]{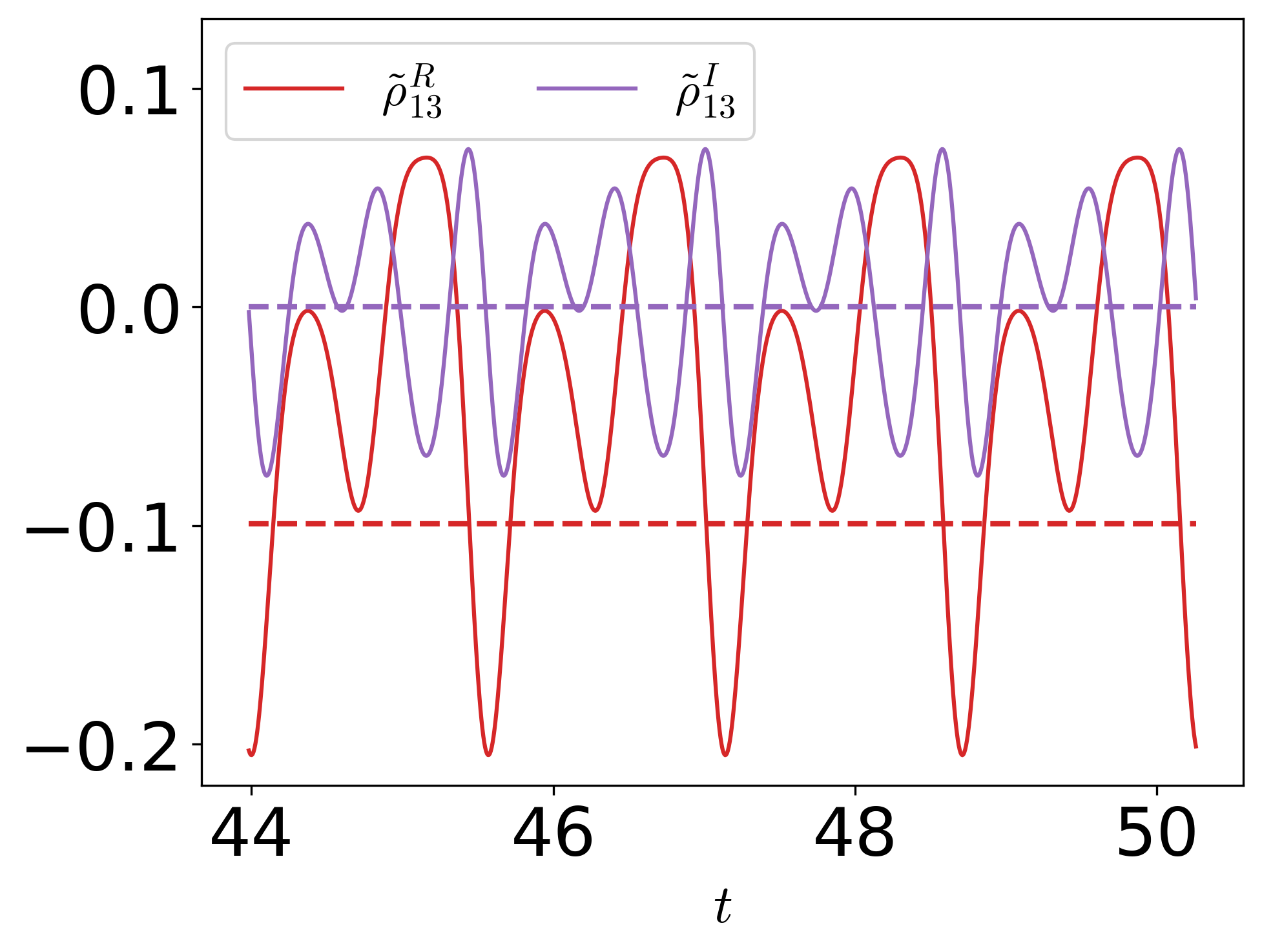}\\
 	(d)\makebox[0.8\columnwidth]{ }\\
 	\includegraphics[width=0.4\columnwidth]{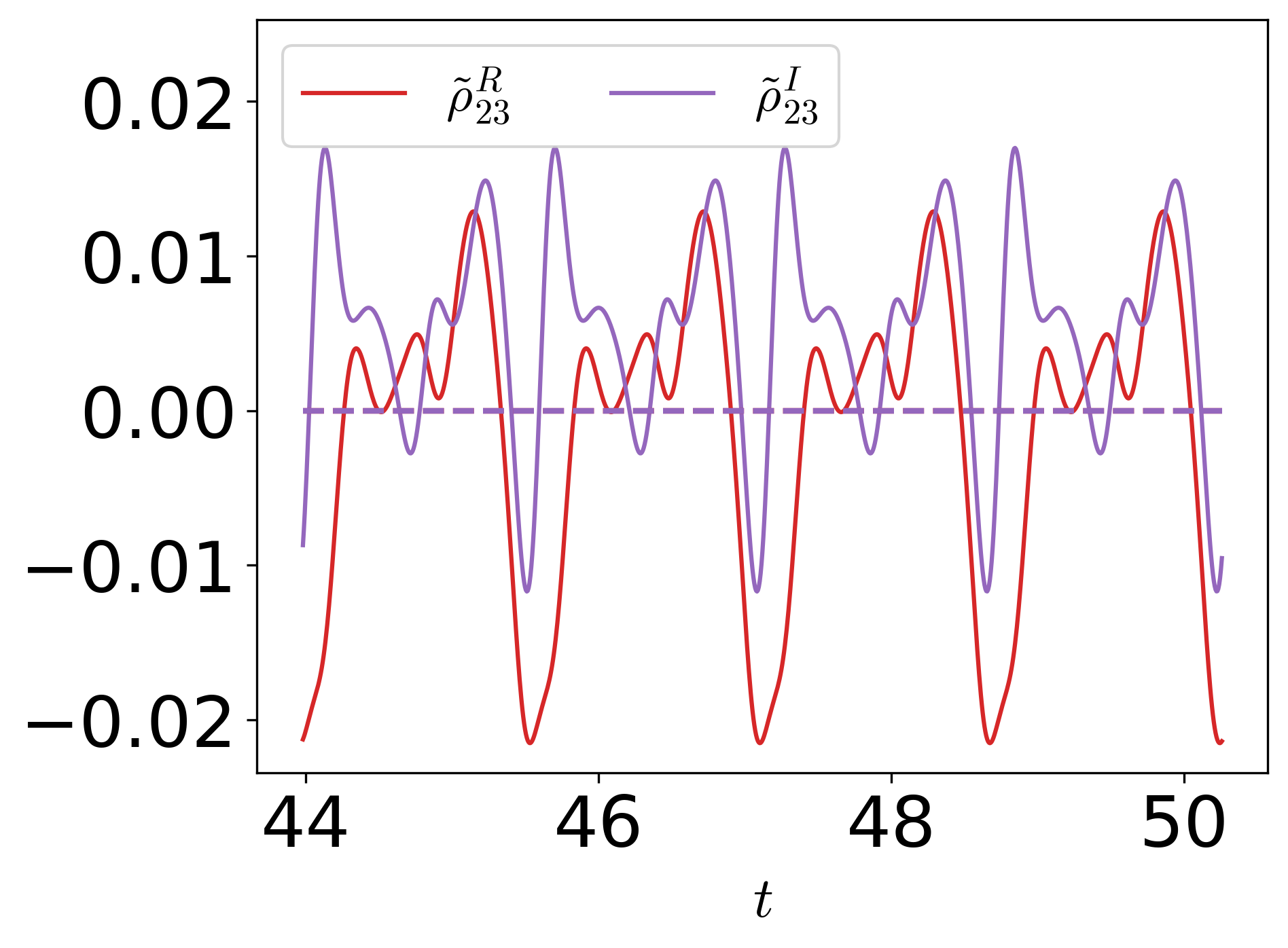}
 	\caption{Elements of the time dependent system density operator in the long time limit for the case C-II in Table I of the main text, an open system non-unitary dynamics, with full Hamiltonian (solid) and the RWA(dashed), for which the Hamiltonians are respectively $\hat H(t)$ and $\hat H^{\rm \small RWA}(t)$.  Both were calculated using the 4th order ME-propagator with commutator, Eq. (11) in the main text. }
 	\label{fig:open_both_steadystate_C}
 \end{figure}

\begin{figure}
	\centering
	(a)\makebox[0.4\columnwidth]{ }(b)\makebox[0.4\columnwidth]{ }\\
	\includegraphics[width=0.4\columnwidth]{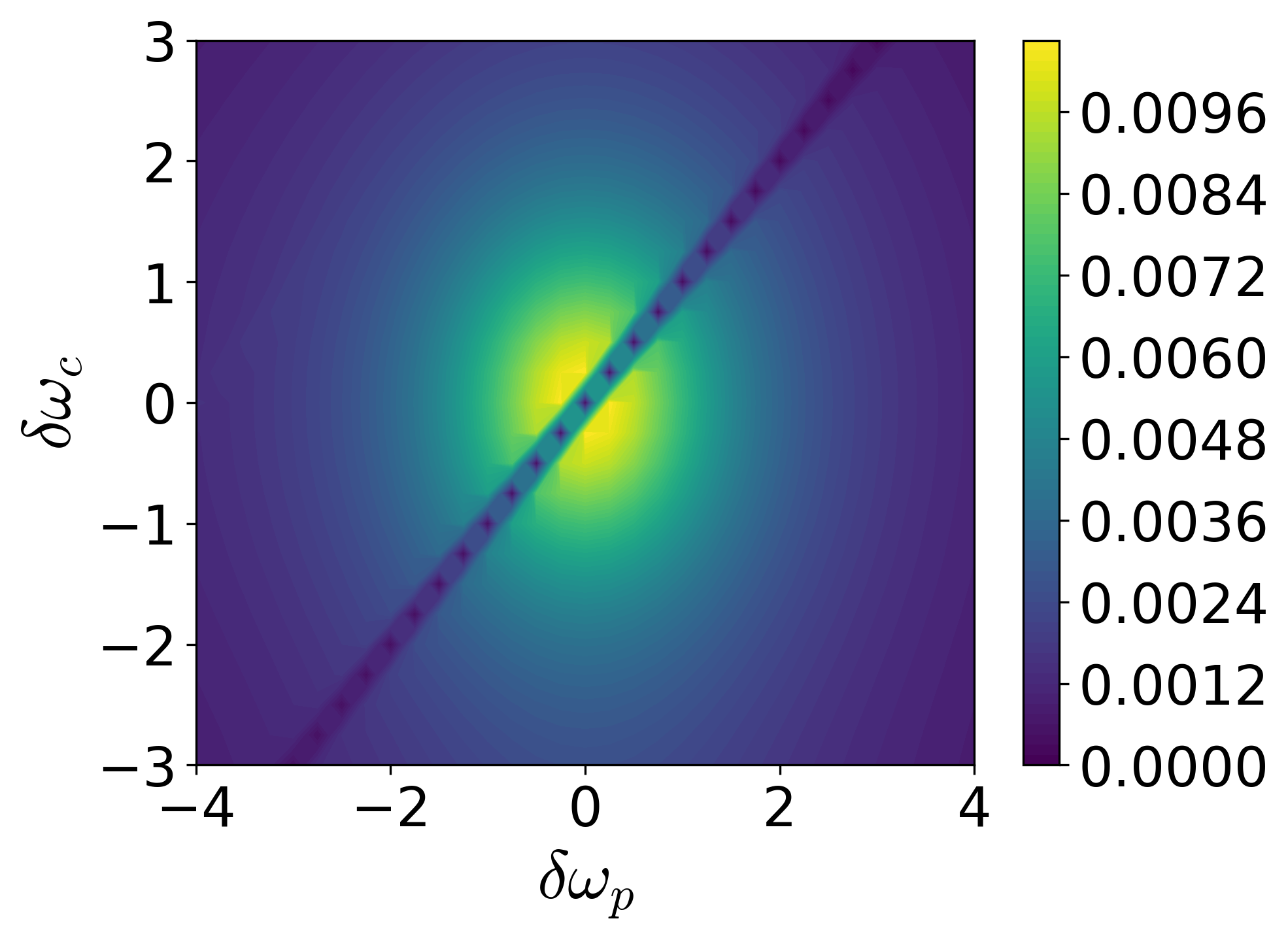}\vspace{.2in}\includegraphics[width=0.4\columnwidth]{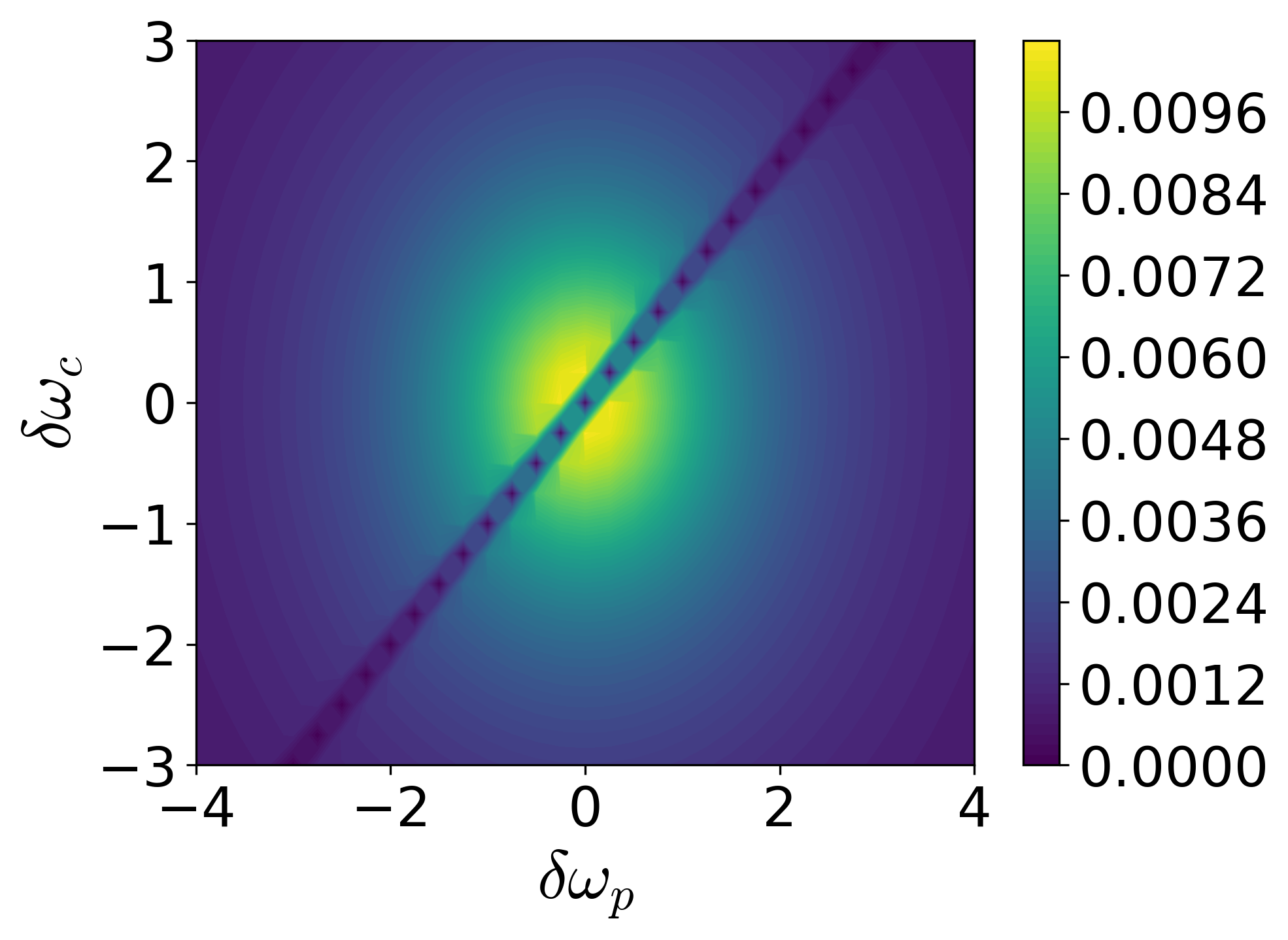} \hspace{.1in}\\
	(c)\makebox[0.4\columnwidth]{ }(d)\makebox[0.4\columnwidth]{ }\\
	\includegraphics[width=0.4\columnwidth]{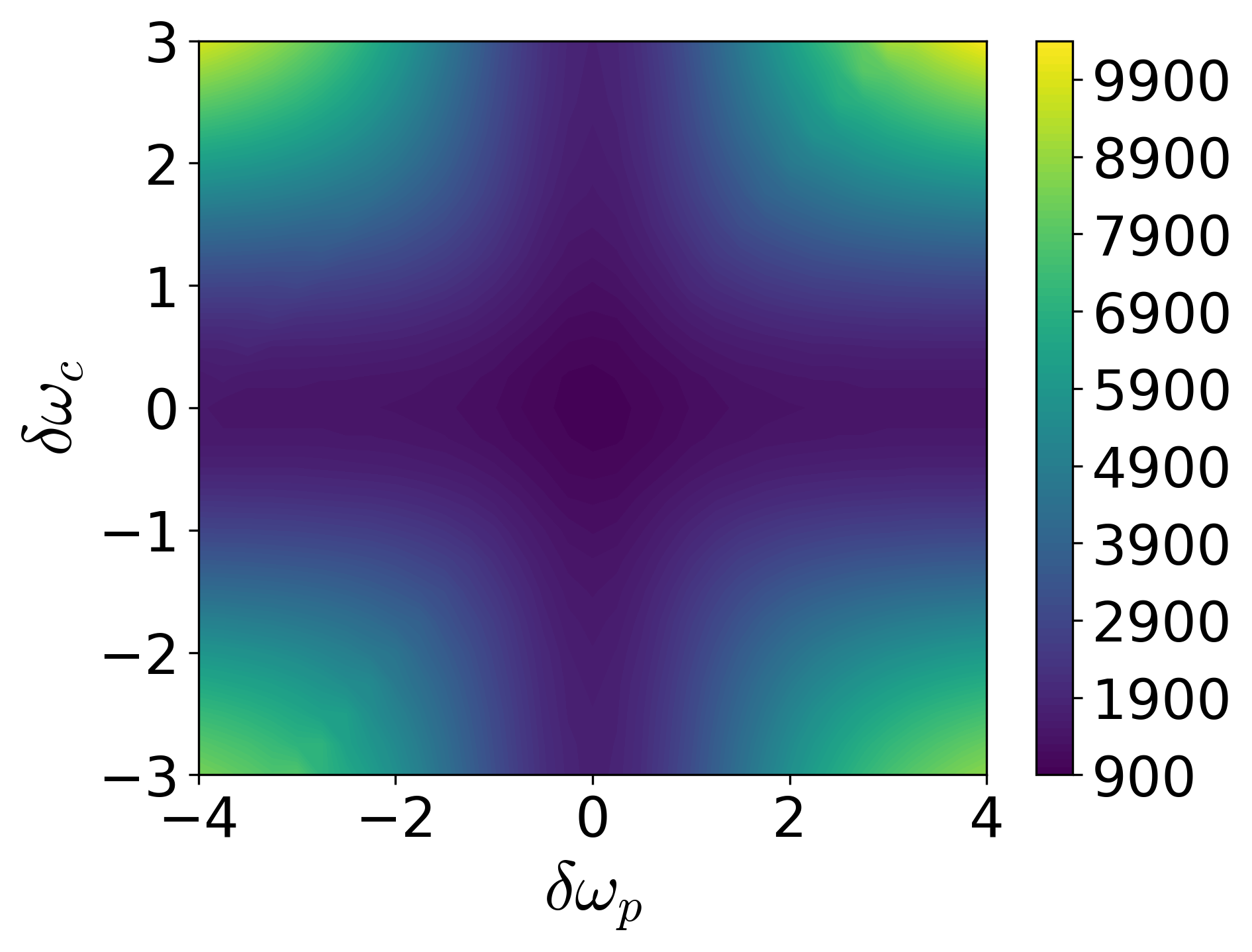}\vspace{.2in}\includegraphics[width=0.4\columnwidth]{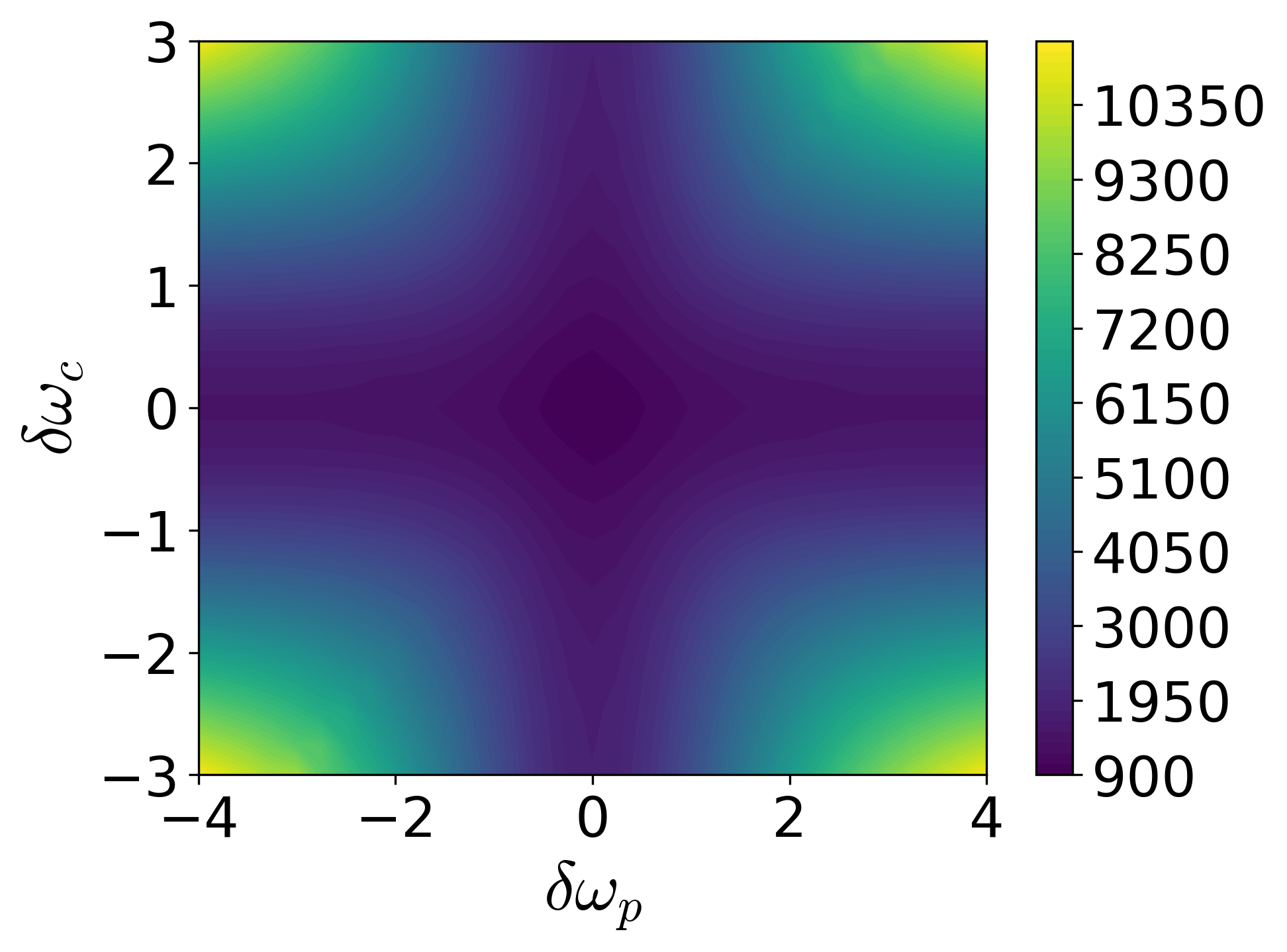} 
	\caption{Steady state limits of populations at state 2 ($\rho_{22}$) for the open system of case A-II with full Hamiltonian (a) and RWA (b), for which  times required for convergence are respectively plotted in (c) and (d).}
	\label{fig:3D_B}
\end{figure}

\begin{figure}
	\centering
	\includegraphics[width=0.85\columnwidth]{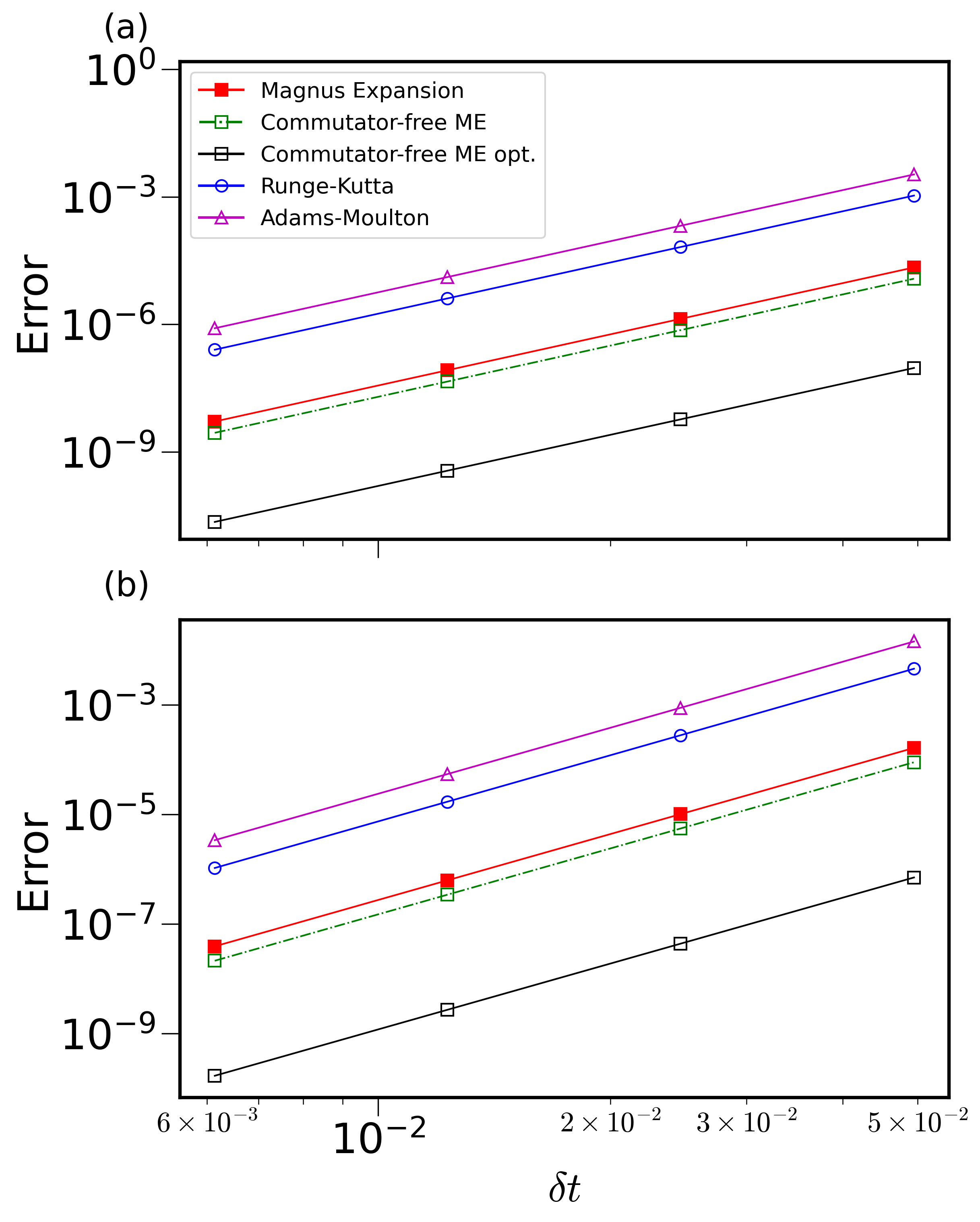}\\
	\caption{Comparison of fourth-order Magnus Expansion, Runge-Kutta, and Adams-Moulton methods for the closed-system dynamics of case A-I (a) for short time evolution ($t = 2\pi$) and (b) for long time evolution $16\pi$. The error was calculated using Eq. (19) in the main text. } \label{fig:A_closed_error}
\end{figure}
\begin{figure}
	\centering
	\includegraphics[width=0.85\columnwidth]{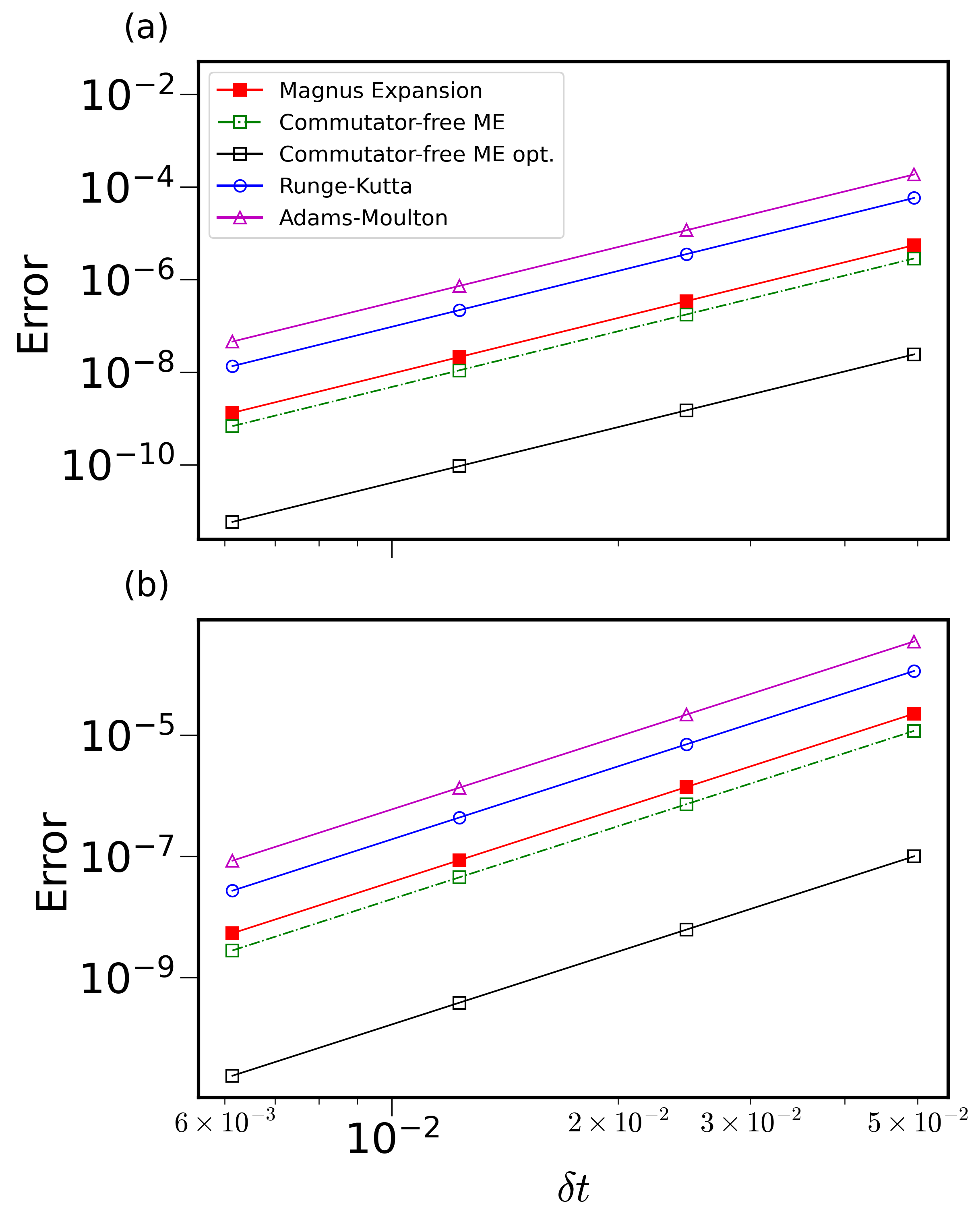}\\
	\caption{Comparison of fourth-order Magnus Expansion, Runge-Kutta, and Adams-Moulton methods for the open-system dynamics of case A-II (a) for short time evolution ($t = 2\pi$) and (b) for long time evolution $16\pi$. The error was calculated using Eq. (19) in the main text. } \label{fig:A_open_error}
\end{figure}

\begin{figure}
	\centering
	\includegraphics[width=0.85\columnwidth]{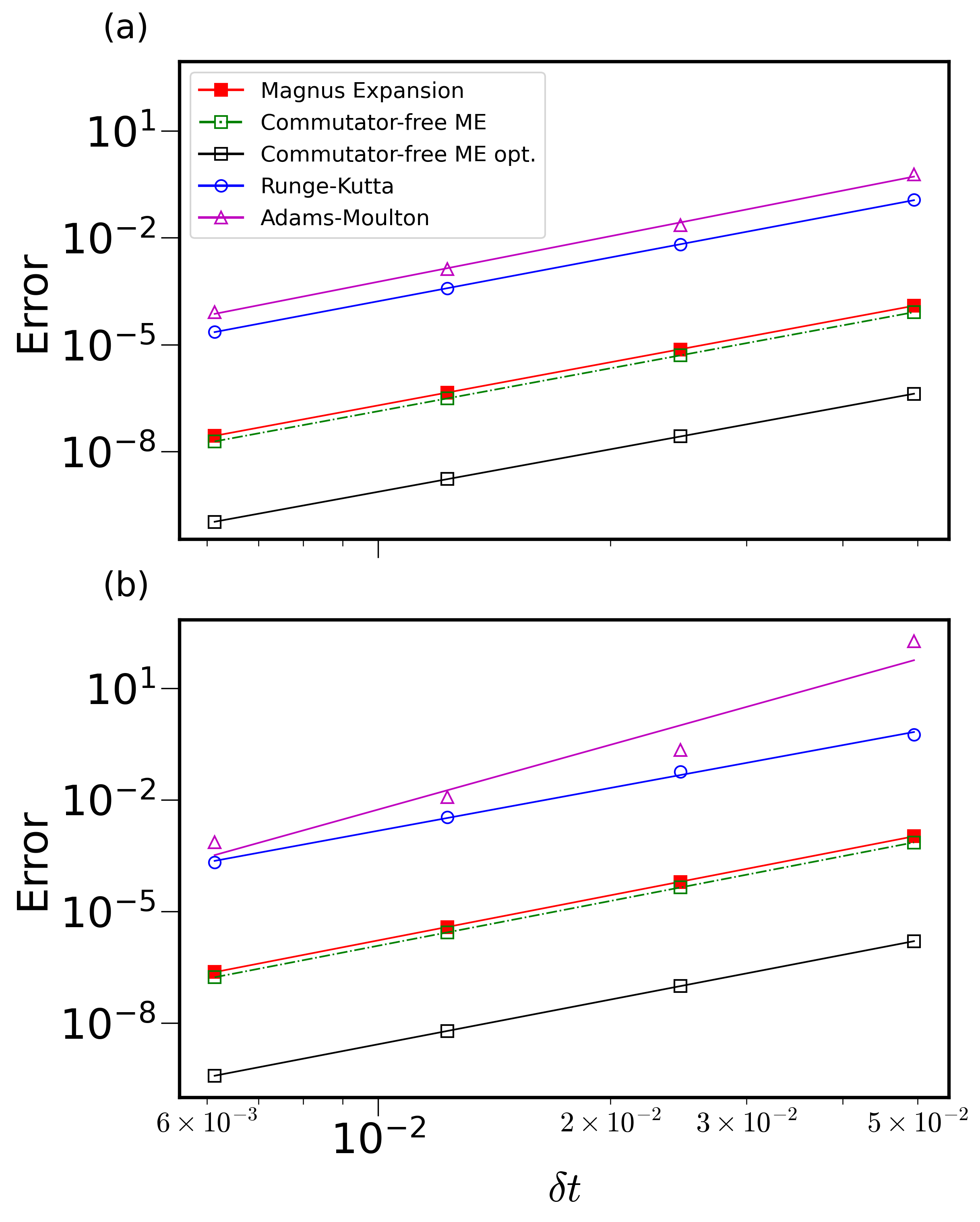}\\
	\caption{Comparison of fourth-order Magnus Expansion, Runge-Kutta, and Adams-Moulton methods for the closed-system dynamics of case C-I (a) for short time evolution ($t = 2\pi$) and (b) for long time evolution $16\pi$. The error was calculated using Eq. (19) in the main text. } \label{fig:C_closed_error}
\end{figure}
\begin{figure}
	\centering
	\includegraphics[width=0.85\columnwidth]{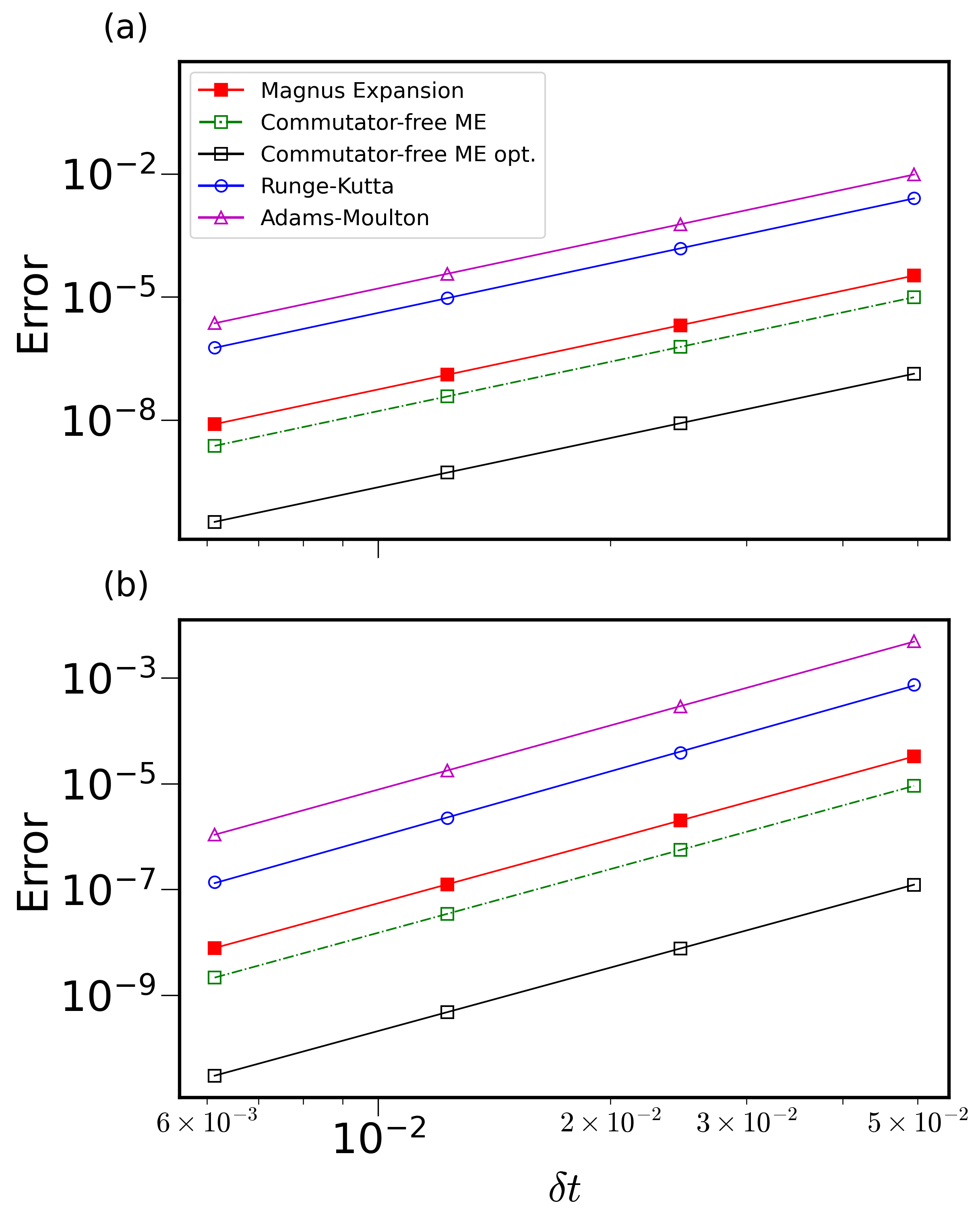}
	\caption{Comparison of fourth-order Magnus Expansion, Runge-Kutta, and Adams-Moulton methods for the open-system dynamics of case C-II (a) for short time evolution ($t = 2\pi$) and (b) for long time evolution $16\pi$. The error was calculated using Eq. (19) in the main text. } \label{fig:C_open_error}
\end{figure}

\begin{figure}
	\includegraphics{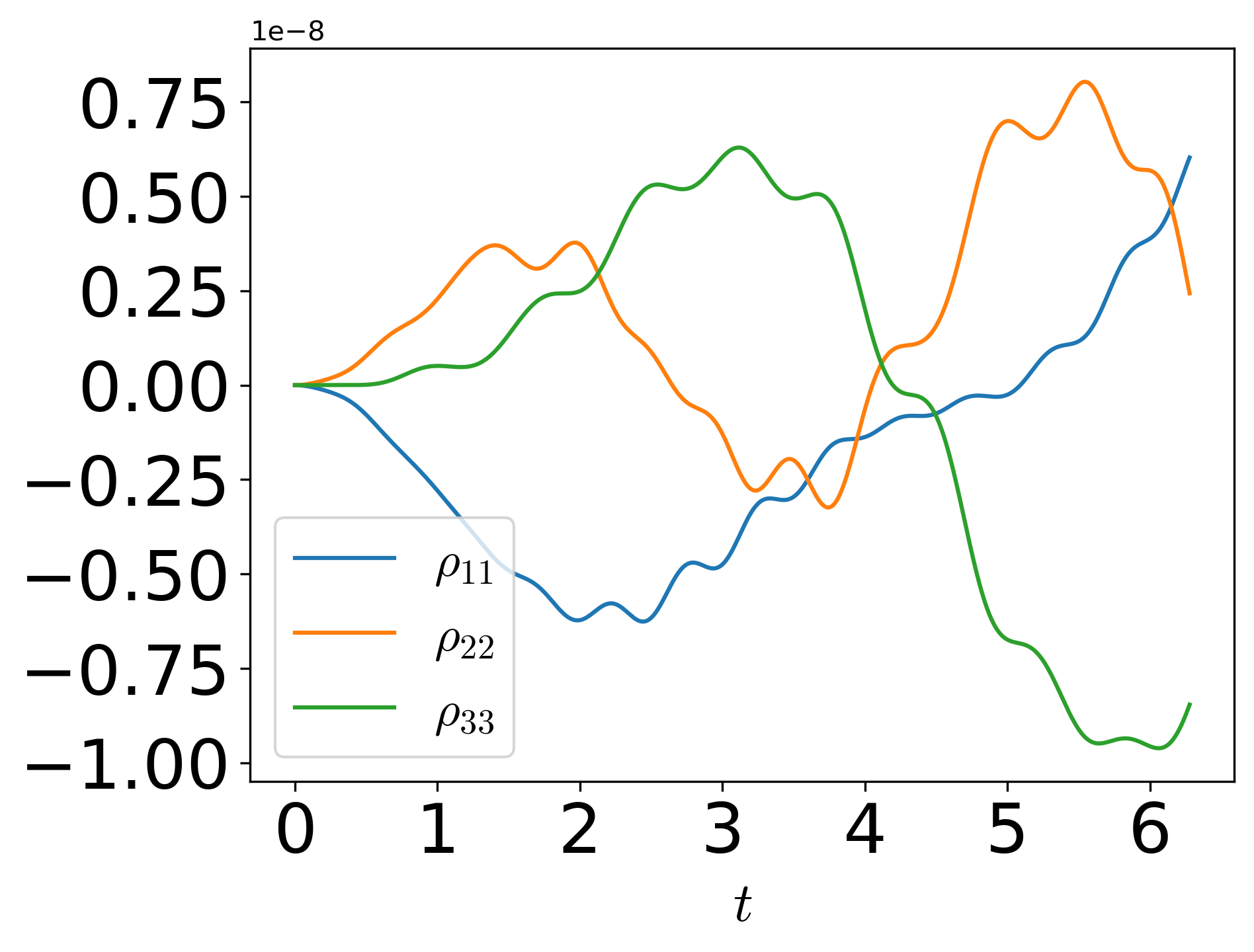}
	\caption{Plot of error (given by sixth-order - fourth-order) for the case B-I of the Table I, a unitary closed system dynamics, in the main text.}
\end{figure}

\begin{figure}
	\includegraphics{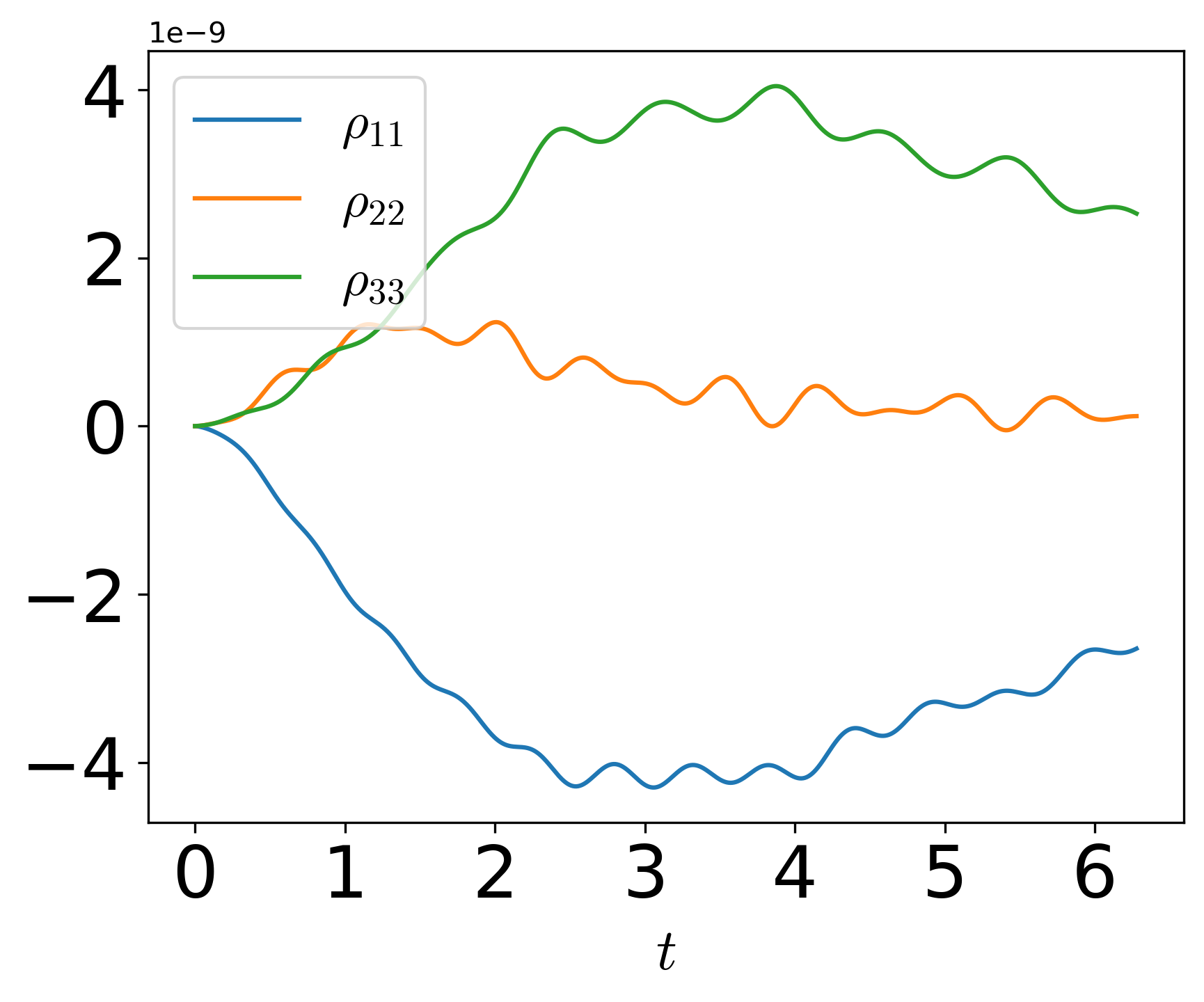}
		\caption{Plot of error (given by sixth-order - fourth-order) for the case B-II of the Table I, a nonunitary open system dynamics, in the main text.}
\end{figure}

\begin{figure}
	\includegraphics{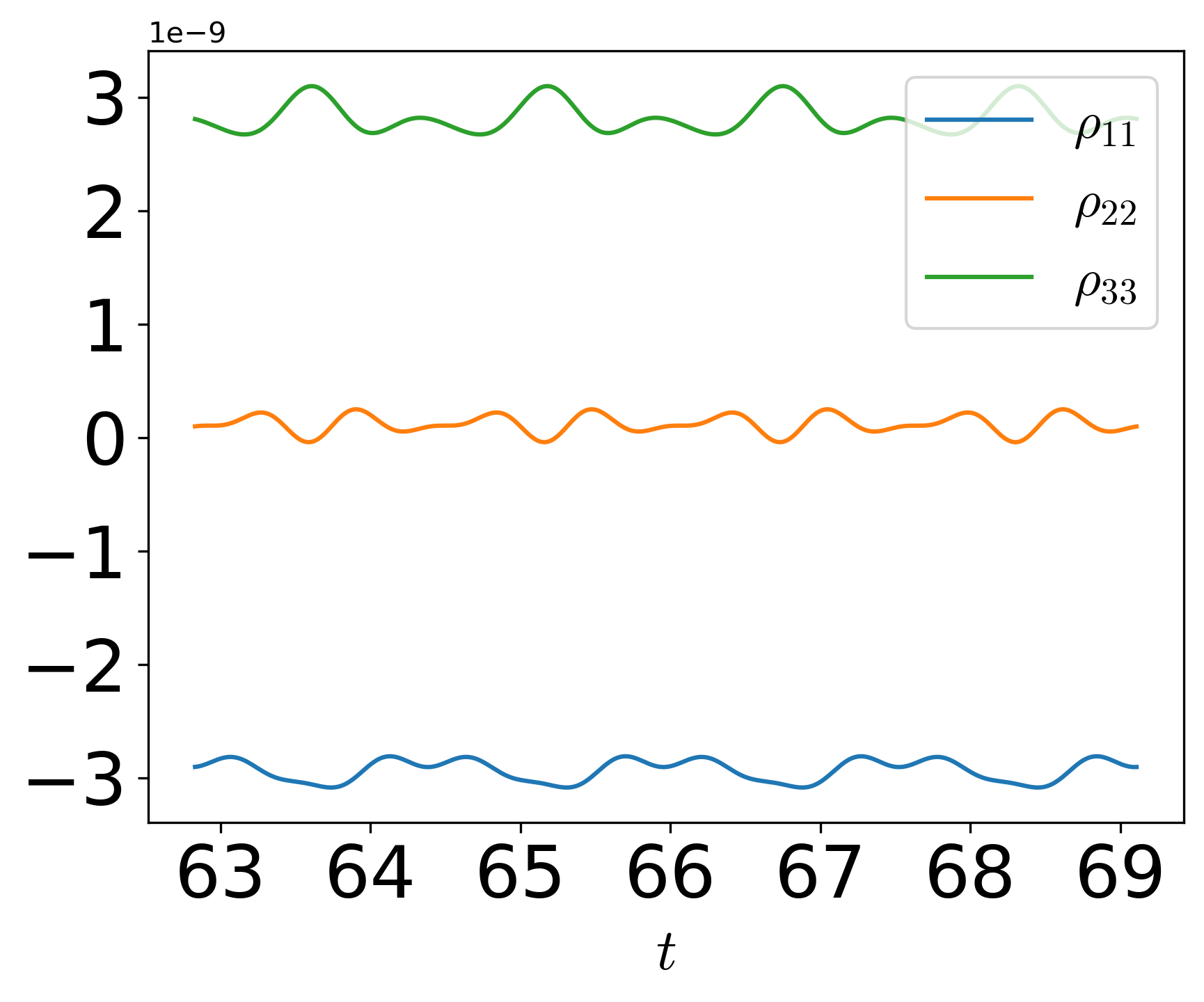}
	\caption{Plot of error (given by sixth-order - fourth-order) for the case B-II of Table I, a nonunitary open system dynamics, in the steady state limit.}
\end{figure}

\end{widetext}

\end{document}